\documentclass[twocolumn]{aastex631}
\usepackage{graphicx}
\usepackage{multirow}

\usepackage{placeins}
\usepackage{amsmath}
\newcommand{\degree}{$^{\circ}$}
\newcommand{\kms}{km s$^{-1}$}

\newcommand{\cisotope}{$^{12}$C/$^{13}$C}
\newcommand{\logeps}[1]{$\log \epsilon$(#1)}
\newcommand{\feh}{[Fe/H]}
\newcommand{\xh}[1]{[#1/H]}
\newcommand{\xm}[1]{[#1/M]}
\newcommand{\xfe}[1]{[#1/Fe]}
\newcommand{\wln}{\texttt{wln}}
\newcommand{\chitwo}{\texttt{chi2}}
\newcommand{\eqw}{\texttt{eqw}}
\newcommand{\syn}{\texttt{syn}}
\newcommand{\intmethod}{\texttt{int}}

\begin{document}

\title{BACCHUS Analysis of Weak Lines in APOGEE Spectra (BAWLAS)}

\author[0000-0003-2969-2445]{Christian R. Hayes}
\affiliation{NRC Herzberg Astronomy and Astrophysics Research Centre, 5071 West Saanich Road, Victoria, B.C., Canada, V9E 2E7}

\author[0000-0002-6939-0831]{Thomas Masseron}
\affiliation{Instituto de Astrof\'{i}sica de Canarias, 38205 La Laguna, Tenerife, Spain}
\affiliation{Departamento de Astrof{\'i}sica, Universidad de La Laguna, E-38206 La Laguna, Tenerife, Spain}

\author[0000-0002-4989-0353]{Jennifer Sobeck}
\affiliation{Department of Astronomy, University of Washington, Box 351580, Seattle, WA 98195, USA}

\author[0000-0002-1693-2721]{D. A. Garc\'{i}a-Hern\'{a}ndez}
\affiliation{Instituto de Astrof\'{i}sica de Canarias, 38205 La Laguna, Tenerife, Spain}
\affiliation{Departamento de Astrof{\'i}sica, Universidad de La Laguna, E-38206 La Laguna, Tenerife, Spain}

\author{Carlos Allende Prieto}
\affiliation{Instituto de Astrof\'{i}sica de Canarias, 38205 La Laguna, Tenerife, Spain}
\affiliation{Departamento de Astrof{\'i}sica, Universidad de La Laguna, E-38206 La Laguna, Tenerife, Spain}

\author[0000-0002-1691-8217]{Rachael L. Beaton}
\altaffiliation{Carnegie-Princeton Fellow}
\affiliation{Department of Astrophysical Sciences, Princeton University, 4 Ivy Lane, Princeton, NJ~08544}
\affiliation{The Observatories of the Carnegie Institution for Science, 813 Santa Barbara St., Pasadena, CA~91101}

\author{Katia Cunha}
\affiliation{Observat\'{o}rio Nacional, 77 Rua General Jos\'{e} Cristino, Rio de Janeiro, 20921-400, Brazil}
\affiliation{Steward Observatory, University of Arizona, 933 North Cherry Avenue, Tucson, AZ 85721, USA}

\author{Sten Hasselquist}
\affiliation{Space Telescope Science Institute, 3700 San Martin Drive, Baltimore, MD 21218, USA}

\author[0000-0002-9771-9622]{Jon A. Holtzman}
\affiliation{New Mexico State University, Las Cruces, NM 88003, USA}

\author[0000-0002-4912-8609]{Henrik J\"{o}nsson}
\affiliation{Materials Science and Applied Mathematics, Malm\"{o} University, SE-205 06 Malm\"{o}, Sweden}

\author{Steven~R.~Majewski} 
\affiliation{Department of Astronomy, University of Virginia, Charlottesville, VA, 22903, USA}

\author[0000-0003-0509-2656]{Matthew Shetrone}
\affiliation{University of California Observatories, University of California Santa Cruz, Santa Cruz, CA, 95064}

\author[0000-0002-0134-2024]{Verne V. Smith}
\affiliation{NSF's National Optical-Infrared Astronomy Research Laboratory, 950 North Cherry Avenue, Tucson, AZ 85719, USA}

\author{Andr\'es~Almeida}
\affiliation{Department of Astronomy, University of Virginia, Charlottesville, VA, 22903, USA}

\email{Christian.Hayes@nrc-cnrc.gc.ca}
\email{chrishayesastro@gmail.com}

\begin{abstract}

Elements with weak and blended spectral features in stellar spectra are challenging to measure and require specialized analysis methods to precisely measure their chemical abundances.  In this work, we have created a catalog of approximately 120,000 giants with high signal-to-noise APOGEE DR17 spectra, for which we explore weak and blended species to measure Na, P, S, V, Cu, Ce, and Nd abundances and $^{12}$C/$^{13}$C isotopic ratios.  We employ an updated version of the BACCHUS (Brussels Automatic Code for Characterizing High accUracy Spectra) code to derive these abundances using the stellar parameters measured by APOGEE's DR17 ASPCAP pipeline, quality flagging to identify suspect spectral lines, and a prescription for upper limits.  Combined these allow us to provide our BACCHUS Analysis of Weak Lines in APOGEE Spectra (BAWLAS) catalog of precise chemical abundances for these weak and blended species that agrees well with literature and improves upon APOGEE abundances for these elements, some of which are unable to be measured with APOGEE's current, grid-based approach without computationally expensive expansions.  This new catalog can be used alongside APOGEE and provide measurements for many scientific applications ranging from nuclear physics to Galactic chemical evolution and Milky Way population studies.  To illustrate this we show some examples of uses for this catalog, such as, showing that we observe stars with enhanced s-process abundances or that we can use the our \cisotope{} ratios to explore extra mixing along the red giant branch.

\end{abstract}


\section{Introduction}

Currently, there are several spectroscopic surveys that have undertaken the difficult task of measuring stellar parameters and abundances for large samples of stars spanning a wide range in stellar parameters, such as the Apache Point Observatory Galactic Evolution Experiment \citep[APOGEE;][]{apogee}, Galactic Archaeology with HERMES \citep[GALAH;][]{galah}, or Large sky Area Multi-Object fiber Spectroscopic Telescope survey \citep[LAMOST;][]{lamost}.  Despite the difficulty of doing so, these surveys have had great success in producing large high-quality samples of stellar parameters and chemical abundances with which to study the large-scale properties of the Milky Way (MW), its stellar populations, and even the stellar populations of MW satellites. 

One limitation of such surveys, however, is that analyzing such large samples is time consuming and resource intensive.  To provide high quality data for large samples, some simplifications or compromises must be made.  This could be in the form of limiting the dimensionality of synthetic grids, synthesizing only one or a few elements at a time, among others.  While these simplifications are often necessary, they typically make it difficult to measure those elements with weak and/or significantly blended lines , where detailed syntheses and a careful treatment of blends is critical to measuring abundances.  Here we perform a careful analysis of APOGEE spectra to complement its exploration of chemical space by measuring weak and blended elements with a simultaneously flexible and detailed methodology.

APOGEE is a dual-hemisphere, high-resolution (R$\sim$22500), H-band (1.5$-$1.7 $\mu$m) spectroscopic survey of the Milky Way and its nearby dwarf galaxies \citep{apogee} that began in as APOGEE-1 SDSS-III \citep{sdss3} and continued into SDSS-IV \citep{sdss4} with APOGEE-2.  In the 17th data release of SDSS \citep[DR17;][]{dr17, Holtzman2021}, APOGEE has produced its final data release, which contains spectra, stellar parameters, and chemical abundances for up to 20 species. In addition to these 20 species, there are some chemical species that have features in APOGEE spectra, but were not satisfactorily measured by the APOGEE pipeline, including P, Cu, Nd, and the \cisotope{} ratio.

In this work, we provide a complementary analysis of APOGEE spectra for high S/N red giants in the APOGEE sample that expands the set of elemental abundances for these stars.  We do so by analyzing the elements and isotopic ratios that are difficult to assess with the APOGEE Stellar Parameters and Chemical Abundances Pipeline \citep[ASPCAP;][]{aspcap} methodology and are therefore not provided by APOGEE. We also reanalyze a few additional elements with weak and blended lines. 

For our analysis, we use the Brussels Automatic Code for Characterizing High accUracy Spectra \citep[BACCHUS][]{masseron2016} that measures line-by-line elemental abundances from on-the-fly spectral synthesis and provides additional quality flags for each line it analyzes (e.g., identifying when a line is severly blended, too weak to be measured, etc.).  We follow up this BACCHUS analysis with a post-processing of the line-by-line abundances to remove erroneous or suspect measurements, and combine the high quality line-by-line measurements to provide sample of chemical abundances for weak and blended elements in APOGEE spectra.  The final data set -- the BACCHUS Analysis of Weak Lines in APOGEE (BAWLAS) catalog -- can be found on the SDSS DR17 website\footnote{The table will be available at \url{https://www.sdss.org/dr17/data_access/value-added-catalogs/}}.

This paper is outlined as follows.  Section \ref{sec_data} describes the APOGEE data we use in this study.  Section \ref{sec:bacchus} provides a brief overview of the BACCHUS code and our specific implementation of it in this work.  Section \ref{sec:line_comb} explains our post-processing analysis, flagging, line combination, and error estimation.  Section \ref{sec:results} explores our derived chemical abundances, their uncertainties, trends, and presents any known caveats.  In Section \ref{sec:comparison} we compare our derived abundances with both APOGEE, for overlapping elements, and high-resolution literature measurements.  Finally in Section \ref{sec:conclusion} we give a short summary of this work.  We also provide some detailed information about our post-processing choices, flagging, and upper limit relations for each element in the Appendices.

\section{Data}
\label{sec_data}

In this work, we use the spectra and stellar parameters from the final data release of the APOGEE survey \citep{apogee} in SDSS-IV's DR17 \citep{sdss4,dr17,Holtzman2021}.  This data release includes all APOGEE spectra taken from both the Northern and Southern hemispheres using the APOGEE spectrographs \citep{wilson2019} on the SDSS 2.5-m \citep{gunn06} and the NMSU 1-m \citep{Holtzman2010} telescopes in the north, and the 2.5-m du Pont \citep{bv73} telescope in the south. Targeting for APOGEE survey is described in \citet{zas13,zas17},  \citet{Beaton2021}, and \citet{Santana2021}.  Details of the data reduction pipeline for APOGEE can be found in \citet{dln15} and \citet{Holtzman2015}, for the main survey and 1-meter data respectively, which have been updated in DR17 to use the {\texttt{Doppler}}\footnote{\url{https://github.com/dnidever/doppler}} code for radial velocity determination \citep{Holtzman2021}.  

APOGEE's stellar parameters and chemical abundances are derived from ASPCAP \citep{aspcap}, based on the {\texttt{FERRE}}\footnote{\url{https://github.com/callendeprieto/ferre}} code \citep{AllendePrieto2006}.  As in prior APOGEE data releases, ASPCAP uses a grid of MARCS stellar atmospheres \citep{marcs,Jonsson2020}, and an H-band line list from \citet{Smith2021} that updates the earlier APOGEE line list presented in \citet{shetrone2015}, which includes the Ce and Nd line identifications from \citet{Cunha2017} and \citet{Hasselquist2016}, respectively.  These atmospheres and the line list are then used to generate a grid of synthetic spectra \citep{zamora2015} using the Synspec code \citep{Hubeny2011} and nLTE calculations for Na, Mg, Ca, and K from \citet{Osorio2020} that are fit to the observed spectra to determine stellar parameters and chemical abundances.

From the full APOGEE survey, we use a sample of 126,362 spectra that we analyze using BACCHUS as described in Section \ref{sec:bacchus}.  To build this sample, we mainly select high signal-to-noise APOGEE spectra, S/N $ > 150$ per pixel of giants with calibrated ASPCAP stellar parameters between $3500 \ {\rm K} < T_{\rm eff} < 5000$ K and $\log g < 3.5$.  We also filter these stars so that we select the ones with the high quality spectra according to their APOGEE {\texttt{STARFLAG}} and {\texttt{ANDFLAG}} flags\footnote{A description of these flags can be found in the online SDSS DR17 bitmask documentation (\url{http://www.sdss.org/dr17/algorithms/bitmasks/})} by selecting stars with {\texttt{STARFLAG}} = 0 and {\texttt{ANDFLAG}} = 0.  We specifically select these stars using the  \texttt{allStar-dr17-synspec.fits}\footnote{details of the different \texttt{allStar} versions can be found in the following links: \url{https://www.sdss.org/dr17/irspec/apogee-libraries/\#Spectralgrids} and \url{https://www.sdss.org/dr17/irspec/caveats/\#wronglsf}} version of APOGEE data.

\begin{figure}
  \centering
  \includegraphics[scale=0.4,trim = 0.in 0.in 0.in 0.in, clip]{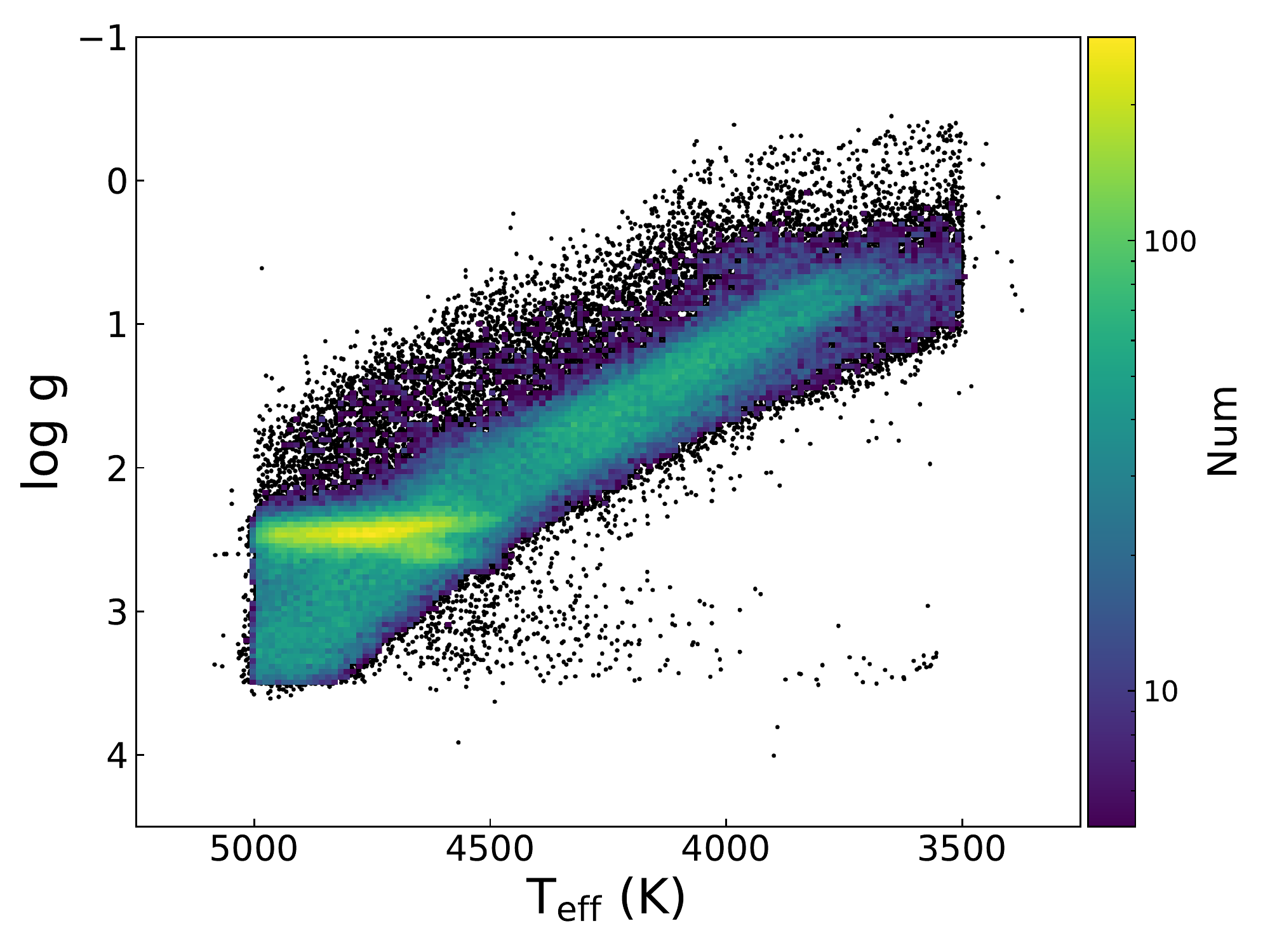}
  \caption{Spectroscopic HR diagram (sometimes called a Kiel Diagram) showing APOGEE calibrated $T_{\rm eff}$ vs. $\log g$, for the BAWLAS sample of the red giant branch.}
  \label{fig:hrd}
\end{figure}

While our sample is mostly driven by these selections, there are two caveats to the final sample regarding stars that have duplicate or repeat entries in APOGEE because they were observed in multiple distinct fields and therefore have multiple unique spectra \citep[i.e., stars that have multiple spectra that were processed separately through ASPCAP; for a further description of these stars see Sections 2 and 5.4 of][]{Jonsson2020}.  1) We have intentionally included a sample of duplicate/repeat spectra (of the same stars) in order to investigate some of the systematic uncertainties in our analysis, as described in Section \ref{sec:final_abund}.  This is a set of 4,876 unique stars with a total of 11,752 observed spectra and corresponding entries in APOGEE's catalogs.  2) In some cases for stars with duplicate entries, an entry/spectrum that does not conform to our initial selections was included in our analysis (e.g., ASPCAP derives parameters outside of the above range, or the spectra do not conform to our S/N selection), this affects $\sim 4\%$ of the final sample (4,668 spectra).  Figure \ref{fig:hrd} shows a spectroscopic Hertzsprung-Russell (HR) diagram of our full sample, which illustrates that in cases where duplicates have parameters outside of our initial selection range they typically fall close to our selection limits.

For each star in our sample, we obtain the 1D combined and resampled APOGEE apStar and asStar spectra (for northern and southern targets respectively), including any duplicate spectra, and the ASPCAP derived, calibrated stellar parameters and chemical abundances from each of these spectra (using the APOGEE summary file version \texttt{allStar-dr17-synspec.fits}).  The spectra are converted to air wavelengths according to the transformations given in \citet{shetrone2015}.  Then, these spectra and calibrated stellar parameters are used in BACCHUS (with APOGEE's chemical abundances as starting guesses) to derive chemical abundances and isotopic ratios of weaker features in APOGEE spectra, as described in the following section.

\section{BACCHUS Processing}
\label{sec:bacchus}
Throughout this analysis, we used the abundance determination module of the latest version of the BACCHUS code ($\sim$ v67). More specifically, this BACCHUS module consists of a shell script code that computes on the fly synthetic spectra for a range of abundances and compares these syntheses to observational data on a line-by-line basis, deriving abundances from different methods (e.g. using equivalent width, line depth, or $\chi^2$). The synthetic spectra are calculated using the 1D-LTE Turbospectrum radiative transfer code \citep{AlvarezPlez1998,Plez2012}, with spherically symmetric radiative transfer for giants, and the MARCS model atmosphere grids \citep{marcs}.

From these synthetic spectra the BACCHUS code then identifies both continuum (or pseudo-continuum) points for normalizing the observed spectra, and the relevant pixels  (i.e., a ``mask'' or ``window'') to use for the abundance determination of each line in each spectrum.  The normalization of observed spectra is performed by selecting continuum wavelengths in the synthetic spectra, and fitting a linear relation across these continuum points within 30 \AA{} around the line of interest in the observed spectra. The spectral window for each line is determined by comparing where changing the elemental abundance causes significant changes in the synthetic spectra with the second derivative of the observed flux (i.e., to identify the local maxima on either side of the line of interest).

The last, crucial and unique feature of this code is that it uses four methods to compare the observed and synthetic spectra within the selected window and provides an abundance measurement for each method.  The four methods are:

\begin{enumerate}
\item The ``\chitwo{}'' method, which determines an abundance by minimizing the squared differences between synthetic and observed spectra.
\item The ``\syn{}'' method, which looks for the abundance that makes the difference between the synthetic and the observed points zero.
\item The ``\eqw{}'' method, which determines the abundance needed to match the equivalent widths of the synthetic spectra to the observations.
\item The ``\intmethod{}'' method, which measures abundances by matching the line core in the synthetic and observe spectra. 
\end{enumerate}

Each of these methods has benefits and limitations.  For example, the \eqw{} method is not sensitive to uncertainties in spectral broadening (such as, instrumental or macroscopic velocity broadening), but it is more affected by blends in the wings of lines, whereas the inverse is true for the \intmethod{} method. Each of these methods also have diagnostic flags that indicate the quality of their fits which are listed in Table~\ref{tab:methodflagdesc} and are described in more detail in Section \ref{sec:method_flags}.

\subsection{New features}
Given the difficulties imposed by analyzing weak and strongly blended lines, we have also implemented some new features. One improvement is the incorporation of the latest version of Turbospectrum (v19.1.3), which includes Stark broadening (important for a Mg blend with one Nd line; see Appendix \ref{app:Nd}). Another improvement is the addition of the extended MARCS model atmosphere grid that has been specifically built for APOGEE that notably includes C and $\alpha$ abundances dimensions \citep[see][]{Jonsson2020}.  

While BACCHUS does interpolate atmospheric structure in T$_{\rm eff}$, $\log g$ and metallicity, for C and $\alpha$ it uses the atmospheric model with the closest value to the input C and $\alpha$ abundance.  This interpolation first attempts to select the closest grid points in T$_{\rm eff}$, $\log g$ and metallicity, and when it encounters a hole in the grid it will try an alternative search in T$_{\rm eff}$ first, then $\log g$, and finally metallicity to find a set of grid points that are fully populated.  If all of these attempts fails, i.e., there are too many missing grid points in the atmosphere grid around the stellar parameters of a given star, the star will be rejected and not processed by BACCHUS.

We have also implemented a new, fifth method for deriving abundances in BACCHUS, ``\wln{}'' method. This method interpolates the synthetic spectra as a function of elemental abundance at the exact wavelength of the input line. This method has been particularly useful for the abundance determination of one of the Cu lines, because it bypasses the automatic mask determination, which is heavily biased in the strong blends affecting this line. However this method has the disadvantage of relying on one single pixel, and, therefore, is more sensitive to random or systematic errors in that pixel. 

\subsection{Specific features for the BACCHUS-APOGEE run}

For the APOGEE spectra, we use the APOGEE DR17 Turbospectrum linelist, version \emph{180901t20}, with one modification (to reduce the strength of one C$_2$ line, the 15710.665 \AA{} line, changing its loggf from -1.502 to -2.502).  To select a model atmosphere for each star, we interpolate the MARCS model atmosphere grid at the calibrated APOGEE T$_{\rm eff}$, $\log g$, and [M/H] assuming a microturbulence relation of \citep{Masseron2019}:
\begin{displaymath}
\rm \mu_t (km/s) = 2.488 -0.8665*\log g + 0.1567*\log g^2
\end{displaymath}

To determine the convolution parameter, which is used to encompasses the instrumental, rotational and macroturbulence broadening with a Gaussian profile, we use Si lines rather than more typical Fe lines because the former are generally cleaner and stronger in the APOGEE spectra, even at low metallicities.

Because they are responsible for blending and accurate continuum adjustment, we first determined the abundances for C, O, N and Mg in each star, in this specific order to establish a proper chemical equilibrium.  We then measure the carbon isotopic ratio, V, S, Na, Ce, Nd, P, Yb, Zn and Cu, while locally adjusting possible blending features for each line.  

The code iterates whenever the measured abundance is out of the initial synthesis range, so when possible we use the ASPCAP calibrated abundances as a starting guess for each elemental abundance to help accelerate the convergence of the code.  When ASPCAP calibrated abundances are not available for a given element \citep[typically because the star is in a temperature range where the abundances of that element are suspect abundances, see][for more details]{Holtzman2021}, we use an $\alpha$-scaled starting guess for $\alpha$-elements (O, Mg, Si, S, etc.) or a solar-scaled abundance for all other elements.  While the code is expected to automatically iterate over an element until convergence, we force two iterations of the overall procedure to ensure self-consistency between the elements and also to avoid artificial stratification in the abundance diagrams when values approach the edge of BACCHUS search range that can occur for elements with multiple lines and enhancements over solar-abundances, such as Ce and Nd.

\section{Line Combination and Flagging}
\label{sec:line_comb}

The BACCHUS processing produces line-by-line measurements of the \logeps{X} $\equiv \log_{10}({\rm N_X/N_H})$ abundance of each element, X, or the \cisotope{} ratio, and quality flags of that measurement from the five different methods mentioned in Section \ref{sec:bacchus}.  This allows us flexibility in how to flag and remove individual line measurements when combining our line-by-line measurements for the highest quality results.

Our process for combining the variety of line-by-line measurements and flags is as follows:
\begin{enumerate}
    \item Identify which measurement method to use for each line (Section \ref{sec:meas_method}).
    \item Evaluate the quality of each line-by-line measurement using BACCHUS measurement flags (Section \ref{sec:method_flags}), flags designed to indicate the quality of the spectra around each line (Section \ref{sec:spectra_flags}), and upper limit relations to indicate measurements below our upper limit threshold (Section \ref{sec:upper_limits_relation}).
    \item Flag stars whose stellar parameters may be of suspect quality for deriving chemical abundances (Section \ref{sec:param_flags}).
    \item Derive and apply line-by-line zero-point abundance calibrations (Section \ref{sec:zero_points}).
    \item Finally, we average the zero-point calibrated abundances for the lines that pass all of our quality cuts to arrive at our final abundances and derive two separate estimates of the uncertainties on these values (Section \ref{sec:final_abund}).
\end{enumerate}

In each of the sections below, we describe our general methodology for the above steps, and we give specific details for what our flagging and combination choices are for each element in Appendix \ref{app:comb_settings}.

\subsection{Measurement Method Choices}
\label{sec:meas_method}

BACCHUS uses several methods of measuring the abundance of each line in each star, from which we then select one method to use across all stars for a given line.  
Because BACCHUS returns abundance measurements from several different methods, we are able to select the optimal method for each line in our post-processing.  This flexibility is beneficial, since some methods are more or less sensitive to different issues that may arise, such as, how badly blended a line is, if there are problems with the continuum around that line, etc.  

For all of the lines that we measure in this work, we either use the \chitwo{} method or the \wln{} method.  In general, because it uses multiple pixels, the \chitwo{} method gives higher precision measurements and is less sensitive various effects that can impact the spectrum at the level of individual pixels.  But, the \chitwo{} method can also provide poor measurements when a line has a poorly fit blend, especially on the wings.  

The \wln{} method however, because it is determined using a single pixel is more susceptible to noise and may be somewhat less precise, but performs better in cases where there are poorly fit blends on the wings of the line, or when lines are weak and only the central pixels on a line provide measurable signal. While we typically use the \chitwo{} method to take advantage of multiple pixels across a line, for particularly weak or blended lines we use the \wln{} method as discussed in more detail in Appendix \ref{app:comb_settings}.

\subsection{BACCHUS Method Flags}
\label{sec:method_flags}

Each of the five abundance measurement methods employed in our BACCHUS analysis have integer flags that indicate the quality of the spectral line fit from that method.  Every line of every star that is measured will be given one of the flags in Table \ref{tab:methodflagdesc} for each method (note that the \syn{} and \eqw{} method flags have the same general description).

\begin{deluxetable}{r p{6cm}}
\tabletypesize{\scriptsize}
\tablewidth{0pt}
\tablecolumns{2}
\tablecaption{BACCHUS Method Flag Description \label{tab:methodflagdesc}}
\tablehead{\colhead{Flag} & \colhead{Description}}
\startdata
\cutinhead{\syn{}/\eqw{}}
0 & Indicates when BACCHUS is only measuring an upper limit with this method\\
1 & The measurement from this method is okay\\
2 & The measurement is an extrapolation outside of the synthesized range of abundances\\
3 & Large (wavelength) offset between the observed flux minimum (nominally the line core) and the synthetic flux minimum \\
\cutinhead{\intmethod{}}
0 & Indicates when BACCHUS is only measuring an upper limit with this method\\
1 & The measurement from this line is okay\\
2 & Large (wavelength) offset between the observed and synthesized line cores or strong line (line intensity is below 0.4 of the continuum)\\
\cutinhead{\chitwo{}}
0 & The measurement is beyond the synthesized range of abundances\\
1 & The measurement from this line is okay\\
2 & The \chitwo{} method crashed\\
\cutinhead{\wln{}}
0 & The syntheses are too close in flux to provide a reliable interpolation or extrapolation\\
1 & The measurement from this line is okay\\
2 & The measurement is an extrapolation outside of the synthesized range of abundances\\
\enddata
\end{deluxetable}

We use these flags to determine the quality of the BACCHUS fits to each line in a given star and we can improve this assessment by combining the information from the flags of multiple methods.  For each line that we measure we have a list of method flags that we consider reflective of a ``good'' BACCHUS fit.  In general, our default is to require that all methods have flag = 1 for a line's measurement in a star to be considered ``good''.  So with this as an example if all method flags = 1 for a specific line in a star we will use the measurement of that line in our line combination (assuming said measurement passes our other quality criteria), and if any of the method flags $\neq$ 1, we will not use that line for the abundance measurement in that star (and consider it ``flagged,'' removing it from consideration).  

However, for some lines or elements we consider other combinations of these method flags on a case-by-case basis.  For example, some lines lie on the wings of deeper blends (e.g., Cu 16006 \AA{} or Nd 16053 \AA{}).  This can cause BACCHUS to give a \syn{} or \eqw{} flag = 3 because of the large wavelength difference between the input line center and the local minimum in the spectrum (i.e., the deeper line center of the blending line), despite providing a good fit to the line of interest.  So, we would want to consider measurements with \syn{} and \eqw{} flag = 1 or 3 for these lines.

In cases such as this example, we have used a modified selection of quality flags to appropriately treat the line in question.  Appendix \ref{app:comb_settings} gives a full summary of our flag choices for each element and their lines as well as individual illustrations of the line profiles and blends.  
  
\subsection{Additional Spectra Flags}
\label{sec:spectra_flags}
  
In addition to the native flags produced by BACCHUS we supplement these with an additional set of quality flags defined to indicate the condition of the spectrum around the lines of interest that we are measuring.  These flags are roughly categorized as those that indicate the state of any blends that may affect our line of interest (blend) and those that describe the quality of the continuum around each line of interest (cont).  Both flags are defined such that flag = 1 means the line in a specific star is okay, whereas flag = 0 warrants further investigation (and typically we remove lines in stars with blend/cont flag = 0)

The blend flags indicate poorly fit blends with other spectral lines in the star or a blend with a sky line or telluric feature that may have an imperfect subtraction, either of which can lead to incorrect abundance measurements.  The continuum flags vary in meaning from line-to-line.  For some lines this flag is set to 0 when the spectrum is consistent with being only continuum, i.e., the line of interest is too weak to be reasonably measured.  In other lines this flag is defined to be 0 when the pseudo-continuum around a line is a poor match to observations, typically because the wings of a nearby blend are poorly fit.

A detailed description of the spectra flags that we have implemented in our analysis can be found in Appendix \ref{app:spectra_flags}.

\subsection{Upper Limit Relations}
\label{sec:upper_limits_relation}

\begin{figure*}
  \centering
  \includegraphics[scale=0.5,trim={1.5cm 1cm 1cm 0.5cm},clip]{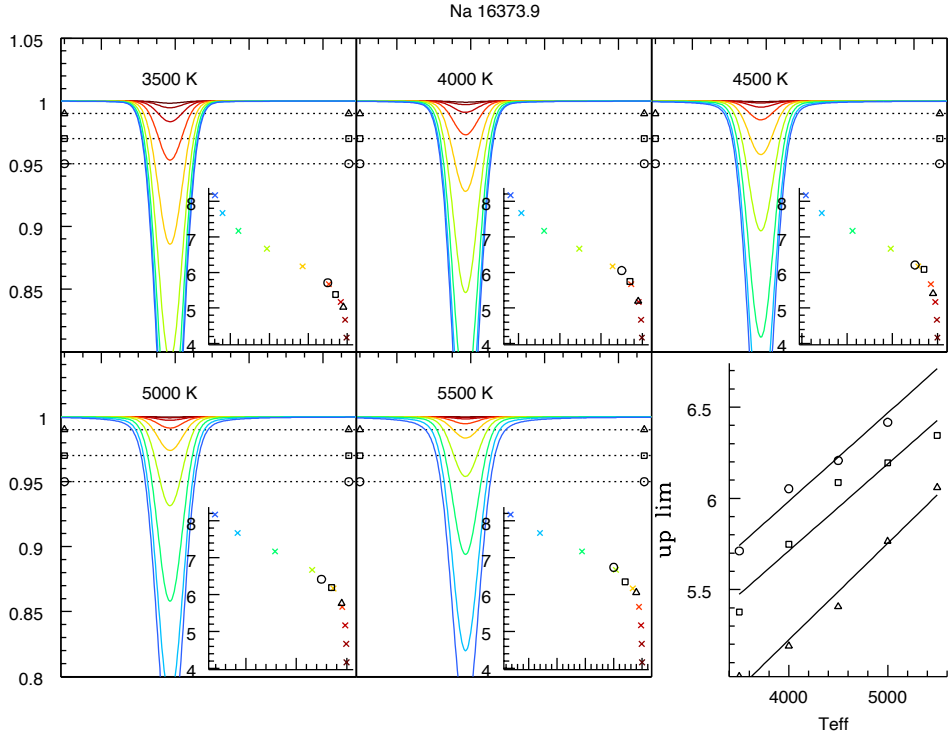}
  \caption{Illustration of the computation of upper the limit relations of the Na 16374 \AA{} line. Each of the first five panels shows the synthetic Na line for different abundances. The dotted lines represent flux thresholds of 1, 3 and 5\% of the continuum flux. The respective inlaid plots show the Na abundance as a function of the line center flux. The triangle, square and circle mark the line flux level and abundance corresponding respectively to 1, 3 and 5\% of the continuum flux. Each panel is for a different temperature except for the summary plot (bottom right) that shows the \logeps{Na} abundance corresponding to these percentage thresholds as a function of temperature, from which we derive our upper limit relations by linear regression (black lines).}
  \label{fig:Na_upperlimit}
\end{figure*}
To assess BACCHUS's fits in a programmatic way, we have also included a process for distinguishing detections from upper limit measurements. BACCHUS inherently provides an evaluation of upper limits based on a continuum threshold computed from the variance or the signal-to-noise of the observed spectrum. However, this procedure is not valid for blended lines, such as the lines used in this study, because the blends prevent the line growth from reaching the continuum threshold at very low abundances. 

Instead, we developed a procedure based on synthetic spectra computed with Turbospectrum. We synthesize individual elemental lines without any blending features  in 9 steps of \logeps{X} from \logeps{X} = $-$2.0 to $+$2.0 dex around solar at five temperatures from 3500 K to 5500 K in steps of 500 K along the giant branch (with $\log$ g adjusted to have the syntheses appropriately lie on the RGB). At each temperature step we interpolate between these syntheses to determine the abundances when line depths are 1, 2, 3, 4, or 5$\%$ of the continuum.  Thus, these are the upper limit of abundances we could measure with a t = X\% threshold.  For each threshold, we then derive an upper limit relation for each line as a linear function of temperature following: \logeps{X}$_{\rm lim} = A_{\rm t} \cdot T_{\rm eff} + B_{\rm t}$, where t is the \% threshold level used. Figure \ref{fig:Na_upperlimit} illustrates this process for the Na 16374 \AA{} line, and Table \ref{tab:upper_limits} in Appendix \ref{app:upper_limits} gives the derived constants $A$ and $B$ for each of the values of t mentioned above for each element.

Therefore, with a chosen threshold we can predict the abundance upper limit that is measurable for a given temperature for each line.  If a line is measured with an abundance less than that limit for a star, we don't use that line in our combination of line-by-line abundance measurements.  Instead, when considering the final, combined abundances in a star, if all of the lines for a given element in a star that pass the flagging criteria (i.e., the BACCHUS fit is considered okay) are upper limits, we record the minimum of the line-by-line upper limits as the abundance upper limit on said star. Note, that we did not compute limits in this way for 
\cisotope{}, because it is a more complex function of temperature and carbon abundance. Instead, we use a simple line-by-line lower limit for \cisotope{} based on the synthesis range we have used (see Appendix \ref{app:comb_settings}), but also include line-by-line detection threshold flag based on the observed normalised flux, and a more complicated lower limit relation set for the final combined \cisotope{} measurements (see Appendix \ref{app:spectra_flags}).

As a default we have chosen to use a threshold of 1\%, requiring that lines should have abundances that produce at least a 1\% line depth.  With a S/N ratio of $>$ 150, as in our sample, one may expect to use a threshold below 1\% as the detection limit, however because these relations do not account for the effect of blends, uncertainties in stellar parameters, etc., we have used this slightly more conservative upper limit threshold.  Furthermore, for individual lines that are especially blended or sensitive to continuum placement, we have adjusted the threshold to even higher levels based on visual inspection of abundance measurements near the upper limit boundary to ensure that we have bona fide detections (see Section \ref{app:comb_settings} for more case-by-case details). 

\subsection{Star Flags}
\label{sec:param_flags}

In addition to line-by-line quality flags, we also flag some stars based on how they have been processed by BACCHUS, and therefore we do not report abundances for these stars.  Specifically we flag stars that may have suspect fits to critical molecular blends and stars that require particularly high convolution value to explain their spectral features.  

\subsubsection{Missing C, N, and O Updates}

\begin{figure*}
  \centering
  \includegraphics[scale=0.29,trim = 0.in 0.in 0.in 0.in, clip]{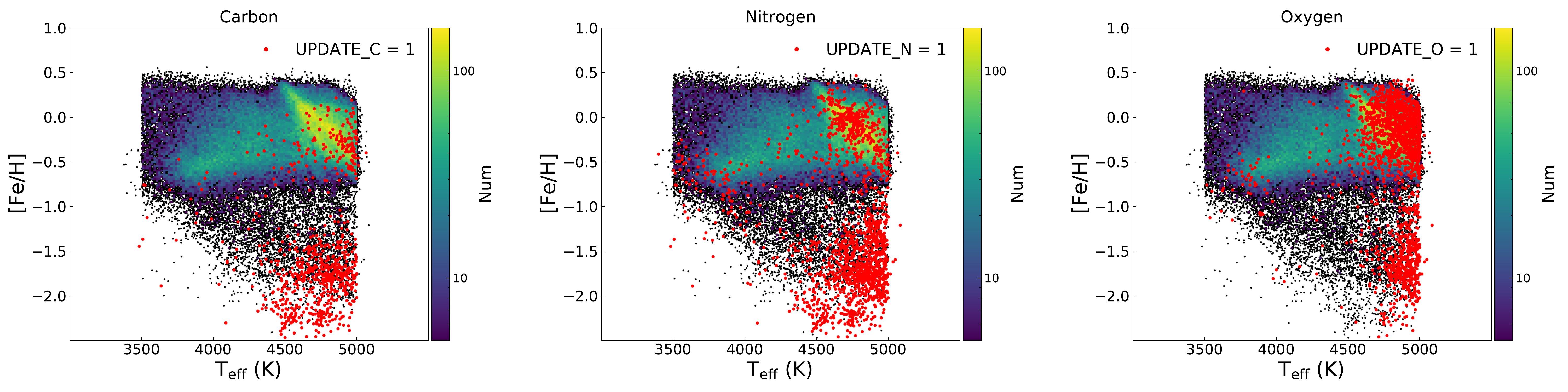}
  \caption{Effective temperature, T$_{\rm eff}$, vs. metallicity, \xh{Fe}, for the full BAWLAS sample (black points and 2D histogram), and for the stars that have {\texttt{UPDATE\_C}}, {\texttt{UPDATE\_N}}, or {\texttt{UPDATE\_O}} = 1 (red points) for carbon (left), nitrogen (middle), and oxygen (right), respectively.  Stars that do not have their C, N, and O updated are typically those at low metallicities or warm temperatures where lines are weak.}
  \label{fig:cno_update}
\end{figure*}

Because the C, N, and O abundances of a star have a significant impact on the H-band spectrum and are critical for fitting blends with our weaker lines, we have used the ASPCAP C, N, and O abundances as an initial guess.  But, with BACCHUS we rederive these abundances, given that their corresponding molecular features are highly temperature sensitive.  This allows us to find the best solution for the C, N, and O abundances using the calibrated stellar parameters from ASPCAP (whereas the ASPCAP C, N, and O are derived using the uncalibrated stellar parameters).  

However, the C, N, and O abundances are not always updated with new values from BACCHUS and are fixed to their initial APOGEE DR17 values when fitting other elements (i.e., when C, N, and O features may be blended with the lines of an element of interest), which can occur for a variety of reasons, e.g., their lines are too weak for BACCHUS to provide a good measurement.  To track this, we record which stars have had their C, N, and O abundances updated by BACCHUS measurements in integer flags {\texttt{UPDATE\_C}}, {\texttt{UPDATE\_N}}, and {\texttt{UPDATE\_O}} respectively, where flag = 1 indicates stars that were not updated and flag = 2 indicates stars that have been updated with BACCHUS measured values for that element.

Figure \ref{fig:cno_update} shows the effective temperatures and metallicities of the BAWLAS sample and the stars that do not have their C, N, or O updated.  These elements are updated in almost all of the stars in our sample.  Those stars that do not have updated C, N, or O are preferentially at low metallicities or warm temperatures, where these features are weaker and less important to fit. 

While we have not removed any stars that have not had their C, N, or O abundances updated by BACCHUS, we have provided these flags so that stars whose molecular blends may be more poorly fit can be tracked.

\subsubsection{Stars with Large Line Broadening}

One of the parameters that we fit for each star with BACCHUS is a Gaussian convolution broadening parameter, {\texttt{CONVOL}}, that consolidates observational and stellar effects from multiple sources, including:  instrument resolution, macroturbulent velocity, rotation, etc.  In giants, most of this broadening (in our spectra) comes from the instrument resolution and macroturbulent velocities.  Combined, these effects produce a median convolution broadening of 14.9 \kms{} across all of the stars in our sample with stars typically having a convolution broadening ranging between 14.1 \kms{} and 15.8 \kms{}, the 16$^{\rm th}$ and 84$^{\rm th}$ percentiles respectively.

However, the broadening has a tail to even higher values, which could indicate that there are additional broadening effects, such as rotation, larger macroturblent velocities at stellar surfaces, or that APOGEE's stellar parameters do not adequately describe the observed spectra (e.g., red supergiants observed in the Magellanic Clouds, see Section \ref{sec:lmc_rsg}).  Giants are not treated with rotation or other broadening parameters (other than macroturbulence) by ASPCAP, so stars that exhibit additional broadening may have erroneous stellar parameters, which can lead to incorrect abundances from BACCHUS as well.  In order to flag these suspect stars we have removed final combined abundances for stars with {\texttt{CONVOL}} $> 18$ \kms{} (around 2\% of the BAWLAS sample), but we have retained their line-by-line abundances for the possibility of more detailed investigation.

\subsection{Zero Point Calibration}
\label{sec:zero_points}

Similar to APOGEE DR17, we have chosen to calibrate our chemical abundances to a solar zero point derived from solar neighborhood samples (except for C, N and \cisotope{}), instead of using literature solar zero points or abundances derived from solar spectra by BACCHUS.  Errors in loggf values or other linelist parameters, non-Local Thermodynamic Equilibrium (nLTE) effects, or systematic errors in stellar parameters or blend fitting can all lead to different line-by-line zero points, which would not be captured by using solar zero points from literature.  

\begin{figure*}
  \centering
  \includegraphics[scale=0.85,trim = 0.in 0.15in 0.in 0.in, clip]{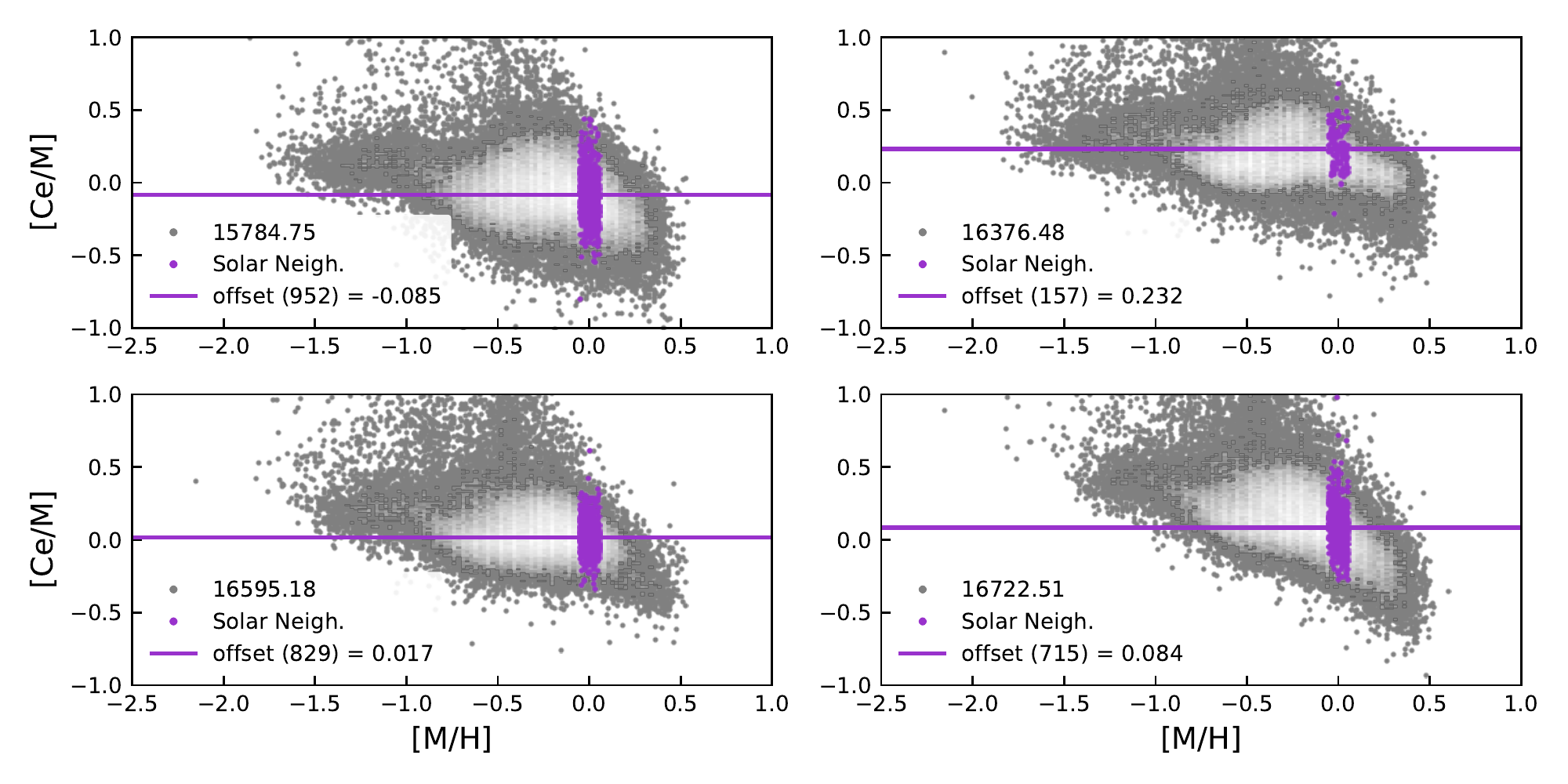}
  \caption{\xm{Ce} vs \xh{M} for the BAWLAS sample (gray points and 2D histogram) and our solar neighborhood sub-sample (purple points) from each of the four Ce lines used in this study.  The \xm{Ce} values have been adjusted to a solar zero point of \logeps{Ce}$_{\odot} = 1.70$ \citep{Grevesse2007}, yet the median abundance of the solar neighborhood (purple line) differs, sometimes significantly, between each line.  Note that the vertical stripes in the 2D histogram are due to aliasing of the rounded APOGEE metallicity used in the BACCHUS processing.}
  \label{fig:ce_zeropoint}
\end{figure*}

Indeed, Figure \ref{fig:ce_zeropoint} illustrates that the abundances we derive can have even significant offsets from line to line, even when we narrow to a solar neighborhood sample (defined below) that should nominally have solar abundance ratios on average.  Therefore, by using a single zero point between all lines we may still end up with systematic offsets in our abundances.

In addition, because these sources of line-by-line zero point offsets may conceivably have trends with stellar parameters (such as nLTE effects, systematic errors in stellar parameters, etc.), analyzing a solar spectrum with BACCHUS and using that as a zero point for a red giant sample may not be appropriate.  Instead, we have chosen to empirically calibrate our line-by-line abundance measurements such that the solar neighborhood stars in our sample have a solar ([X/M] $= 0$) abundances in each line.  

To establish our solar neighborhood sample, we use APOGEE stellar parameters and radial velocities combined with {\it Gaia} early data release 3 \citep[EDR3][]{gaiaedr3} parallaxes and proper motions.  We calculate Galactocentric positions using inverted parallax distances from {\it Gaia} EDR3 parallaxes and assuming the sun is at $R_{\rm GC,\odot} = 8.122$ kpc \citep{gravity}.  We also use these inverted parallax distances to convert APOGEE radial velocities and {\it Gaia} EDR3 proper motions to Galactocentric velocities using the prescription from \citet{JohnsonSoderblom1987} and assuming a solar motion of $(V_{r}, \ V_{\phi} \ V_{z})_{\odot} = (14, 253, 7)$ km s$^{-1}$ in right-handed notation \citep{Schonrich2010,Schonrich2012,Hayes2018c}.  Our solar neighborhood sample is then defined, as those stars in our BAWLAS sample with:
  \begin{itemize}
      \item {\it Gaia} EDR3 relative parallax uncertainty below 10\% ($ 0 < \sigma_{\varpi}/\varpi < 0.1$)
      \item Metallicities, $-0.05 <$ [M/H] $< 0.05$
      \item Galactocentric (cylindrical) radii, $8 < R_{\rm GC} < 9$ kpc
      \item Height above the MW midplane, $|z| < 0.5$ kpc
      \item Total space velocity within 40 km/s of the local standard of rest, assuming $(V_{R}, V_{z}, V_{\phi})_{\rm LSR} = (0, 0, 229)$ km/s using the value of $V_{\phi,\ {\rm LSR}} = 229$ km/s from \citet{Hayes2018c}
  \end{itemize}
  
For each line of each element, we apply the quality flag and upper limit cuts described in the previous sections and calculate a zero-point from this solar neighborhood sample by taking the median abundance in that line.  We use the median to limit the effects of outliers, either in the case of random scatter or in the case of elements like Ce and Nd to reduce the effect of stars with enhanced abundance ratios from processes other than Galactic chemical evolution (for more information about these stars see Section \ref{sec:cend_results}).

\begin{deluxetable}{l c c c}
\tablewidth{0pt}
\tablecolumns{12}
\tablecaption{Line-by-Line Abundance Zero Points \label{tab:zero_points}}
\tablehead{\colhead{Element} & \colhead{Line (\AA{})} & \colhead{\logeps{X}$_{\rm zpt}$} & \colhead{Grev07 offset\tablenotemark{{\scriptsize a}}}}
\startdata
O & 15373.5 & 8.800 & 0.140 \\
O & 15391.0 & 8.902 & 0.242 \\
O & 15569.0 & 8.880 & 0.220 \\
O & 15719.7 & 8.928 & 0.268 \\
O & 15778.5 & 8.964 & 0.304 \\
O & 16052.9 & 8.814 & 0.154 \\
O & 16055.5 & 8.854 & 0.194 \\
O & 16650.0 & 8.798 & 0.138 \\
O & 16704.8 & 8.874 & 0.214 \\
O & 16714.5 & 8.888 & 0.228 \\
O & 16872.0 & 8.863 & 0.203 \\
O & 16909.4 & 8.938 & 0.278 \\
Na & 16373.9 & 6.367 & 0.197 \\
Na & 16388.8 & 6.399 & 0.229 \\
P & 15711.6 & 5.700 & 0.340 \\
P & 16482.9 & 5.474 & 0.114 \\
S & 15403.5 & 7.159 & 0.019 \\
S & 15422.3 & 7.151 & 0.011 \\
S & 15469.8 & 7.157 & 0.017 \\
S & 15475.6 & 7.242 & 0.102 \\
S & 15478.5 & 7.133 & -0.007 \\
V & 15924.8 & 4.022 & 0.022 \\
V & 16137.3 & 4.347 & 0.347 \\
V & 16200.1 & 4.209 & 0.209\\
Cu & 16005.5 & 4.228 & 0.018 \\
Cu & 16006.0 & 4.147 & -0.063 \\
Ce & 15784.8 & 1.615 & -0.085 \\
Ce & 16376.5 & 1.932 & 0.232 \\
Ce & 16595.2 & 1.717 & 0.017 \\
Ce & 16722.5 & 1.784 & 0.084 \\
Nd & 15368.1 & 1.769 & 0.319 \\
Nd & 16053.6 & 1.531 & 0.081 \\
Nd & 16262.0 & 1.860 & 0.410 \\
\enddata
\tablenotetext{a}{Zeropoint offset from the \citet{Grevesse2007} solar abundances, reported as (\logeps{X}$_{\rm zpt}$ $-$ \logeps{X}$_{\rm \odot, Grev07}$)}
\end{deluxetable}
  
The zero-points that we calculate for each line are listed in Table \ref{tab:zero_points}, which we then use to calculate our ``bracket notation'' abundances e.g., \xh{X}$_{*}$ $=$ \logeps{X}$_{*}$ - \logeps{X}$_{\rm zpt}$ for each line.  The exception to this is the C and N abundances and the \cisotope{} ratios that we derive.  Because these three values are expected to change along the giant branch due to dredge-up, we do not necessarily expect that the solar neighborhood should have solar abundance or isotopic ratios.  Instead, for C and N we have used the \citet{Grevesse2007} solar abundances for our zero point, and for \cisotope{} we use only the raw ratios that we calculate without applying any calibration.

\subsection{Combined Abundances and Uncertainties}
\label{sec:final_abund}

Our final combined abundances are derived by averaging all of the ``good,'' unflagged, zero-point calibrated line measurements for each star (following the detailed, element-by-element and line-by-line selections given in Appendix \ref{app:comb_settings}).  While we consider most of the lines that were attempted by BACCHUS, a few lines were rejected for all stars because they showed strong trends with temperature or their abundance patterns disagree strongly with the remaining lines (see Appendix \ref{app:comb_settings} for more details).   For stars whose lines of a given element are all identified as upper limits, instead of combined abundances we provide an \xfe{X} upper limit defined as the minimum line-by-line upper limit values as mentioned in Section \ref{sec:upper_limits_relation} and adjusted using the appropriate zero point from Section \ref{sec:zero_points}.

In order to estimate uncertainties on our combined abundance measurements, we use two different methods. (1) We use the line-by-line scatter (and the scatter between two of BACCHUS's measurement methods, \chitwo{} and \wln{}) as an estimate of the measurement uncertainty, and (2) we can estimate the uncertainty empirically from the scatter in repeat observations (and separate reductions/analyses) of the stars with duplicate entries in APOGEE. 

\subsubsection{Measurement Uncertainties}

One way we estimate our abundance uncertainties is by measuring the line-by-line and method-to-method scatter in BACCHUS's abundance measurements.  The ``measurement'' uncertainty that we report for each star is the standard error of the abundances as measured by the \chitwo{} and \wln{} methods of ``good,'' unflagged lines of a given element.  Numerically this is calculated by taking the standard deviation of the abundance measurements from the \chitwo{} and \wln{} methods in each of the good lines and dividing this by $\sqrt{\rm N_{\rm lines}}$.

This estimate of the uncertainty only includes the uncertainty that comes from line choice and measurement methodology, and does not account for more systematic errors, such as the uncertainties or errors in the input stellar parameters, linelist, choice of code, etc.   For cases where only one or two lines of an element were measured in a star, we do note that these uncertainties may be underestimated, so on a star-by-star basis these uncertainties may need to be considered carefully.  However, we do provide an additional estimate of the abundance uncertainties below, that is less sensitive to star-by-star cases.

\subsubsection{Empirical Uncertainties}

\begin{figure*}
  \centering
  \includegraphics[scale=0.45,trim = 0.in 0.15in 0.in 0.in, clip]{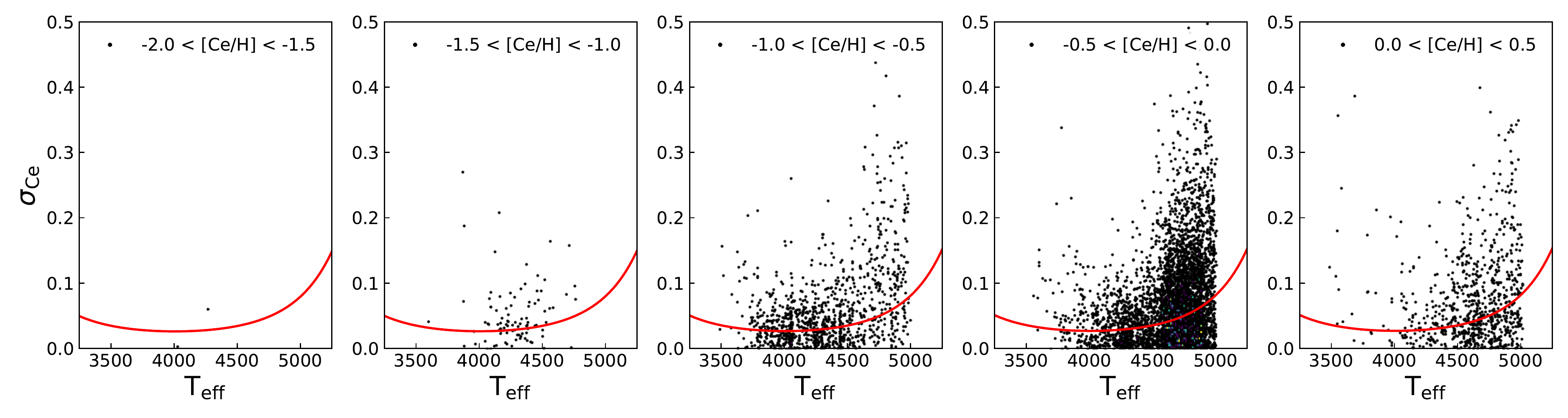}
  \caption{Standard deviation estimates of \xh{Ce} versus stellar effective temperature, $T_{\rm eff}$ for our repeat sample in 0.5 dex bins of mean \xh{Ce}, with estimators ($\sigma$) calculated as described in the text.  Overplotted (red line) is the best fit empirical abundance uncertainty relation we calculate at \xh{Ce} of the bin center (this is shown for illustration; the relation, however, fit as a function of individual \xh{Ce} abundances rather than in bins of \xh{Ce}).}
  \label{fig:ce_emp_errs}
\end{figure*}

Motivated by what is done by APOGEE to estimate its abundance uncertainties, we also provide empirical uncertainty estimates determined from stars that have been observed multiple times in APOGEE but have their multiple spectra processed and analyzed separately.  These stars may have ``random'' observational effects in their spectra, different stellar parameters from APOGEE, etc., so by processing them separately we can understand how these differences impact the derived abundances.  This allows us to understand some of the impact of varying the input stellar parameters and observational noise on the abundances we measure.

We take our sample of $\sim 4,900$ stars with $\sim 12,000$ different spectra and ASPCAP results, process them separately through BACCHUS and our post-processing and then compare the resulting abundances that we measure.  Each pair of absolute differences in \xh{X} provides an estimate of the standard deviation in the abundance measurements for an individual star when multiplied by $\sqrt{\pi}/2$ for an unbiased estimator.  Each star then provides an estimate, $\sigma_{\rm X} = \sqrt{\pi}/2 \ |\, \Delta {\rm [X/H]} \, |$, of the typical random abundance errors for that stars parameters. Therefore we can fit the ensemble distribution of these standard deviation estimators to derive an empirical relation for the typical uncertainties on our measurements as a function of various parameters.  For this work we use a simple relationship to fit the uncertainty distributions, using the following equation (and substituting \cisotope{} for \xh{X} when calculating \cisotope{} uncertainties):
   \begin{displaymath}
     \ln{\sigma_{\rm [X/H]}} = A + B \cdot T^{'}_{\rm eff} + C \cdot (T^{'}_{\rm eff})^2 + D \cdot {\rm [X/H]}
   \end{displaymath}
where $T^{'}_{\rm eff} = T_{\rm eff} - 4500$ K.

We use a slightly different form than used by APOGEE (in ASPCAP) for their empirical uncertainties.  Here we drop the S/N dependence since our sample is high S/N, and we use the [X/H] abundance instead of [M/H] in our formulation, because the abundance uncertainty should depend primarily on the amount of that element rather than the total metallicity\footnote{Although one can imagine that this could be refined by including metallicity and elements with dominant molecular features whose blends might also affect the errors on our derived abundances.  Including more terms in this relation may be a promising way to improve this kind of uncertainty estimation in the future.}.  As an example of our derivation of these relations, Figure \ref{fig:ce_emp_errs} shows the distribution of differences in Ce abundances for our repeat sample and the best fit empirical error relation for this element.

\begin{deluxetable}{l c c c c}
\tablewidth{0pt}
\tablecolumns{5}
\tablecaption{Empirical Error Relation Parameters \label{tab:uncertainties}}
\tablehead{\colhead{Element} & \colhead{A} & \colhead{B} & \colhead{C} & \colhead{D} \\ \colhead{} & \colhead{} & \colhead{(10$^{3}$ K)$^{-1}$} & \colhead{(10$^{3}$ K)$^{-2}$} & \colhead{}}
\startdata
C & -3.528 & 0.894 & 0.214 & -0.110 \\
N & -2.864 & 0.942 & 0.639 & 0.516 \\
O & -3.841 & 1.405 & 2.666 & 0.356 \\
Na & -3.357 & 0.371 & -0.550 & -0.670 \\
P & -2.738 & -0.361 & -0.798 & 0.747 \\
S & -3.265 & -0.721 & -0.568 & 0.196 \\
V & -3.441 & 1.640 & 0.613 & -0.834 \\
Cu & -3.233 & 0.360 & 0.831 & -0.597 \\
Ce & -3.344 & 1.113 & 1.126 & 0.018 \\
Nd & -2.973 & 0.175 & 0.307 & 0.893 \\
\cisotope{} & -1.224 & 1.287 & 0.707 & 0.137 \\
\enddata
\end{deluxetable}

The coefficients of our best fit relations for each element can be found in Table \ref{tab:uncertainties}, which we then use to calculate the empirical uncertainties for our full sample.  Specifically we report $\sigma_{\rm [X/Fe]}$ by summing the $\sigma_{\rm [X/H]}$ uncertainties we calculate in quadrature with the ASPCAP reported \xh{Fe} uncertainties.\\

We report both the measured and empirical uncertainties that we calculate so they can be applied as desired.  The measured uncertainties have the benefit of capturing the variation in measurement from line to line, but do not account for possible variation in stellar parameters (or other systematics like the analysis pipeline, or observations with different instruments/wavelengths of course).  On the other hand, the empirical uncertainties may not account for the conditions or line-to-line variations in an individual star, but do allow us to see what the general effect of different spectra and input stellar parameters has on the derived abundance variability.

\section{BAWLAS Chemical Abundance Patterns and Trends}
\label{sec:results}

Our BAWLAS catalog of input parameters, calculated abundances, upper or lower (for \cisotope{}) limits, and errors (as well as the line-by-line abundances and flags) can be found on the SDSS DR17 Value Added Catalog (VAC) page\footnote{\url{https://www.sdss.org/dr17/data_access/value-added-catalogs/}}.  Here we present some of the overall results.

\subsection{Individual Abundance Patterns}

\begin{figure*}
  \centering
  \includegraphics[scale=0.29,trim = 0.in 0.in 0.in 0.in, clip]{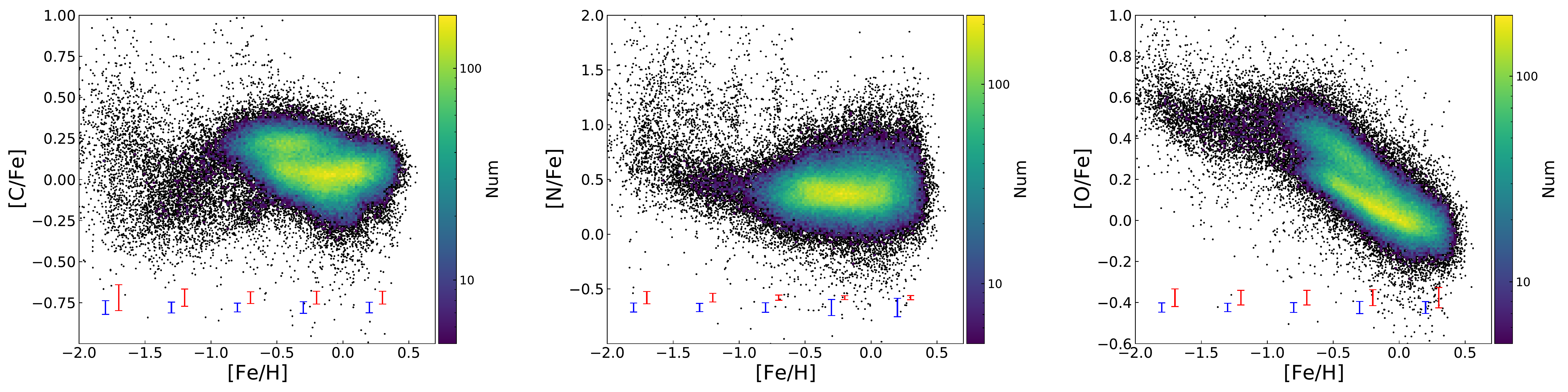}
  \caption{Elemental \xfe{X} abundances vs metallicity, \xh{Fe} of the BAWLAS results for C, N and O (black points and 2D histogram). Error bars show the median ``measured'' uncertainties from line-to-line scatter in measurements (red error bars) and the median ``empirical'' uncertainties from repeat measurements (blue error bars) in bins of 0.5 dex.}
  \label{fig:cno_abund}
\end{figure*}

\begin{figure*}
  \centering
  \includegraphics[scale=0.35,trim = 0.in 0.in 0.in 0.in, clip]{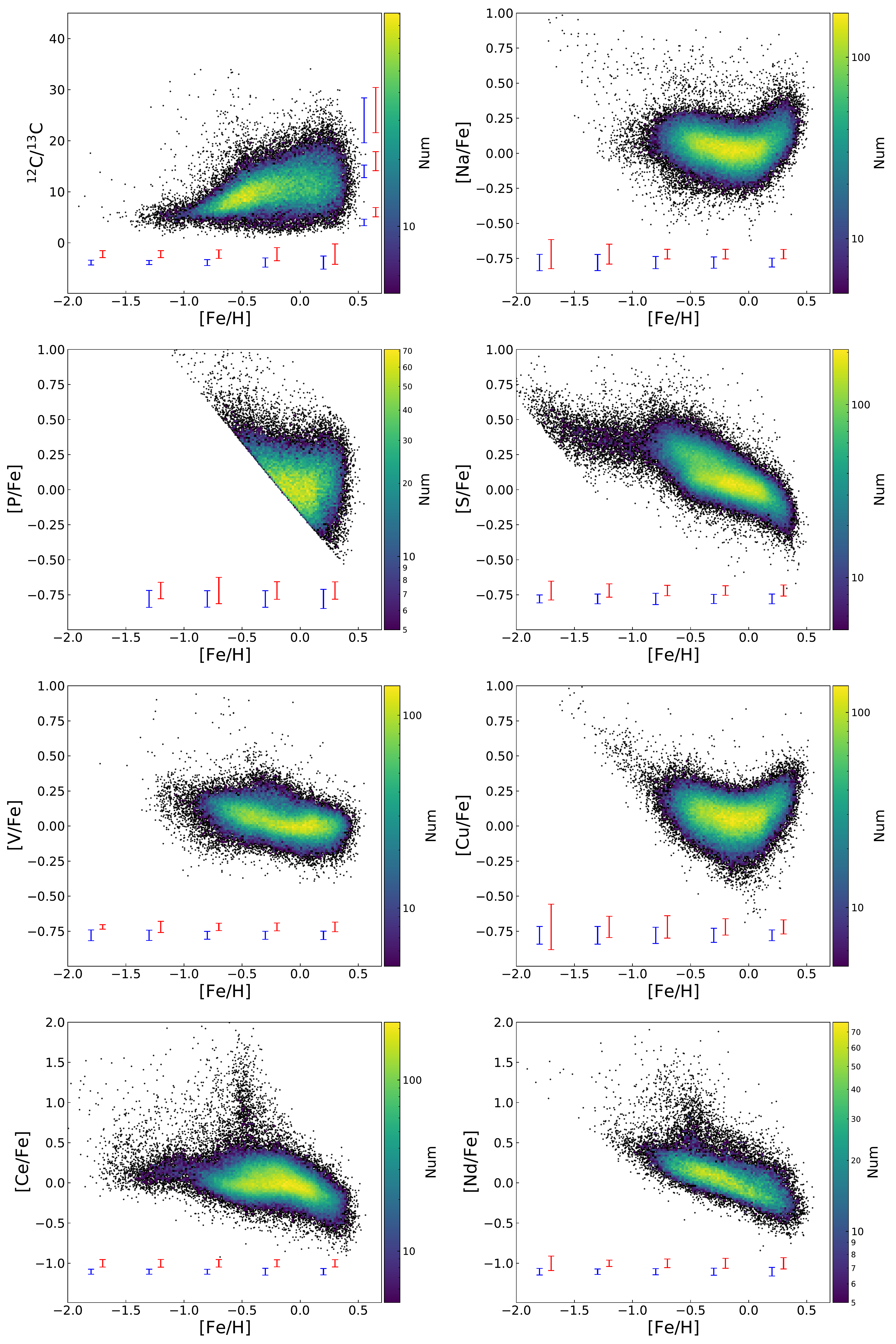}
  \caption{\cisotope{} ratio (top left) and elemental \xfe{X} abundances vs metallicity, \xh{Fe} of the BAWLAS results for our main elements of interest, \cisotope{}, Na, P, S, V, Cu, Ce, and Nd (black points and 2D histogram).  Error bars show the median ``measured'' uncertainties from line-to-line scatter in measurements (red error bars) and the median ``empirical'' uncertainties from repeat measurements (blue error bars) in bins of 0.5 dex.  For \cisotope{} we also show the median uncertainties in bins of \cisotope{} ratio (with bin size of 10), since the uncertainties grow significantly with increasing \cisotope{}. }
  \label{fig:xfe_abund}
\end{figure*}
  
In Figures \ref{fig:cno_abund} and \ref{fig:xfe_abund} we show the combined abundances from BAWLAS.  Figure \ref{fig:cno_abund} shows the C, N, and O abundances that were calculated primarily for the purposes of fitting blends, and Figure \ref{fig:xfe_abund} shows the goal elements that we measure:  \cisotope{}, Na, P, S, V, Cu, Ce, and Nd.  Before delving into each element (discussed below), we point out some of the key features seen in these abundance distributions.

Many of the elements do not cover the full metallicity range probed by our sample, which extends down to APOGEE's lower metallicity limit of \feh{} $= -2.5$.  Instead, for many elements the number of stars for which we measure that element begins to drop off quite rapidly around a metallicity of \xh{Fe} $\sim -1$ to $-1.5$.  This is in part because the underlying density of stars does begin to decrease at these metallicities.  But, this also occurs because most of the elements we examine have relatively weak (and few) lines that become increasingly difficult to measure at low metallicities. 

Essentially only S, Ce, and \cisotope{} are measured below \xh{Fe} $\sim -1.5$, but even these measurements  may be biased towards detecting stars with stronger spectral features, e.g, higher abundances for S and Ce, or lower values of \cisotope{}.  Indeed the imprint of upper limit flagging for S appears as the diagonal line at low metallicities below which we do not populate any S abundances.  P provides another example of a clear hard upper limit flagging, however upper limit flagging also affects the distribution of V, Cu, Ce, and Nd abundances in a less obvious manner.  

\begin{figure*}
  \centering
  \includegraphics[scale=0.29,trim = 0.in 0.in 0.in 0.in, clip]{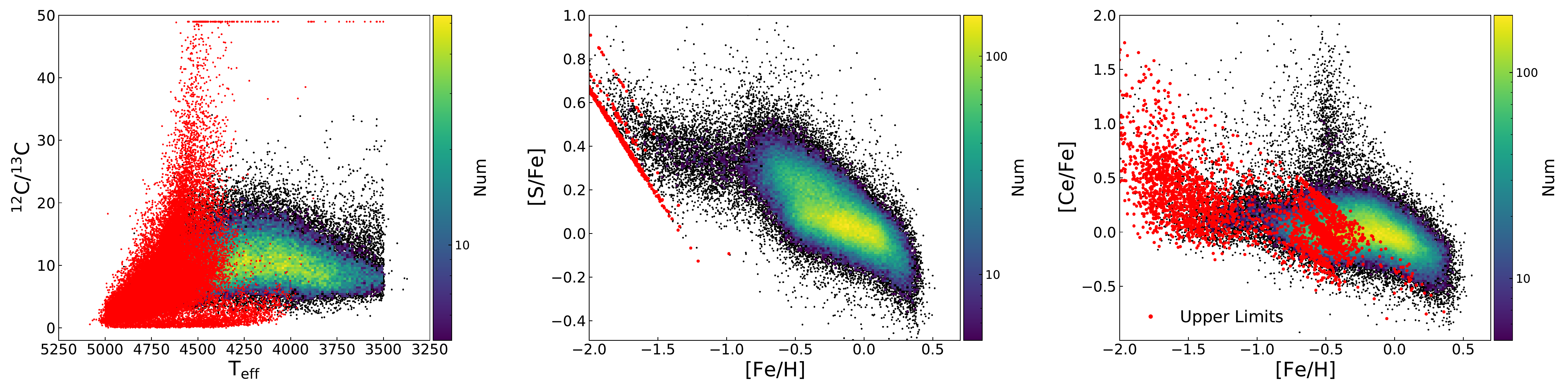}
  \caption{\cisotope{} (left) versus T$_{
\rm eff}$, and \xfe{S} (middle) and \xfe{Ce} (right) versus \feh{} for the BAWLAS sample (black points and 2D histogram) with upper limits overplotted for stars with well-determined upper limits (red points).}
  \label{fig:sce_upperlimits}
\end{figure*}

Figure \ref{fig:sce_upperlimits} shows an example of these upper limits for S and Ce.  The temperature insensitivity of S is apparent as the very narrow spread of upper limits, which come in three tiers (of nearly constant \xh{S}), corresponding to the three S lines that are used, with the two weaker lines providing upper limits when the quality of the spectrum around the strongest line is too poor to measure or place upper limits on.  Ce instead has more temperature sensitive upper limits, which means that \xh{Ce} upper limit as a function of temperature turns into a less localized spread of upper limits as a function of \feh{}.

\subsubsection{Carbon and Nitrogen (C and N)}

While C, N, and O were measured to fit blends, we can also examine the BAWLAS abundance patterns in these elements.  In the C abundances of Figure \ref{fig:cno_abund} we see some internal structure at intermediate metallicities, \feh{} $\sim -0.5$, that appears to the superposition of the thin and thick disk populations.  Whereas the N abundances do not show clear substructure in its chemical abundance pattern, instead showing a large spread, in part due to the larger uncertainties in warm stars.  At low metallicities, C begins to decrease before showing a large scatter, whereas N shows a slightly rising trend with decreasing metallicity.

In addition to the general trends at low metallicities some stars are found with particularly enhanced \xfe{C} or \xfe{N} ($\gtrsim 0.5-1.0$). In the case of N, many of these stars belong to globular clusters which are known to have C-N anti-correlations with high N abundances \citep[e.g.,][]{Smith1996, Gratton2001, Briley2004, Meszaros2015, Masseron2019, Meszaros2020}.

As for the stars with high C abundances there are a few ways that stars may get C-enhancements.  For example, carbon stars will often show higher C abundances, and can be formed in a few ways.  Carbon stars can either be intrinsically enhanced in C, e.g., through the dredge-up of carbon rich material fused in a star's interior, or extrinsically enhanced by C-rich material accreted from an evolved companion, altering its surface chemistry \citep[for a more in depth discussion see][]{Wallerstein1998, LloydEvans2010}.  At low metallicities there are also known to be carbon enhanced metal-poor (CEMP) stars (notable for their enhanced \xfe{C} ratios), which come in a number of astrophysical varieties \citep[discussed more in][]{Beers2005, Masseron2010, Frebel2015}.

\subsubsection{Oxygen and Sulfur (O and S)}

Examining O in Figure \ref{fig:cno_abund} and S in Figure \ref{fig:xfe_abund} we see abundance distributions that are fairly typical for $\alpha$ elements, with a nearly flat plateau at low metallicities that decreases at higher metallicities with a knee occurring around \feh{} $\sim -0.7$ to $-0.5$.  In O we see a bifurcation in \xfe{O} at metallicities $> -0.7$ into high- and low-$\alpha$ sequences that are commonly attributed to the thick and thin disks, respectively.  The \xfe{O} plateau appears to be quite flat, although we note that it seems to widen around metallicities between $-1.5$ and $-1$, which, as previously observed, suggests the presence of stars that have been accreted from dwarf galaxies \citep{Nissen2010,Hawkins2015,Hayes2018a}. 

In some ways S appears qualitatively similar to O.  While there is not quite a bimodality in the \xfe{S} distribution, at metallicities around $-0.5$ (where the thin and thick disk are most chemically distinct in other $\alpha$-elements), the spread of the \xfe{S} distribution exceeds what would be expected from the reported errors alone.  Therefore, this is likely a true astrophysical spread in the S abundances at these metallicities.  Similar to the \xfe{O} bifurcation, this spread would seem to be tied to differing $\alpha$-abundances in the thin and thick disk.  At low metallicities we see a slightly sloped plateau in \xfe{S}, but the upper limit flagging also begins to impact the completeness of our S abundance measurements below \feh{} $\sim -1.5$, so this sloped appearance may be somewhat artificial.

\subsubsection{\cisotope{} Isotopic Ratio}
\label{sec:c12c13_results}

\begin{figure}
  \centering
  \includegraphics[scale=0.4,trim = 0.in 0.in 0.in 0.in, clip]{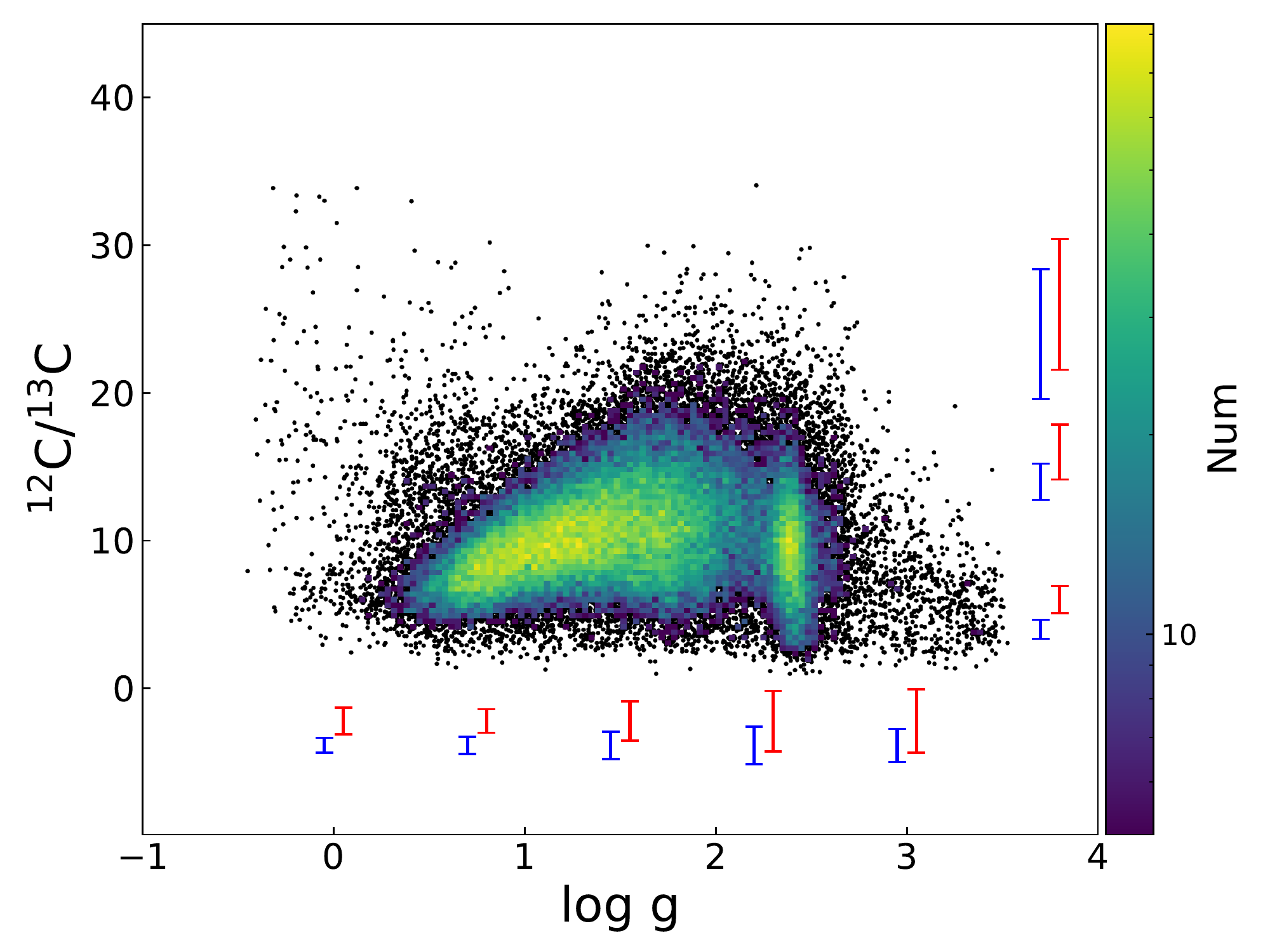}
  \caption{\cisotope{} isotopic ratio vs. surface gravity, $\log g$ of the BAWLAS sample, with color-coding and error bars as in Figure \ref{fig:xfe_abund}.  At $\log g < 2.5$ we see the \cisotope{} ratios decrease and spread of \cisotope{} ratios narrows up the giant branch with decreasing $\log g$. The overdensity at $\log g \sim 2.5$ comes from the red clump that appears to have lower average \cisotope{} than red giant stars at slightly lower $\log g$.}
  \label{fig:c12c13_logg}
\end{figure}

Looking at Figure \ref{fig:xfe_abund} we see that at high metallicities the \cisotope{} ratios measured in BAWLAS range from a few up to $\sim 20$.  This range then narrows and the average \cisotope{} ratio decreases with decreasing metallicity.  These \cisotope{} ratios are also expected to change as a function of $\log g$ due to mixing along the giant branch.  Examining this distribution with $\log g$ in Figure \ref{fig:c12c13_logg}, we see two features, the red clump at $\log g \sim 2.5$ and a distribution of red giant stars at $\log g < 2.5$.  We measure \cisotope{} ratios for very few $\log g \gtrsim 2.5$ stars, because the $^{13}C$ features become too weak at the warm temperatures of these stars, and therefore at low $\log g$ we predominantly report upper limits.

For the stars where we do measure \cisotope{} ratios, the red clump spans a range of \cisotope{} ratios, but sits at slightly lower \cisotope{} ratios, than the red giants at slightly lower $\log g$.  Along the giant branch, the \cisotope{} ratios appear to decrease with decreasing $\log g$.

\begin{figure*}
  \centering
  \includegraphics[scale=0.35,trim = 0.in 0.in 0.in 0.in, clip]{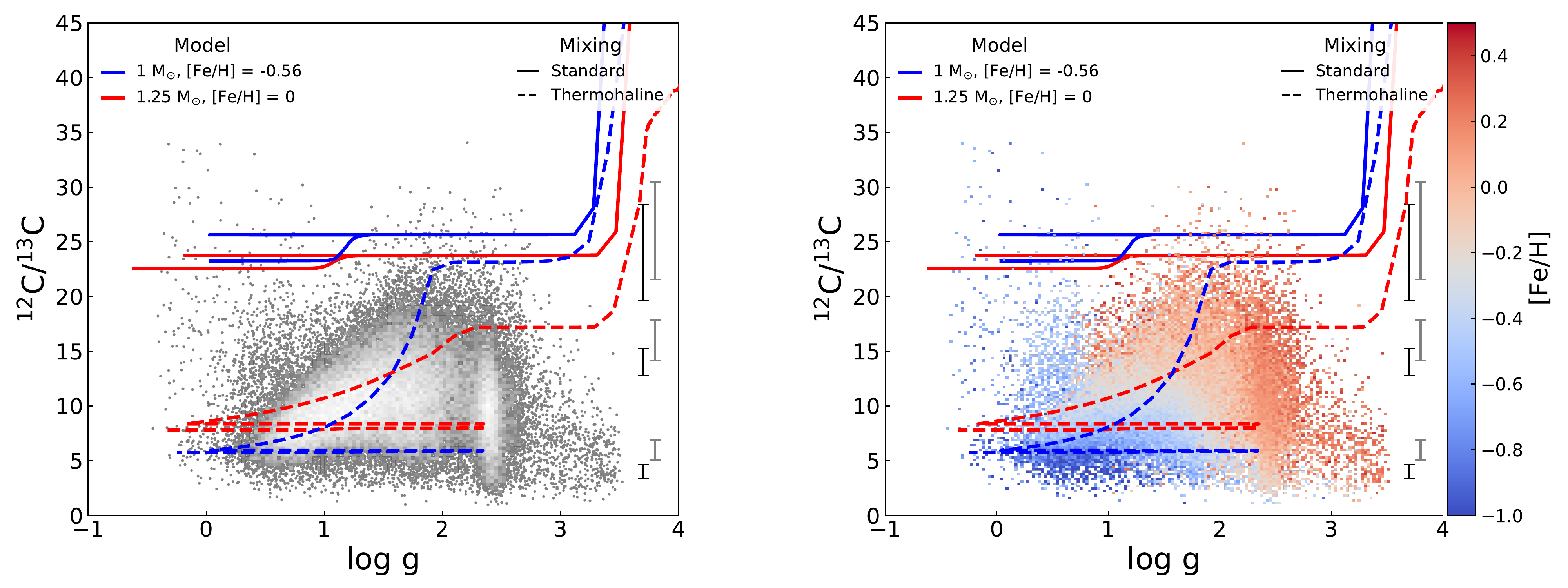} 
  \caption{(Left panel) \cisotope{} isotopic ratio vs. surface gravity, $\log g$ of the BAWLAS sample (in grayscale, with median ``measured'' and ``empirical'' uncertianties as gray and black respectively).  Overplotted are stellar evolution models from \citet{Lagarde2012} for standard and thermohaline mixing (solid and dashed lines, respectively) at two characteristic stellar mass-metallicity combinations ($M$, \feh{}) $=$ (1.0 $M_{\odot}$,$-0.56$) and (1.25 $M_{\odot}$,$0.0$) (blue and red, respectively), which represent typical values expected in this sample.  (Right panel)  \cisotope{} isotopic ratio vs. surface gravity, $\log g$ of the BAWLAS sample in bins that are color-coded by the average metallicity in that bin, with the same models overplotted.  The models illustrate how the \cisotope{} evolves in a star as it ascends the red giant branch (following a given track along decreasing $\log g$), jumps to the red clump at $\log g \sim 2.5$, after reaching the tip of the RGB, and then ascends the asymptotic giant branch at slightly lower \cisotope{} ratios.}
  \label{fig:c12c13_logg_models}
\end{figure*}

This is interesting because these trends generally do not match predictions from standard mixing as shown in Figure \ref{fig:c12c13_logg_models}.  Instead, as has been noted in the past \citep[e.g.,][]{Szigeti2017}, we see that \cisotope{} changes along the red giant branch.  This feature indicates that extra-mixing occurs along the red giant branch following the red giant branch ``bump,'' and, while the dominant source of this extra-mixing has yet to be settled, many different theoretical models have been proposed \citep[such as thermohaline mixing, rotational mixing, gravity waves, or magnetic fields][and references therein]{Charbonnel1998, Denissenkov2000, Charbonnel2007, Busso2007, Karakas2010, Lattanzio2015}.

Here we compare with one of these extra-mixing models, thermohaline mixing, which (at least qualitatively) matches with the \cisotope  ratios that we measure.  We have selected two mass-metallicity combinations of stellar evolution models from \citet{Lagarde2012} to illustrate what stellar evolution trends we might expect for stars of different mass and metallicity.  We show model tracks for mass-metallicity combinations of ($M$, \feh{}) $=$ (1.0 $M_{\odot}$,$-0.56$) and (1.25 $M_{\odot}$,$0.0$).  These values are chosen to be reasonable for typical high-$\alpha$ thick disk stars, and low-$\alpha$ thin disk stars, given the distribution of thin and thick disk RGB star masses found by \citet{Pinsonneault2018}\footnote{\citet{Pinsonneault2018} found that thick disk RGB stars in the {\it Kepler} field are typically around a solar mass, while their thin disk sample spans a larger range in masses, but is more massive than one solar mass on average.}.  

While standard mixing models imply relatively constant \cisotope{} along the upper giant branch (with only significant changes in \cisotope{} as a star begins to ascend the giant branch at $\log g \sim 3.8$ and again slightly as it passes the tip of the giant branch), thermohaline mixing models predict that the surface \cisotope{} ratio should decline as a star evolves up the upper RGB post RGB-bump (occuring around or above $\log g \sim 2$ depending on mass and metallicity).  Our \cisotope{} ratios also show a correlation with metallicity in red clump stars (at $\log g \sim 2.5$), which as shown in the right panel of Figure \ref{fig:c12c13_logg_models}, causing the large spread in \cisotope{} at these surface gravities, and again qualitatively agreeing with thermohaline mixing predictions.  This is generally what we see in our \cisotope{} measurements, but one must take care when interpreting the \cisotope{} ratios exclusively as a function of $\log g$ or metallicity because of the biases in our sample and our detection limits.

For example our sample is biased towards including metal-poor stars at lower $\log g$ and excluding those at high $\log g$.  This comes both from the observational biases to observe more luminous, low $\log g$ metal-poor stars, and from our temperature cuts that remove warm, higher $\log g$ metal-poor stars.  In addition, our detection limits, bias our sample to the stars that have lower \cisotope{} ratios (because that implies $^{13}C$ is more abundant and has stronger features), higher C abundances, or in cool stars, whose molecular features are stronger.  So the trends we observe may be a combination of physical trends and observational and measurement selection biases.

\begin{figure*}
  \centering
  \includegraphics[scale=0.35,trim = 0.in 0.in 0.in 0.in, clip]{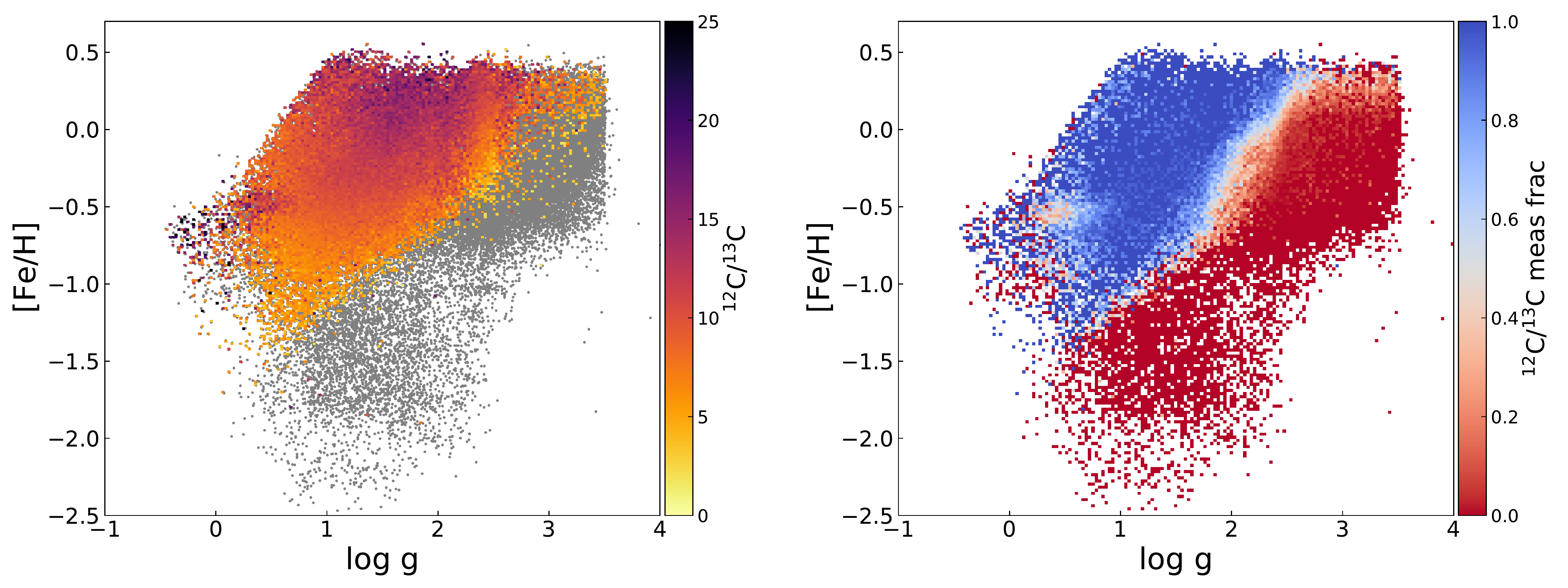}
  \caption{Metallicity, \feh{}, vs. surface gravity, $\log g$, of the BAWLAS sample binned and color-coded to show the average \cisotope{} (left; stars that do not have \cisotope{} ratios measured are shown as gray points) and the fraction of stars which have \cisotope{} measurements (right) within each bin.  \cisotope{} gradients can be see with respect to both \feh{} and $\log g$, even at parameter ranges where the completeness is nearly $100\%$ (e.g., $\log g$ between 1 and 2 and metallicities between $-1$ and solar).  Clear regions of incompleteness can be seen (right), particularly at high metallicities, from flagging stars whose O abundances were not updated by BACCHUS (see Section \ref{sec:param_flags}), and in stars with low metallicities or high $\log g$, because the $^{13}$C features become weak these stars.}
  \label{fig:c12c13_completeness}
\end{figure*}

To investigate the significance of biases that might be present in our sample, Figure \ref{fig:c12c13_completeness} shows the average \cisotope{} ratio in bins of $\log g$ and metallicity in the left panel, and the fraction of stars which have measured \cisotope{} ratios within each of those bins in the right panel.  This clearly shows that there are some biases in our sample.  For example, at low metallicities, $\lesssim -1$, we only measure \cisotope{} in some of the lowest $\log g$ stars.  It also shows that in the red clump ($\log g \sim 2.5$) at metallicities, \feh{} $\sim -0.5$, we measure very low \cisotope{} ratios ($\lesssim 5$) on average, but we are also measuring \cisotope{} ratios in less than 50\% of the stars with similar stellar parameters.  In regions such as this we are only able to measure \cisotope{} ratios in the stars with the lowest ratios, and all other stars are upper limit measurements.

Figure \ref{fig:c12c13_completeness}, however, shows that we are nearly complete in measuring \cisotope{} in our sample at metallicities between $-1$ and solar, and $\log g$ between 1.5 and 0 (with a small drop at low $\log g$ around \feh{} $ \sim -0.5$, because of luminous LMC stars that have been flagged with high convolution values that may have erroneous APOGEE stellar parameters).  Further, this suggests that the gradients seen in the \cisotope{} ratios in this parameter range in the left panel are not driven by selection effects and are real trends in our data.  

Therefore, the decreasing \cisotope{} ratios with decreasing $\log g$ and metallicity in this range are qualitatively consistent with the expectations of the thermohaline mixing models shown above from \citet{Lagarde2012}.  Future works should examine these \cisotope{} ratios more carefully to compare with theoretical predictions, ideally examining stars whose mass, metallicity, and stage of evolution are known such as stars in this sample that have astroseismology from {\it Kepler} or cluster stars, whose masses and ages can be relatively easily measured.  This could provide a more detailed comparison with stellar evolution models.

\subsubsection{Sodium (Na)}

For Na we see a slightly decreasing trend in \xfe{Na} as a function of increasing metallicity, that begins to rise at super-solar metallicities.  The other noticeable feature in the Na distribution are a cloud of stars with high \xfe{Na} ($\gtrsim 0.5$).  { While high Na abundances are seen in some literature studies of field stars \citep[e.g.,][]{Duong2019} and globular clusters \citep{Gratton2001, Briley2004, Meszaros2015, Masseron2019, Meszaros2020} and some of the high Na measurements may be real, others may be suspect, because some of these stars are affected by an inaccurately subtracted sky line.  In many cases we have attempted to flag stars where this sky line subtraction affects the Na abundance measurement (see Appendix \ref{sec:na16388blendflag} for more information), but in some of the more marginal cases, these stars have passed our flagging criteria and remain in our sample with erroneously high Na abundances.  }

In general these high \xfe{Na} stars should be treated cautiously.  Many of these measurements are of only one line, the Na 16388 \AA{} line, which is affected by this sky subtraction issue particularly in stars with radial velocities between $\sim -110$ and $-60$ \kms{}.  The stars that are truly enhanced in Na should have strong enough Na lines that one may expect Na to be measurable in both of the lines we use.  So those interested in exploring stars enhanced in Na may want to examine stars that have both Na lines measured and preferably have radial velocities that are not in the range of $\sim -110$ and $-60$ \kms{}.

\subsubsection{Phosphorus (P)}

Most prominent in the P abundances we measure is the considerable upper limit on the \xh{P} values we are able to measure.  Given the extent of the upper limit flagging on P, we are likely incomplete in our \xfe{P} coverage at nearly all metallicities.  Nonetheless this does provide us the largest sample of stars with P measured present in the literature, and there is some preliminary evidence for a rising \xfe{P} with decreasing metallicity judging from the upper envelope of our observed sample.  We can also see that there is a possible hint of an increase in the \xfe{P} ratios at super-solar metallicities.

While our P distribution is limited in it's metallicity coverage, some of these features may still be able to provide further constraints to chemical evolution models and nucleosynthetic yields for P.  Most of the theoretical P yields have previous been shown to poorly match observed P abundances in the MW \citep[e.g.,][]{Maas2019, Kobayashi2020}.  These predictions typically find sub-solar \xfe{P} ratios at or above solar metallicities, which also seem to be ruled out by our measurements (which is unsurprising given that our P abundance pattern seems to agree reasonably well with past observations, see Section \ref{sec:lit_comp2}).  This disagreement suggests that most theoretical yields do not produce enough P and more modeling is needed to reconcile these differences.  However our data is qualitatively consistent with some of the features produced by predictions, such as the decreasing \xfe{P} ratios with increasing metallicity at sub-solar metallicities and the slight increase in \xfe{P} at super solar metallicities seen by \citet{Kobayashi2020}.

\subsubsection{Vanadium and Copper (V and Cu)}

V and Cu both show a similarly shaped abundance distribution pattern at sub-solar metallicities, with slightly elevated \xfe{X} at low metallicities and decreasing with metallicity.  While V flattens out around solar metallicity, Cu instead begins to increase.  In addition, the \xfe{V} abundances show a tighter distribution than Cu, however Cu also has larger uncertainties because the Cu lines are significantly weaker than the V lines that we use, likely inflating the Cu distribution.

While the production of Cu is somewhat more complicated, it is close to the Fe-peak and produced similar to bona fide Fe-peak elements such as V.  Both of these elements show chemical abundance patterns like that of other Fe-peak elements measured by APOGEE, such as Co and Ni, although here we don't see a rise in \xfe{V} at super solar metallicities, as seen in APOGEE's Co and Ni abundances and our BAWLAS Cu abundances.  Some studies have suggested that several Fe-peak elements may have metallicity dependent nucleosynthetic production in Type Ia SNe at high metallicities \citep[including V][]{Weinberg2021}, {whereas other studies that have seen increases in Cu abundances at super-solar metallicities, like the increases seen here, have suggested that it is due to metallicity dependent yields of Cu in massive stars through the weak s-process \citep{Johnson2014, McWilliam2016, Xu2019}.  In either case,} the stronger rise in Cu abundances at high metallicities may suggest that Cu has a stronger metallicity dependence in its nucleosynthetic production than V, but such modeling is beyond the scope of this work.  

\subsubsection{Cerium and Neodymium (Ce and Nd)}
\label{sec:cend_results}

The two neutron-capture elements in APOGEE spectra with the strongest lines, Ce and Nd, are both produced by a mix of the r- and s-process in the sun.  Ce is a predominantly s-process element in the sun with percentages of 19\%/81\% r-/s-process, whereas Nd has a more even mix, with a slight s-process lean at percentages of 42\%/58\% r-/s-process \citep{Prantzos2020}.

The Ce abundances measured here show one of the more complex abundance patterns seen in this study (partly because it is one of the most precise, non-CNO elements that we measure).  At low metallicities, between $-2$ and $-1$ we see a slightly rising \xfe{Ce}.  Then at higher metallicities, the bulk of the stars have an arched pattern with slightly lower \xfe{Ce} at \feh{} $\sim -0.7$ rising to peak at \feh $\sim -0.25$ before decreasing with increasing metallicity.  This is similar to the \xfe{Ba} abundance pattern \citep[another predominantly s-process element in the sun,][]{Prantzos2020} seen by GALAH in their GALAH$+$ DR3 \citep{Buder2021}.

On the other hand, while Nd is not measured to as low of metallicities, it does show a different chemical abundance pattern, likely because of its greater production in the r-process.  The bulk of the stars with measured Nd form a relatively simple pattern of decreasing \xfe{Nd} abundance ratio with increasing metallicity.  Qualitatively this is indeed between the abundance patterns of Ce and the predominantly r-process element, Eu (97\% r-process in the sun) seen with GALAH \citep[and, reassuringly, similar to their Nd abundance pattern][]{Buder2021}.

\begin{figure}
  \centering
  \includegraphics[scale=0.4,trim = 0.in 0.in 0.in 0.in, clip]{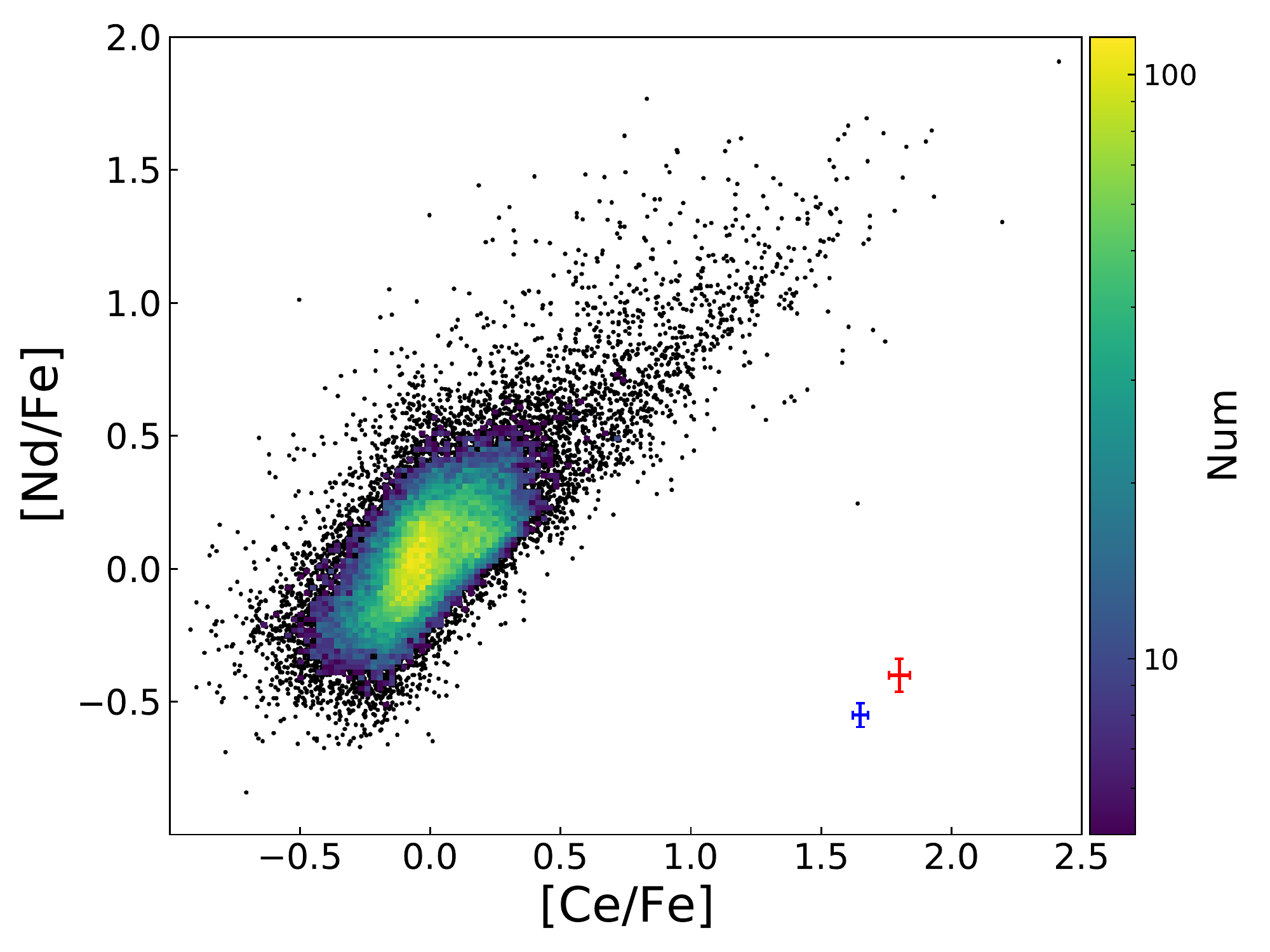}
  \caption{Nd abundances, \xfe{Nd}, vs. Ce abundances, \xfe{Ce} for stars in the BAWLAS sample where Ce and Nd are both measured, and excluding the LMC red supergiants described in Section \ref{sec:lmc_rsg}.  The error bars show the median ``measured'' (red) and ``empirical'' (blue) abundance uncertainties across the sample, as representative uncertainties.  While there is a significant scatter in the \xfe{Ce} and \xfe{Nd} of stars with Ce and Nd enhancements, the enhancement in these elements is correlated and most of the stars enhanced in Ce are also enhanced in Nd and vice versa.}
  \label{fig:cefendfe}
\end{figure}

We also find a population of stars with enhancements in Ce and Nd, extending in some cases to \xfe{Ce} or \xfe{Nd} of $+2.0$.  Figure \ref{fig:cefendfe} shows that these enhancements are generally correlated, so that Ce-enhanced stars typically show Nd-enhancements and vice-versa.  While Ce and Nd are measured from singly ionized transition lines and could be similarly sensitive to systematic parameter errors, \citet{Hasselquist2016} and \citet{Cunha2017} showed that the expected errors in Ce and Nd abundances should be $< 0.1$ dex for Temperature errors of $\sim 100$ K and gravity errors of $\sim 0.2$ dex (and the Ce and Nd errors are anti-correlated for $\log g$ errors).  So systematic errors in $\rm T_{\rm eff}$ or $\log g$ don't seem to be able to account for the enhancements we see (even if there were quite large systematic errors).

Instead, many of these stars with Ce- and Nd-enhancements are likely s-process enhanced stars which can occur because they have dredged up s-process rich material during their evolution \citep[such as N-type carbon stars;][]{LloydEvans2010} or have accreted s-process rich material from a companion, e.g., Ba stars, CH stars, or CEMP-s stars \citep[][and references therein]{Mcclure1984, Masseron2010, LloydEvans2010}.  But it is also possible that stars with r-process enhancements may be present in this sample too (although there are fewer methods of producing r-process enhancements at these metallicities).  

It has long been seen that roughly $\sim 1\%$ of RGB stars are Ba/CH stars \citep{BoehmVitense1984}, and indeed with a very rough calculation we find a similar percentage of our sample is Ce-enhanced (we consider Ce over Nd, since Ce is measured in a larger fraction of our sample).  The density of the bulk sample of stars with Ce measurements begins to drop off rapidly above \xfe{Ce} $> 0.5$, so if we take this as an indication of where we begin to see these various classes of s-process enhanced stars, we find about 1,500 of the 106,000 stars with measured Ce abundances would be considered s-process rich, about 1.4$\%$ of our red giant sample (all of this excluding LMC red supergiants for the reasons mentioned below).  While this calculation can and should be done more carefully (though doing so is beyond the scope of this work), this simple order of magnitude estimate suggests that the number of Ce-enhanced stars (and Nd) agrees well with past predictions of the number of s-process enhanced red giants.

There is also a notable overdensity of Ce-enhanced stars (and Nd-enhanced stars) at a metallicity of -0.5, which appears to be predominantly due to red supergiants in the LMC.  While LMC stars at higher metallicities may in general be enhanced in s-process elements \citep[e.g.,][]{Hasselquist2021}, we suspect the LMC red supergiants have suspect stellar parameters, leading to erroneous chemical abundances not only in Ce and Nd but other elements too, as discussed below in Section \ref{sec:lmc_rsg}.

\subsection{Effective Temperature Trends}
\label{sec:teff_trends}

\begin{figure*}
  \centering
  \includegraphics[scale=0.35,trim = 0.in 0.in 0.in 0.in, clip]{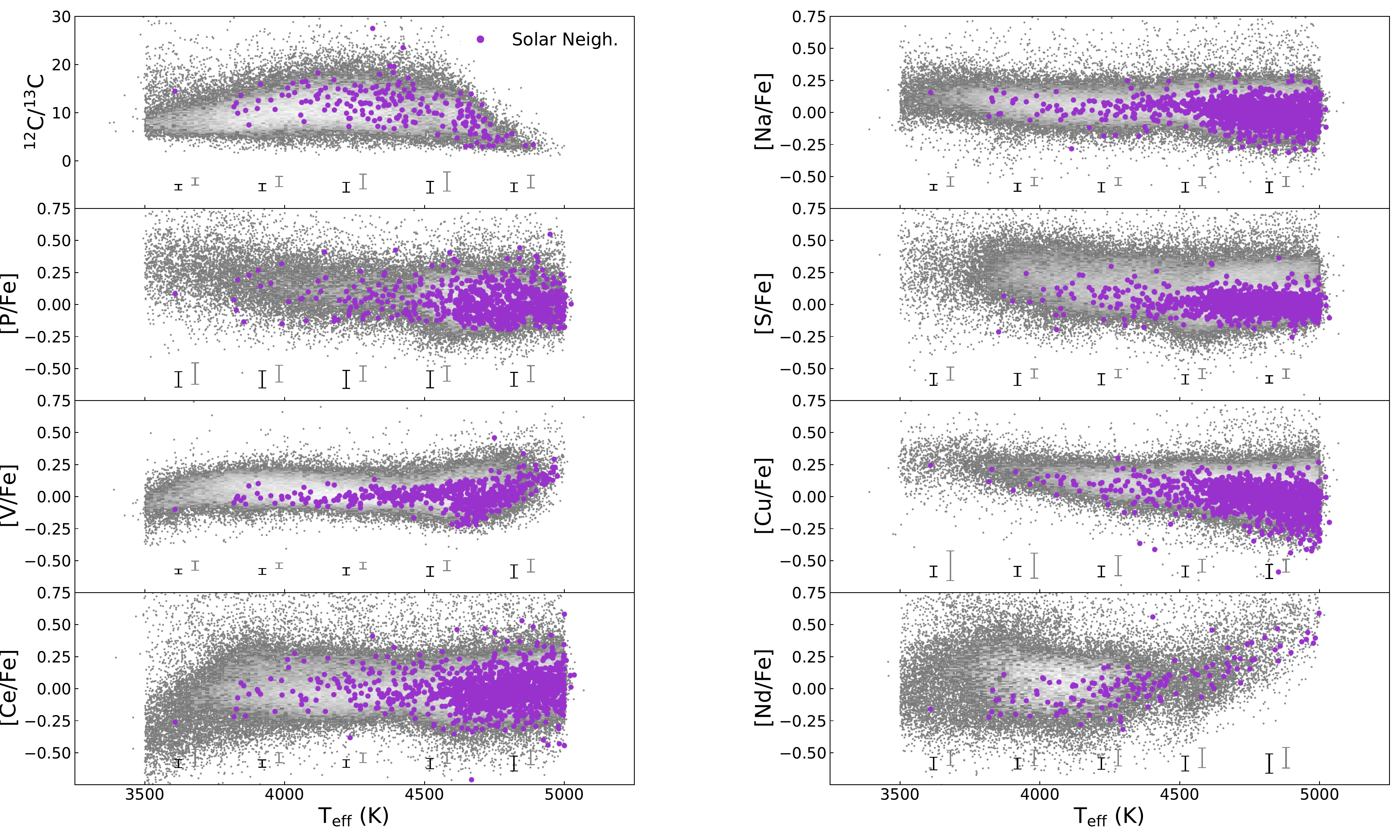}
  \caption{\cisotope{} ratio (top left) and elemental \xfe{X} abundances vs. effective temperature, $T_{\rm eff}$ for our main elements of interest, \cisotope{}, Na, P, S, V, Cu, Ce, and Nd (full BAWLAS sample in gray points and 2D histogram).  The error bars show the median uncertainties of stars in bins of 300 K (color-coded as in Figure \ref{fig:c12c13_logg_models}).  Overplotted is our solar neighborhood sample (purple points), a near mono-abundance population that we would nominally expect to have \xfe{X} $ = 0$, but illustrates temperature trends in some elements.}
  \label{fig:xteff_abund}
\end{figure*}

As with all spectroscopic studies, systematic trends with stellar parameters can be difficult to avoid, particularly when analyzing weak and blended lines, or may be entirely unavoidable in cases where lines are affected by non-LTE effects for example.  To investigate what kinds of trends may be present in this analysis, we show our derived abundances as a function of stellar effective temperature in Figure \ref{fig:xteff_abund}.  Because temperature trends can appear due to selection biases, we have also highlighted our solar neighborhood sample (using the same selection criteria listed in Section \ref{sec:zero_points}, but without the S/N restriction to increase the sample size), which represents a relatively mono-abundance population that should have a solar \xfe{X} abundance ratio at all temperatures.

In the solar neighborhood sample, we do not see particularly significant trends in Na, P, S or Ce, however, the upper giant branch is not well populated because such cool stars are not common in the solar neighborhood.  Looking at the full sample for these elements there may be some hints of temperature trends in the coolest stars, e.g., a gradual rise in P abundances below $\sim 3900$ K, a downward trend in Ce below $\sim 3800$ K, etc.  Because cooler red giants are more luminous and can therefore be seen to larger distances than warmer, fainter red giants (for the same S/N limit), these trends could be coming from selection biases, i.e., picking out cool stars in regions that are not well populated by the rest of the sample such as the MW bulge.  Alternatively, these could be systematic errors because molecular features (the dominant source of blends) become especially strong in cool stars, so any errors in blend treatment could lead to erroneous abundance measurements.

For the remaining elements, we do see some structure or trends in their abundances as a function of temperature.  The impact of our detection threshold and limit flagging can be seen in \cisotope{}, V, and Nd, particularly at warm temperatures $\lesssim 4600$ where only low \cisotope{} ratios ($\lesssim$ 10) and high \xfe{V} or \xfe{Nd} abundances ($\gtrsim$ solar) can be measured.  This compounds with the increasing uncertainties and scatter in V and Nd at low metallicities to cause a decreasing trend in \xfe{X} as a function of temperature below $\sim 4500$ K. 

At lower temperatures, there may be very slight \cisotope{}, V, and Nd trends with temperature, but it is difficult to assess, given the low number of cool stars with measurements in the solar neighborhood.  Looking at the full sample, similar to Ce, we do see a decrease in \xfe{V} at temperatures below $\sim 3800$ K.  We also see a change in the average \cisotope{} ratio of the full sample as a function of temperature, however, as mentioned in Section \ref{sec:c12c13_results}), the likely reflects astrophysical \cisotope{} trends as stars evolve up the giant branch due to internal mixing.

Finally, Cu shows the strongest temperature trend of the elements we measure.  The \xfe{Cu} rises from near solar abundances at 5000 K up to around 0.2 at 3500 K.  This is perhaps unsurprising given that the Cu lines used here are quite blended.  This trend is likely reflecting the difficulty of measuring these lines when they are strongly blended in cool stars.  Therefore, any investigation of the Cu abundances measured here should consider these trends and what impact they may have. 

It may be advisable to use caution even for those elements that do not show obvious temperature trends, since precision and uncertainties also change with stellar parameters.  Comparing stars with similar temperature ranges may provide one way of avoiding these biases.

\subsection{Abundance Uncertainties}
    
In Figures \ref{fig:cno_abund}, \ref{fig:xfe_abund}, and \ref{fig:xteff_abund}, we also show the typical measured and empirical abundance uncertainties in our sample as a function of metallicity and temperature (and as a function of \cisotope{} for its uncertainties).  In general the typical measured and empirical uncertainties are of a similar magnitude, however, the range of measured uncertainties may be larger since they are determined on a star-by-star basis rather than via an empirical relationship.

For some elements, we can see that the measured and empirical uncertainties differ more significantly.  For example, in N the measured uncertainties are, on average, much smaller than the empirical uncertainties at high metallicities.  This is because there are many N lines with which to measure these abundances, so we can very precisely measure the average N abundance in a star.  However, since N is measured from molecular features, it is expected that the measured N abundances will be quite sensitive to the input stellar parameters, so when varying the input stellar parameters in our derivation of empirical abundance uncertainties we find a larger range of resulting N abundances.

In Cu, we see a different case, where the measured uncertainties are significantly larger than the empirical uncertainties in cool stars (see Figure \ref{fig:xteff_abund}).  This occurs because the Cu I 16006 \AA{} is contaminated by a strong nearby blend, and BACCHUS \chitwo{} method, which is used to calculate our measured uncertainties, is biased by this blend. It derives significantly different Cu abundances than either the other Cu line, or the BACCHUS \wln{} method that was also used to determine uncertainties.  

The large difference between line and method measurements for Cu in cool stars leads to these large measured uncertainties, despite the fact that the average abundances are not affected by this issue, since we use the \wln{} method for determining the abundances of the Cu I 16006 \AA{} line, and only included this \chitwo{} method for the purpose of determining uncertainties.  The Cu measurement uncertainty in these cool stars may, therefore, be artificially inflated, but this can also indicate how uncertain the Cu values could be if blends are not properly treated.

\subsection{LMC Supergiant Feature}
\label{sec:lmc_rsg}
    
One of the noticeable features in the abundance patterns of a few of the elements is the overdensity of stars at \feh{} $\sim -0.5$ that cover a wide range of abundance ratios.  This can be seen most clearly in Ce and Nd, and somewhat less obviously in \cisotope{}, but it also occurs in the other elements to a lesser extent, such as S, covering a smaller spread of \xfe{X}.  This feature is primarily from red supergiants (RSGs) in the Large Magellanic Cloud (LMC).  These stars are relatively metal-rich stars for the LMC, and because they are all young, recently formed stars, they only cover a narrow range of metallicity.

\begin{figure*}
  \centering
  \includegraphics[scale=0.35,trim = 0.in 0.in 0.in 0.in, clip]{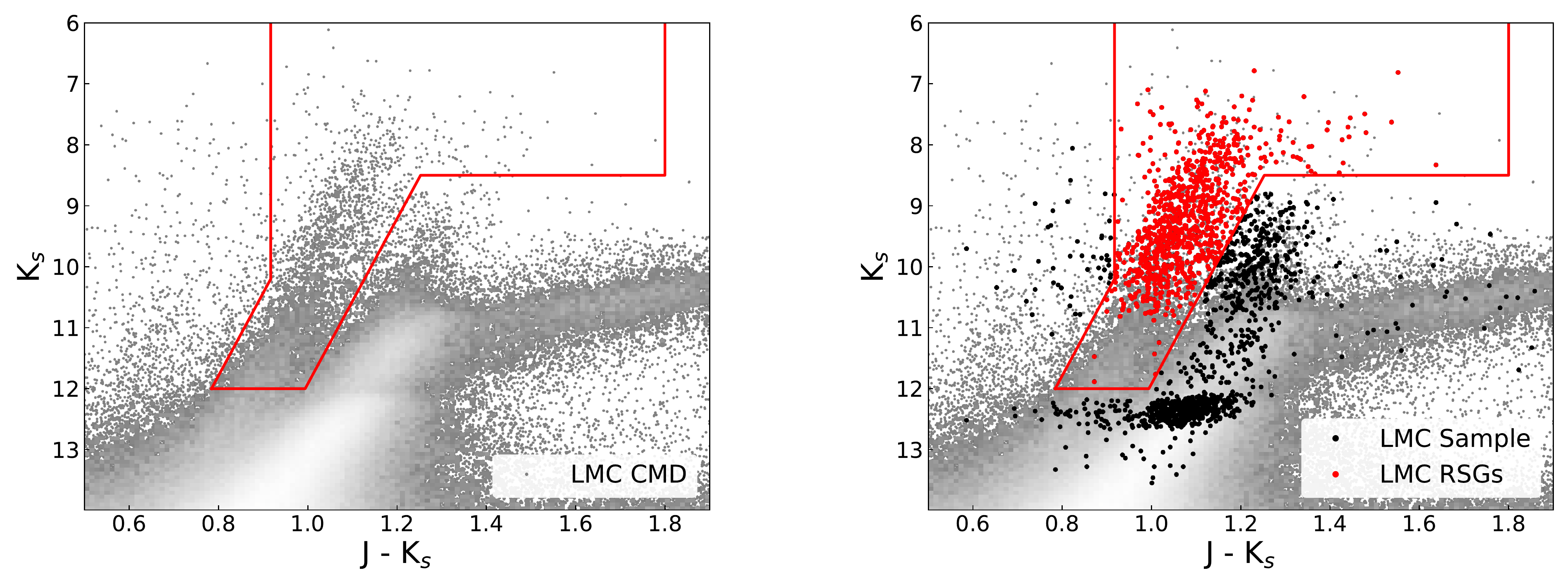}
  \caption{2MASS J vs. J-K$_s$ CMD of proper motion selected LMC stars in a 10\degree{} field centered on the LMC (left panel) overplotted with LMC red supergiants (red points) and the remaining LMC stars (black points) that are in the BAWLAS sample (right panel).  The red supergiant photometric selection from \citet{Neugent2020} that we use to identify these potential red supergiants is also shown for illustration (red box) in both panels.}
  \label{fig:lmc_cmd}
\end{figure*}

Many LMC RSGs have been observed by APOGEE \citep[particularly in the southern TESS continuous viewing zone contributed programs;][]{Santana2021} and analyzed here.  Figure \ref{fig:lmc_cmd} shows a 2MASS CMD of LMC stars in our sample that have been selected according to the spatial and kinematic (proper motion and radial velocity) selections given in \citet{Hasselquist2021}.  

The patchwork distribution of stars in this CMD reflects the variety of programs and subprograms that have targeted LMC stars in APOGEE \citep{Nidever2020,Santana2021}, but for reference the tip of the red giant branch (TRGB) lies around 12 in K$_{\rm s}$ \citep[noting though that the TRGB is actually sloped; see e.g.,][]{Boyer2011,Hoyt2018}.  The stars that are more luminous than the TRGB split into two branches with a redward branch around J-K$_{\rm s}$ $\sim 1.2$  and a blueward branch that is brighter and centered on J-K$_{\rm s}$ $\sim 1.0$.  The redder stars seen here belong to various AGB populations, whereas the bluer stars that have been outlined in Figure \ref{fig:lmc_cmd} are red supergiants that have been selected photometrically following the criteria of \citet{Neugent2020}.

\begin{figure*}
  \centering
  \includegraphics[scale=0.35,trim = 0.in 0.in 0.in 0.in, clip]{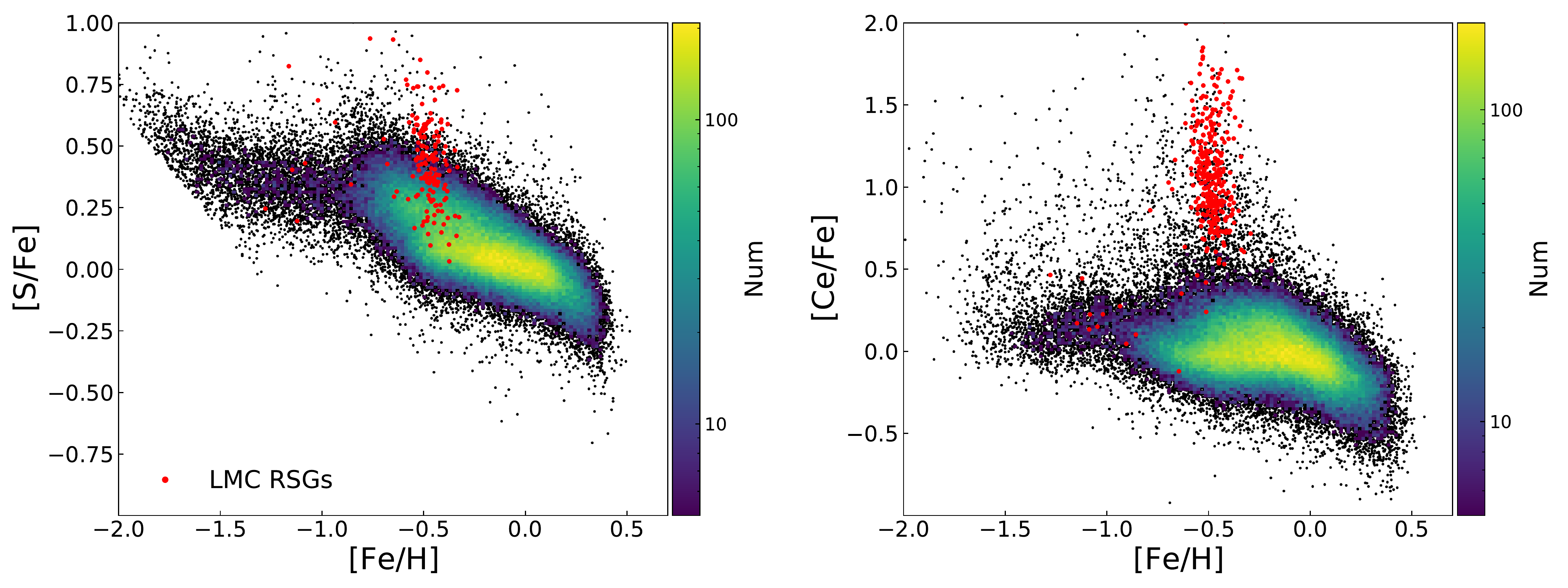}
  \caption{\xfe{S} (left) and \xfe{Ce} abundances versus metallicity, \xh{Fe}, for the BAWLAS sample (black points and 2D histogram), and for the LMC red supergiants (red points) selected as described in the text.  The vertical abundance overdensity at \feh{} $\sim -0.5$ that appears in Figure \ref{fig:xfe_abund} is clearly dominated by LMC RSGs as seen here.}
  \label{fig:lmc_rsg_chemistry}
\end{figure*}

In Figure \ref{fig:lmc_rsg_chemistry}, we show the S and Ce chemistry of these photometrically selected RSGs compared to the rest of the BAWLAS sample.  Indeed the RSGs produce the overdensity of high \xfe{Ce} stars seen in Figure \ref{fig:xfe_abund}.  The source of the Ce enhancement and the large spreads seen in Ce, S and other elements is unclear.  It is possible that the \citet{Neugent2020} photometric selection may have some contamination by massive AGB stars that could have unusual abundances from internal mixing \citep[e.g., ][]{Plez1993}.  However, this photometric selection is expected to have a relatively low contamination from AGB stars \citep[and was designed to be so,][]{Neugent2020}, and because most of this selection has unusual abundances (rather than just a few outliers in this selection), it suggests that this odd feature is related to the LMC RSGs specifically.

To some extent it may be expected that young metal-rich stars in the LMC may have enhancements in s-process elements from many generations of AGB stars.  For instance, \citet{Hasselquist2021} have showed that metal-rich LMC RGB stars (which are thought to be relatively young, but still slightly older than the massive RSGs) do show an enhancement in \xfe{Ce} over MW stars of the same metallicity.  However, these enhancements are at \xfe{Ce} $\sim 0.2 - 0.3$ rather than the \xfe{Ce} $\sim 0.5 - 2.0$ seen in the RSGs here.  

Furthermore, this wouldn't explain the large spread seen in other elements, e.g., S, which extends to high \xfe{S} unexpectedly given that the metal-rich RGB stars in the LMC show relatively a small $\alpha$-element abundance spread at slightly super-solar \xfe{X}.  Therefore, this large spread in multiple elements (beyond just S and Ce shown here) for such a narrow range in metallicity seems unlikely to be a physical abundance pattern and is more likely to be evidence that there are systematic errors for the RSGs.  

This could be in the form of incorrect stellar parameters, improper fitting of blends, heightened nLTE or 3D effects in these stars, etc.  For example, because these stars are very luminous and low $\log g$, spherical radiation transfer might be needed (in addition to the spherical atmosphere models that APOGEE already uses) to properly derive stellar parameters for these stars, rather than the plane-parallel radiation transfer used to derive the DR17 ASPCAP stellar parameters \citep{Holtzman2021}.  This could introduce systematic trends in stellar parameters of these stars that we see propagate into the abundances we derive here.

We note that these stars also exhibit a spread in some APOGEE derived abundances, such as APOGEE's Al or S measurements, which indicates that this feature does not originate exclusively in our BACCHUS analysis.  Fully investigating these stars, however, is beyond the scope of this work, but for completeness we warn that the abundances (and potentially even the stellar parameters of these stars) may be suspect and unphysical, and should be used with caution.

\section{Literature Comparison}
\label{sec:comparison}

\subsection{APOGEE Comparison}
\label{sec:apogee_comp}

While our abundances are derived from the exact same spectra that are used to measure abundances in APOGEE, there are several key differences in our analyses that lead to differences in derived abundances.  In addition to using a different abundance pipeline and spectral synthesis code for our analysis, we also use different input stellar parameters, employ line-by-line and star-by-star flagging that allows us to provide a more cleaned sample of abundances, and, in some cases, use a slightly different selection of lines.

\begin{deluxetable*}{l c c c c c c l}
\tablewidth{0pt}
\tablecolumns{8}
\tablecaption{BAWLAS Abundance Comparison to APOGEE DR17 and High-resolution Literature Measurements\label{tab:lit_comp}}
\tablehead{\colhead{Element} & \multicolumn{3}{c}{APOGEE DR17} & \multicolumn{4}{c}{High-res Lit} \\ & Mean\tablenotemark{{\scriptsize a}} & St Dev & N$_{\rm stars}$ & Mean\tablenotemark{{\scriptsize a}} & St Dev & N$_{\rm stars}$ & References}
\startdata
$\Delta$\xh{C} & -0.09 & 0.10 & 122,425 & -0.16 & 0.22 & 45 & 1, 2 \\
$\Delta$\xh{N} & -0.20 & 0.16 & 122,071 & -0.21 & 0.14 & 41 & 1, 2 \\
$\Delta$\xh{O} & -0.01 & 0.11 & 114,591 & 0.03 & 0.13 & 116 & 1, 2, 3\\
$\Delta$\xh{Na} & -0.02 & 0.11 & 93,893 & 0.11 & 0.15 & 30 & 1, 2\\
$\Delta$\xh{V} & -0.15 & 0.10 & 67,115 & -0.12 & 0.13 & 69 & 1, 2, 4\\
$\Delta$\xh{S} & 0.03 & 0.09 & 104,319 & - & - & - & - \\
$\Delta$\xh{Cu} & - & - & - & 0.03 & 0.18 & 19 & 1\\
$\Delta$\xh{Ce} & -0.10 & 0.14 & 90,191 & 0.04 & 0.18 & 96 & 5\\
$\Delta$\cisotope{} & - & - & - & -0.1 & 4.7 & 23 & 6-14\\
\enddata
\tablenotetext{a}{Calculated as (\xh{X}$_{\rm Ref}$ $-$ \xh{X}$_{\rm BAWLAS}$)}
\tablecomments{References for high-resolution literature measurements:  1 -- \citet{daSilva2015}; 2 -- \citet{Brewer2016}; 3 -- \citet{Jonsson2017}; 4 -- \citet{Lomaeva2019};  5 -- \citet{Forsberg2019}; 6 -- \citet{Briley1994}; 7 -- \citet{Briley1997}; 8 -- \citet{Smith2002}; 9 -- \citet{Pavlenko2003}; 10 -- \citet{Smith2007}; 11 -- \citet{Mikolaitis2012}; 12 -- \citet{Smith2013}; 13 -- \citet{Drazdauskas2016}; 14 -- \citet{Szigeti2017}}
\end{deluxetable*}

These analysis choices can all contribute to differences between the abundances measured by APOGEE and our work here for elements that are analyzed by both.  In Table \ref{tab:lit_comp}, we tabulate some basic statistics on the differences between the APOGEE DR17 and BAWLAS abundances, showing the mean abundance differences and 1$\sigma$ standard deviation of these differences.  A few elements see a moderate shift, such as C, N, V, and Ce, which can be attributed to dependence on input stellar parameters for C and Ce, and the removal of temperature trends in APOGEE data for V, as discussed below.  For N the offset isn't obviously tied to one single source, and is likely a result of changing stellar parameters as well as differences in C and O which have an important in determining N abundances.

The scatter in the differences between BAWLAS and APOGEE are typically larger than the combined uncertainties for most of the elements by $\sim 25-50\%$ (and slightly larger still for the more precisely measured molecular elements C, N and O).  However, the reported uncertainties from BAWLAS and APOGEE are simply estimates of the true uncertainty, and there are some systematic differences between the two analyses, in terms of input stellar parameters, methodology, etc., either of which may account for the difference between random uncertainties and the scatter between the two sets of measurements.  

\begin{figure*}
  \centering
  \includegraphics[scale=0.29,trim = 0.in 0.in 0.in 0.in, clip]{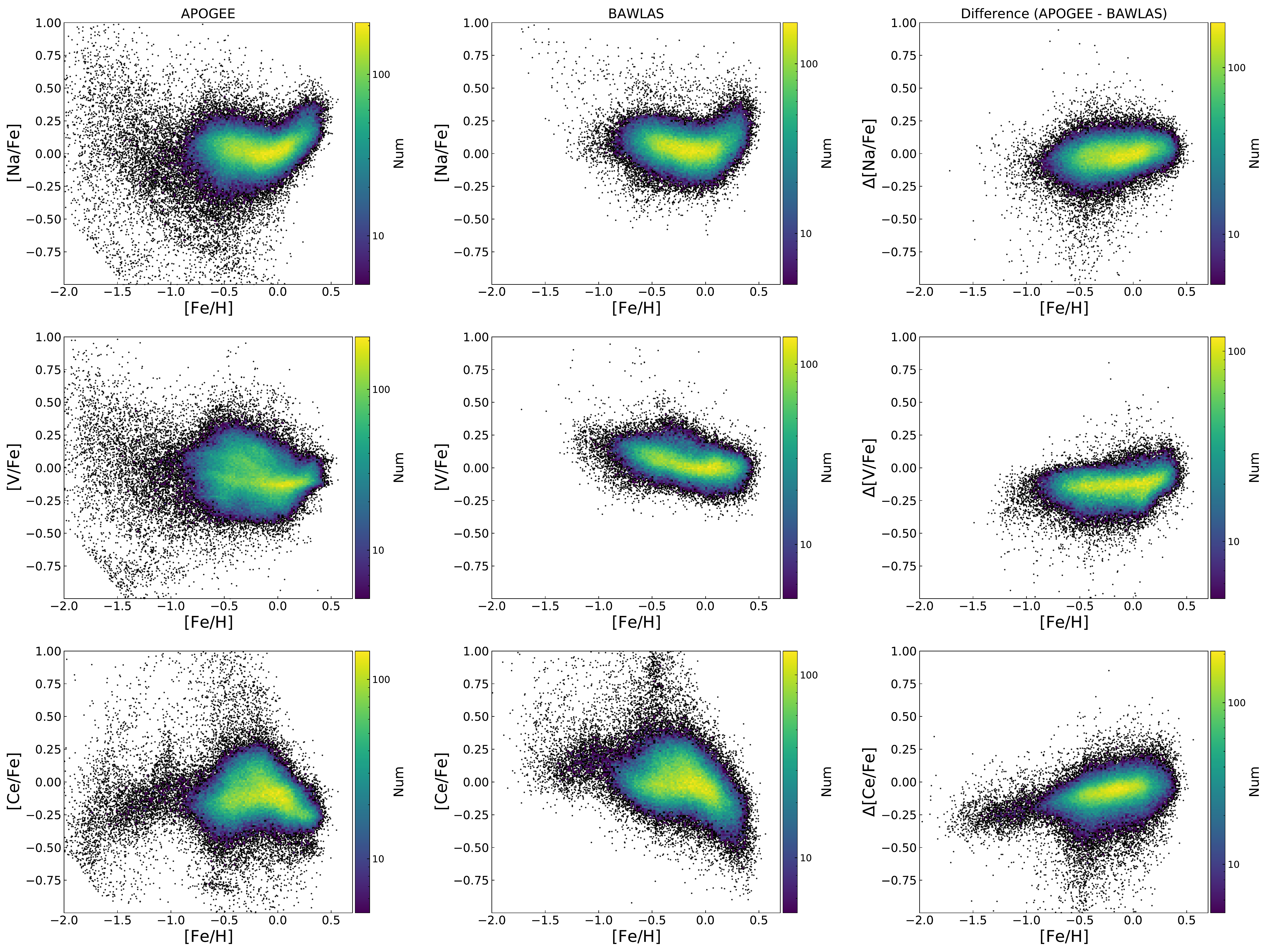}
  \caption{\xfe{X} for Na (top row), V (middle row), and Ce (bottom row) vs. metallicity, \feh{}, for APOGEE DR17 measurements (left column), the BAWLAS measurements presented here (middle column), and the difference between the two (right column) for stars that have reported abundances from both.}
  \label{fig:apogee_diffs}
\end{figure*}

Figure \ref{fig:apogee_diffs} shows a few examples of the ways in which the abundances we derive differ from APOGEE's measurements.  In particular, this shows APOGEE abundance measurements for the sample of stars we analyze, the measured BACCHUS abundances in this sample, and for the stars that are measured in both, we also show the differences between those measurements.

\subsubsection{Systematic Effects of Stellar Parameter Choices}

BAWLAS also differs from APOGEE because we use APOGEE's calibrated stellar parameters to derive abundances, while APOGEE has used its uncalibrated stellar parameters.  The differences between the calibrated and uncalibrated stellar parameters are primarily in the effective temperatures and surface gravities, which are, respectively, calibrated to photometric temperature relations, and astroseismic surface gravities in ASPCAP's post-processing \citep[for giants;][]{Holtzman2021}.  We also re-derive the C, N, and O abundances for fitting blends, since we are using different input stellar parameters, whereas C, N and $\alpha$ abundances (which adjust the O abundance by assuming [O/$\alpha$] = 0) are fit simultaneously by APOGEE in its stellar parameter fits.

Not surprisingly, it appears that the change in stellar parameters, particularly the use of calibrated temperatures and gravities, affects some of the elements that we measure here.  One clear example of this is Ce.  While, in a broad sense, the chemical abundance patterns of Ce from APOGEE DR17 and BAWLAS are relatively similar (as seen in Figure \ref{fig:apogee_diffs}), we find that there is a trend in the differences between the two Ce measurements as a function of metallicity, with BAWLAS finding systematically higher \xfe{Ce} at lower metallicities.

\begin{figure*}
  \centering
  \includegraphics[scale=0.29,trim = 0.in 0.in 0.in 0.in, clip]{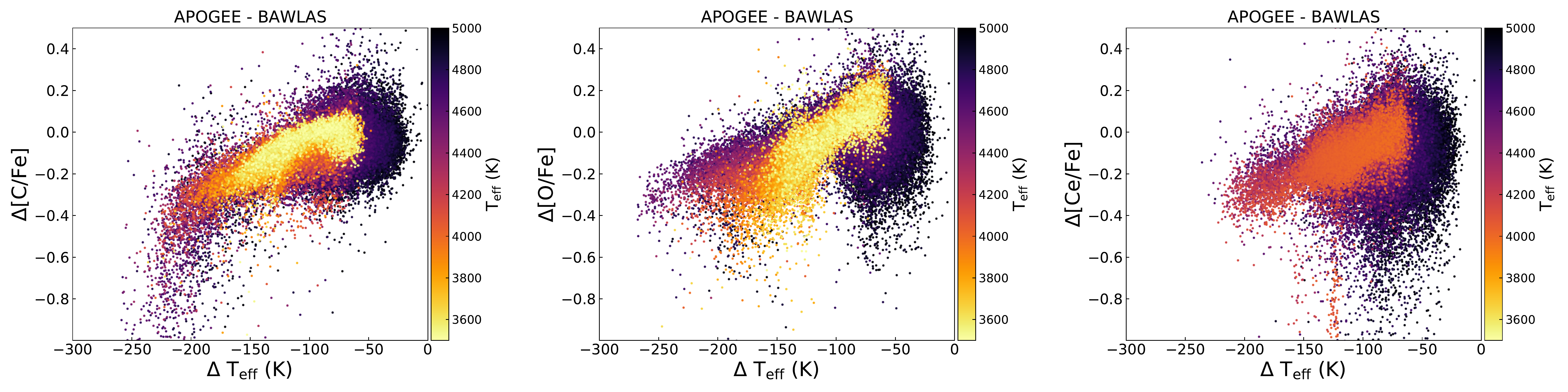}
  \caption{Difference between APOGEE and BAWLAS \xfe{X} measurements for C, O, and Ce as a function of the difference between the $T_{\rm eff}$ used by APOGEE and our BACCHUS analysis for abundance measurement and color-coded by the calibrated $T_{\rm eff}$.}
  \label{fig:apogee_cediff}
\end{figure*}

This metallicity-correlated offset seems to primarily be a result of using different $T_{\rm eff}$ and $\log g$.  The Ce II lines in APOGEE's wavelength range are quite sensitive to surface temperatures and gravities, such that using a hotter $T_{\rm eff}$ or larger $\log g$ results in deriving higher Ce values \citep{Cunha2017}.  This compounds with the APOGEE DR17 $T_{\rm eff}$ and $\log g$ calibrations, both of which are largest at low metallicities, to produce Ce differences that grow with decreasing metallicity.  Figure \ref{fig:apogee_cediff} shows a clear correlation of the difference between APOGEE and BAWLAS \xfe{Ce} as a function of the difference between the uncalibrated and calibrated temperatures used in the respective analyses.  Note that the scatter at low $\Delta T_{\rm eff}$ coming primarily from warm stars where the Ce lines are weaker and the abundance measurements are more uncertain, and there are a lack of cool stars, since APOGEE does not populate its \xfe{Ce} abundances for cool stars where the Ce is deemed unreliable \citep{Holtzman2021}.  Similar trends also appear as a function of the $\log g$ calibration differences.

We also find that C and O abundances appear to be affected by the differences in assumed stellar parameters.  We also show the differences between the BAWLAS and APOGEE abundances for C and O as a function of the APOGEE temperature calibration in Figure \ref{fig:apogee_cediff}.  While there are some differences between the trends for C, O, and Ce, each of them show a correlation with the temperature calibration.  The variety in these correlations reflects the complexities in deriving these abundances and each of their individual dependencies on the input stellar parameters, line choice, the abundance of elements with molecular species (for molecular equilibrium or blending), etc.  

For example the larger spread as $\Delta {T_{\rm eff}}$ approaches zero is from warm stars, which have weaker features, increasing uncertainty and scatter in their resulting abundances.  Another example is the increase in the spread of the O abundance differences around $\Delta {T_{\rm eff}} \sim 175$ K, which appears to be due to correlations between the O differences and the $\log g$ calibration (which is temperature dependent, hence the temperature trend for a fixed $\Delta {T_{\rm eff}}$), which has less of an effect on the C and Ce measurements.

\begin{figure*}
  \centering
  \includegraphics[scale=0.35,trim = 0.in 0.in 0.in 0.in, clip]{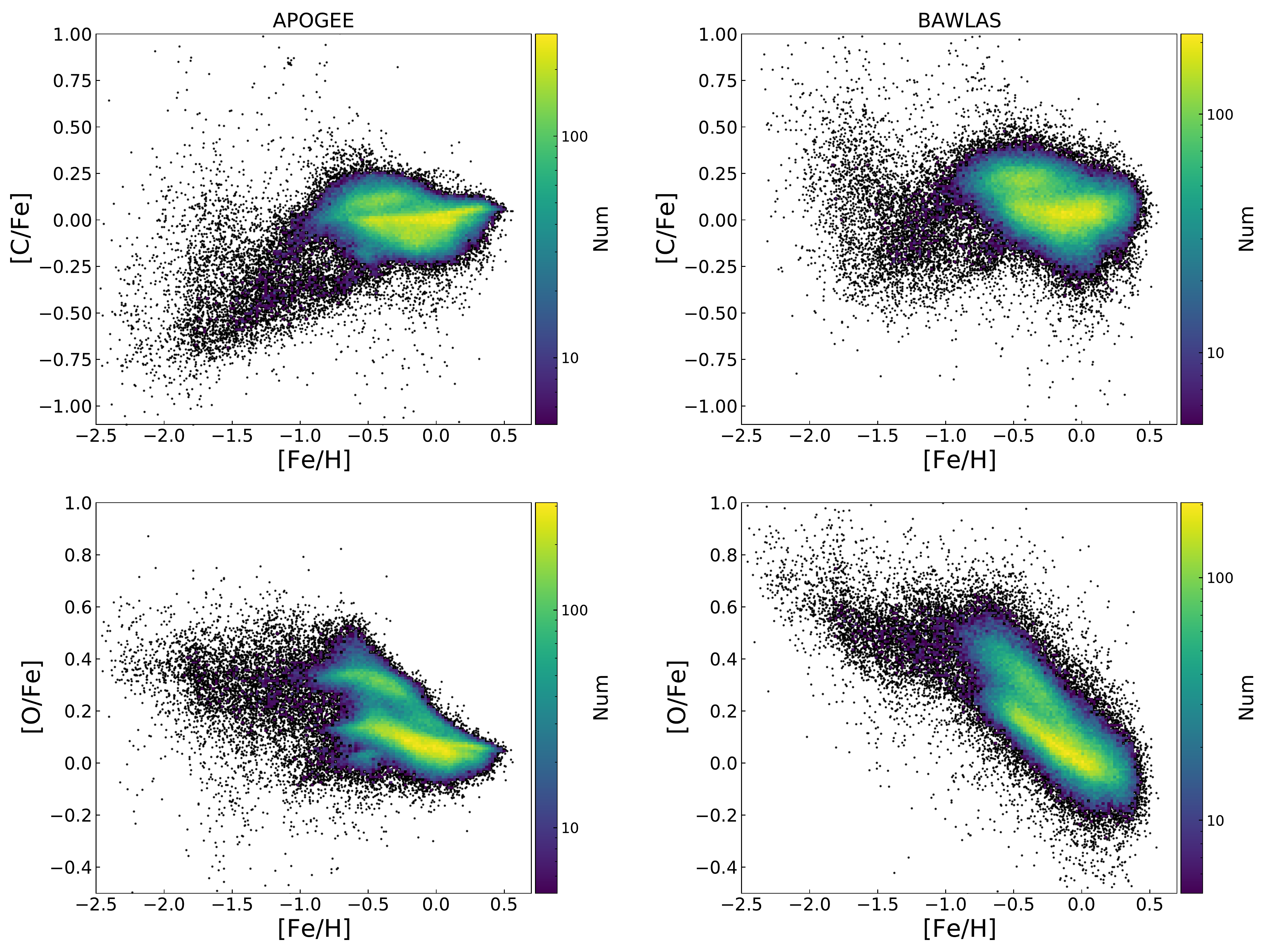}
  \caption{\xfe{X} for C (top row) and O (bottom row) vs. metallicity, \feh{}, for APOGEE DR17 measurements (left column) and the BAWLAS measurements presented here (right column) for stars that have reported abundances from both.}
  \label{fig:apogee_co_comp}
\end{figure*}

These systematic differences can be important to consider because they do have some impact on the chemical abundance patterns of these elements.  Figure \ref{fig:apogee_co_comp} compares the C and O abundances patterns of APOGEE and BAWLAS for the present sample.  While in both cases the BACCHUS measurements have more scatter, because fewer lines have been used than in APOGEE's analysis, we can see they also result in higher \xfe{X} abundances at low metallicities.  

For C this does not have a particularly significant effect on the appearance of the abundance pattern.  However, this systematic difference at lower metallicities is important to consider, because different metallicity trends can affect the interpretation of C for stellar evolution \citep[e.g.,][]{Shetrone2019}.  Interestingly, these changes for C bring the BAWLAS abundances into better agreement with APOGEE's atomic C measurements rather than the molecular ones that are compared to here.  Some of this may be because BACCHUS measures a mix of atomic and molecular C features, but it may be worth investigating if calibrated temperatures and surface gravities bring atomic and molecular C abundances into better agreement than APOGEE's uncalibrated parameters.

Similar to C, in Figure \ref{fig:apogee_co_comp} we see differences between the O abundances in BAWLAS and those in APOGEE, which also seem to be related to the choice of stellar parameters.  In BAWLAS we see a higher \xfe{O} plateau of $\sim 0.5$ at low metallicities, as opposed to the $\lesssim 0.4$ plateau seen in APOGEE.  At higher metallicities we also see a difference in the \xfe{O} abundance patterns of the thick disk/high-$\alpha$ and thin disk/low-$\alpha$ sequences.  Both of these sequences appear flatter and plateau at \feh{} $\sim -0.7$ in APOGEE, whearas BAWLAS finds both of these sequences monotonically increasing towards lower metallicities.

Interestingly, the BAWLAS \xfe{O} abundance patterns are more similar than APOGEE to what has been seen with high-resolution optical studies.  Such studies, typically find steeper \xfe{O} trends in the high- and low-$\alpha$ sequences and a higher plateau at low metallicities than what APOGEE finds \citep[e.g., averaging a plateau value of \xfe{O} $\sim 0.6$][]{Bensby2014}, and instead more similar to what we see in BAWLAS. The difference in the \xfe{O} plateau is likely a result of using calibrated stellar parameters, but it is not clear what causes the differences between the abundance patterns of the high- and low-$\alpha$ sequences.  These may be in part affected by using different parameters or may be a result of separating O into an individual dimension instead of changing it along with other $\alpha$-elements as is done in APOGEE.  It is also possible that the lower precision of BAWLAS O abundances blur out the fine details that are apparent in APOGEE's O abundances.

While we have used the calibrated APOGEE DR17 parameters, it is not entirely clear which set of stellar parameters might be best to use for abundance analysis.  The calibrated stellar parameters used here may more accurately reflect the true stellar parameters of the stars we observe and help remove systematic trends with uncalibrated stellar parameters.  

On the other hand the spectroscopic stellar parameters are derived simultaneously and (theoretically) provide the best fit to the observed spectra.  So the use of spectroscopic stellar parameters may reduce the impact of any systematic errors or unaccounted for physics in the model atmospheres or synthetic spectra that might lead to poorly fit blends, line shapes, etc.  Therefore we simply note that there may be differences that arise from using different input stellar parameters, and that by comparing the differences in derived abundances, we may be able to explore some of these effects.

\subsubsection{Cleaning Suspect Abundance Measurements}

One significant difference between our analysis here and APOGEE's is our more detailed flagging that helps identify lines or stars whose abundances may be erroneous or highly uncertain.  In APOGEE, there is no explicit flagging for lines that may be significantly affected by blends or for measurements that should be below upper limits.  In cases such as these, the abundances APOGEE reports may be coming from poor fits or measuring noise.  An example of this can be seen in Na.

While the Na measurements agree reasonably well (within uncertainties) for stars measured in both APOGEE and BAWLAS, the effect of flagging and upper limits can clearly be seen in Figure \ref{fig:apogee_diffs}.  At low metallicities, $\lesssim -0.5$ APOGEE reports abundances for many stars, however these stars show a large scatter, whereas with BAWLAS they are flagged as upper limits or as having poor quality fits (typically because the lines are weak and dominated by spectral noise).  While we perform similar to APOGEE when stars are measured in BAWLAS, the benefit of our analysis is that we are able to flag stars with suspect measurements even at the level of indivdual lines on a star-by-star basis.  Some systematic differences are also seen between APOGEE and BAWLAS as a function of metallicity (BAWLAS reports slightly higher \xfe{Na} at lower metallicities), which is similar to other elements as discussed in the previous section.  

The flagging and upper limits can then improve or simplify the interpretation of the chemical abundances such as Na. S, and to some extent N (aside from its systematic offset), are similar to Na, in that the chemical abundance patterns do not significantly change, but by removing suspect abundances we present a clearer picture of these patterns.

\subsubsection{Holistic Treatment of Weak and Blended Features}

The last element that overlaps with APOGEE in DR17 is V, which is also shown in Figure \ref{fig:apogee_diffs}.  The V abundance pattern is significantly different between APOGEE and BAWLAS, and appears to be improved in the BAWLAS analysis here.  In APOGEE the V abundances show a large spread (which is even larger at low metallicities where there is no clear pattern), whereas in BAWLAS we see a tight distribution at higher metallicities and that much of the low metallicity scatter seen in APOGEE has been cleaned by our flagging and upper limits at low metallicities.  

\begin{figure*}
  \centering
  \includegraphics[scale=0.29,trim = 0.in 0.in 0.in 0.in, clip]{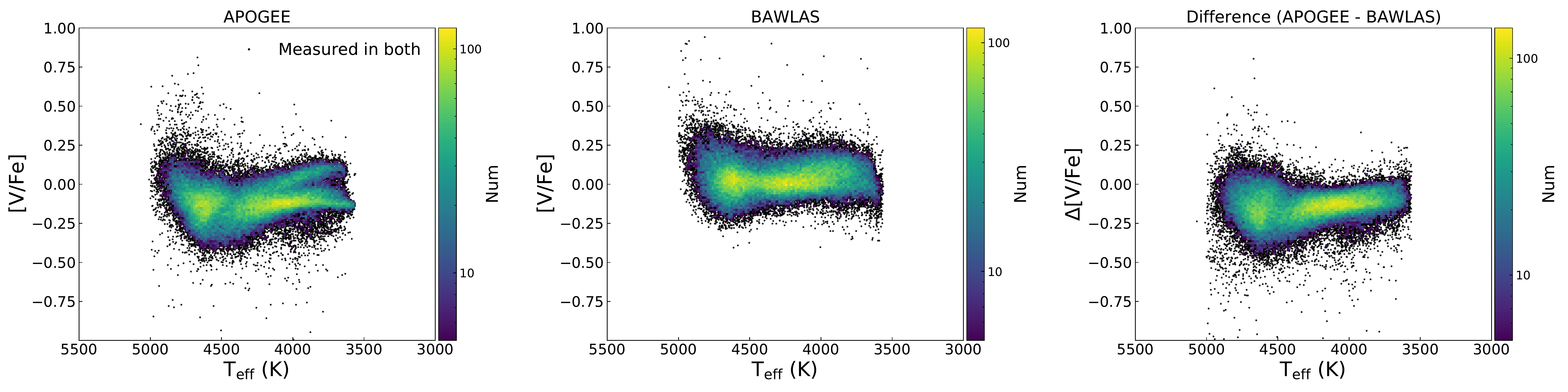}
  \caption{\xfe{V} vs. stellar effective temperature, \feh{}, for APOGEE DR17 measurements (left column), the BAWLAS measurements presented here (middle column), and the difference between the two (right column).}
  \label{fig:apogee_vdiff}
\end{figure*}

Comparing the V abundances as a function of temperature for stars that were measured by both APOGEE and BAWLAS in Figure \ref{fig:apogee_vdiff}, we see that there are clear systematic trends in APOGEE's V abundances as a function of temperature.  This produces the large range of \xfe{V} and complex abundance pattern seen at higher metallicities in APOGEE.  In BAWLAS, while we do see slight temperature trends (see Section \ref{sec:teff_trends}), they are significantly reduced, leading to a much tighter \xfe{V} pattern.
       
This improvement is not clearly tied to any one difference between the APOGEE and BAWLAS analyses and is likely a result of a variety of improvements in localized blend treatment, line choice and flagging, upper limits, etc.  This allows us to more precisely measure elements like V that are heavily blended, and in some cases only present themselves in weak lines, such as P and Nd as well as the \cisotope{} ratio, illustrating the strength the BAWLAS methodology here.

\subsection{High-resolution External Literature Comparison}
   
In addition to comparing with APOGEE, we have compiled high-resolution literature measurements for our elements of interest to compare with our derived abundances.  

\subsubsection{Star-by-Star Comparison}

As with \citet{Jonsson2020,Holtzman2021} we compare star-by-star measurements of BAWLAS to those of \citet{daSilva2015} and \citet{Brewer2016} for C, N, O, Na, V and Cu (with Cu coming from \citealt{daSilva2015} only), and to the sample from \citet{Jonsson2017,Lomaeva2019,Forsberg2019} for O, V, and Ce respectively.  For \cisotope{} we compare with values measured from several literature studies \citep{Briley1994, Briley1997, Smith2002, Pavlenko2003, Smith2007, Mikolaitis2012, Smith2013, Drazdauskas2016, Szigeti2017}; this sample may be relatively small and inhomogenous, but it can be used to indicate the quality of our \cisotope{} measurements.  Our overlap with the literature for P, S, and Nd is too small (a couple stars if any) to draw any meaningful conclusions from.

Table \ref{tab:lit_comp} gives the mean difference between our measurements and these high-resolution studies, and the $1\sigma$ standard deviation scatter around that mean.  The mean differences are relatively small for most elements, and the scatter is similar to or smaller than APOGEE comparisons with literature values for the same elements \citep[as of DR16,][]{Jonsson2020}.  The one exception is C, which has slightly larger scatter and is likely a reflection of the lower precision that we are able to achieve using only a handful of C lines.

\begin{figure}
  \centering
  \includegraphics[scale=0.4,trim = 0.in 0.in 0.in 0.in, clip]{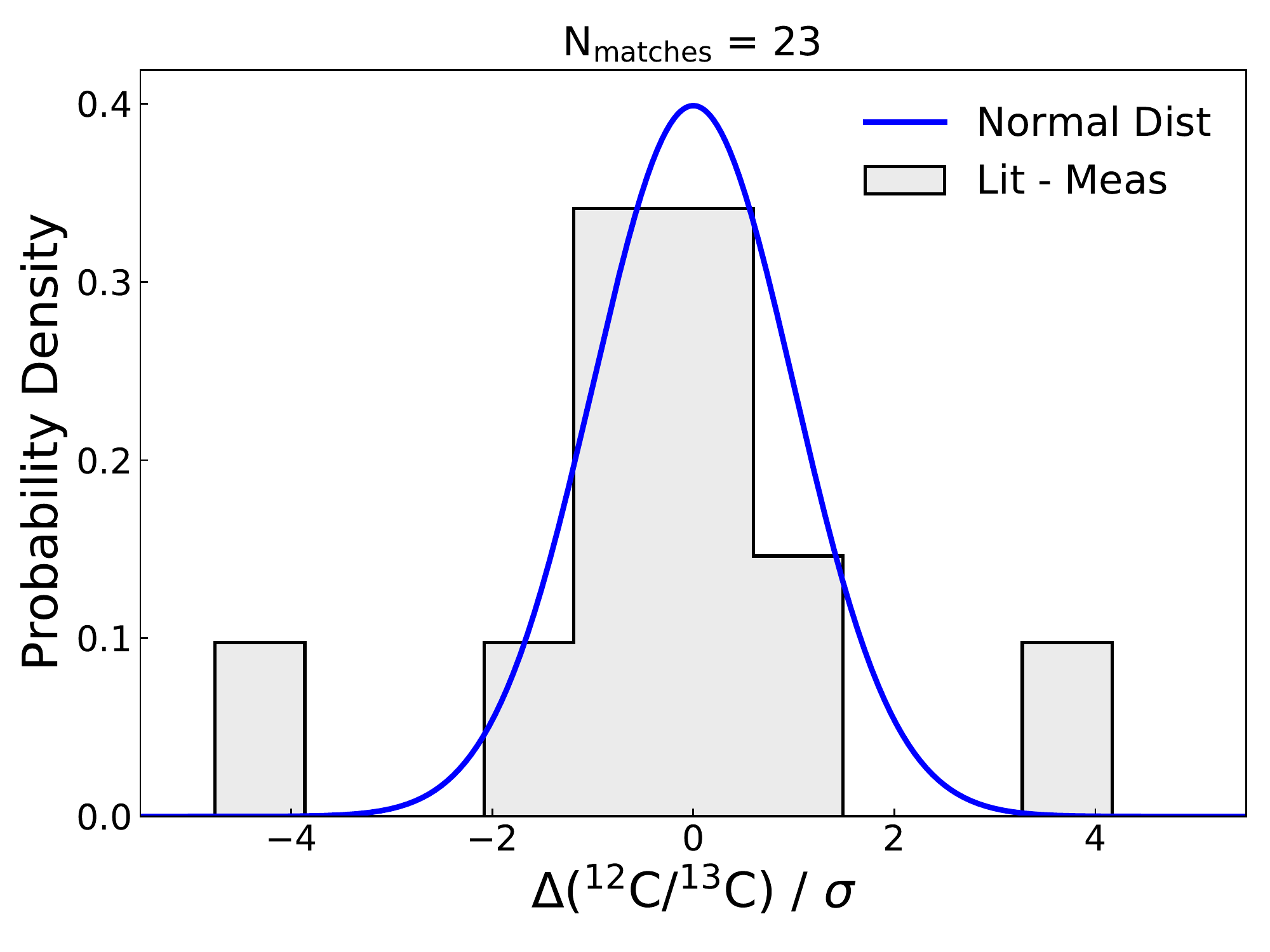}
  \caption{Area-normalized histogram of \cisotope{} differences between BAWLAS and literature measurements, scaled by their combined uncertainties (using the measured uncertainty from our BAWLAS sample), compared with a normal distribution (blue curve).  If the differences between our measurements and literature match within their combined uncertainties, we would expect the distribution of our cross-matched sample to match the blue curve.}
  \label{fig:c12c13_litcomp}
\end{figure}

Because the high \cisotope{} ratios are more uncertain, this can skew the scatter in our literature comparison to follow the differences in stars with high \cisotope{} values.  So, to compare our \cisotope{} ratios in another way, in Figure \ref{fig:c12c13_litcomp} we show a normalized histogram of the difference between BAWLAS and literature measurements of \cisotope{}, divided by the uncertainty on that difference (combining the reported literature uncertainty, where available, with the measurement uncertainty on our \cisotope{} ratios).  This effectively shows how many $\sigma$ away from agreement these differences are.  We also indicate the distribution we would expect to see if the differences were normally distributed according to the combined uncertainties.

We find that the differences are close to within what we would expect from random errors alone, although there are a few outliers, which may be true outliers or have underestimated uncertainties (either in our study or in the literature studies that they were drawn from).

\subsubsection{Chemical Abundance Pattern Comparison}
\label{sec:lit_comp2}

The above comparisons are for relatively small samples, so to expand our comparison with literature, we have also compiled a larger sample of high-resolution studies to compare their general chemical abundance patterns with those we measure here.  This sample is made of measurements from the following studies:
   \begin{itemize}
       \item Na:  \citet{Chen2000, Reddy2006, Adibekyan2012, Bensby2014, Roederer2014, daSilva2015, Brewer2016, Duong2019} 
       \item P:  \citet{Caffau2011, Roederer2014p, Maas2017, Caffau2019, Maas2019}
       \item S:  \citet{Chen2002, Caffau2005, Nissen2007, CostaSilva2020}
       \item V:  \citet{Chen2000, Reddy2006, Battistini2015, Roederer2014, daSilva2015, Brewer2016, Duong2019, Lomaeva2019} (we exclude the V measurements of \citealt{Adibekyan2012} since their V show strong systematic temperature trends)
       \item Cu:  \citet{Reddy2006, Roederer2014, daSilva2015, Duong2019} 
       \item Ce:  \citet{Reddy2006, Mishenina2013, Roederer2014, Battistini2016, DelgadoMena2017, Forsberg2019} 
       \item Nd:  \citet{Mashonkina2004, Reddy2006, Mishenina2013, Roederer2014, Battistini2016, DelgadoMena2017, Duong2019} 
   \end{itemize}

\begin{figure*}
  \centering
  \includegraphics[scale=0.35,trim = 0.in 0.in 0.in 0.in, clip]{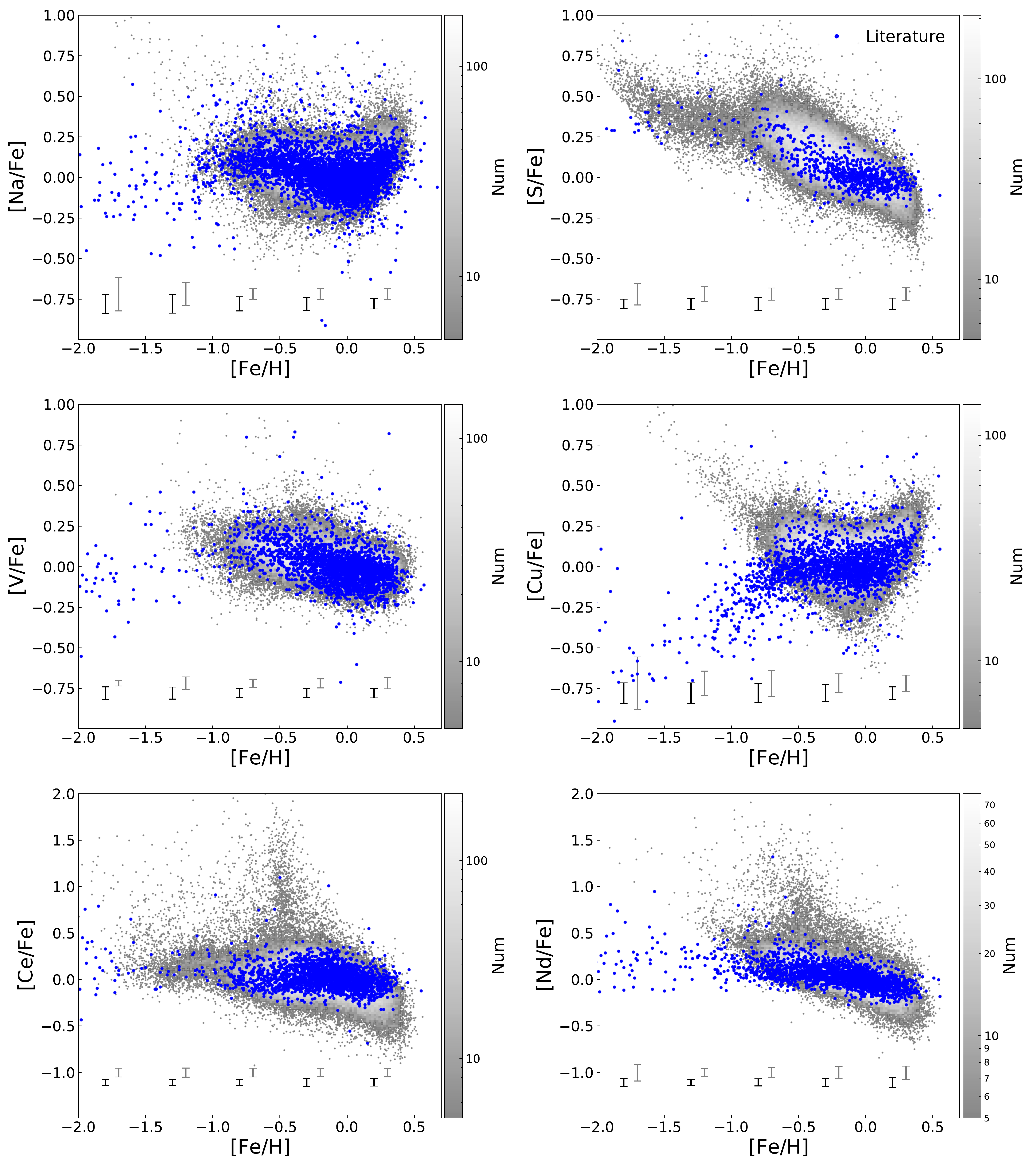}
  \caption{\xfe{X} for Na, S, V, Cu, Ce, and Nd vs. metallicity, \xh{Fe}, for the entire BACCHUS sample (gray points and 2D histogram), and for high resolution literature measurements of MW stars (blue points; see the text for literature references). }
  \label{fig:xfe_litcomp}
\end{figure*}

\begin{figure}
  \centering
  \includegraphics[scale=0.4,trim = 0.in 0.in 0.in 0.in, clip]{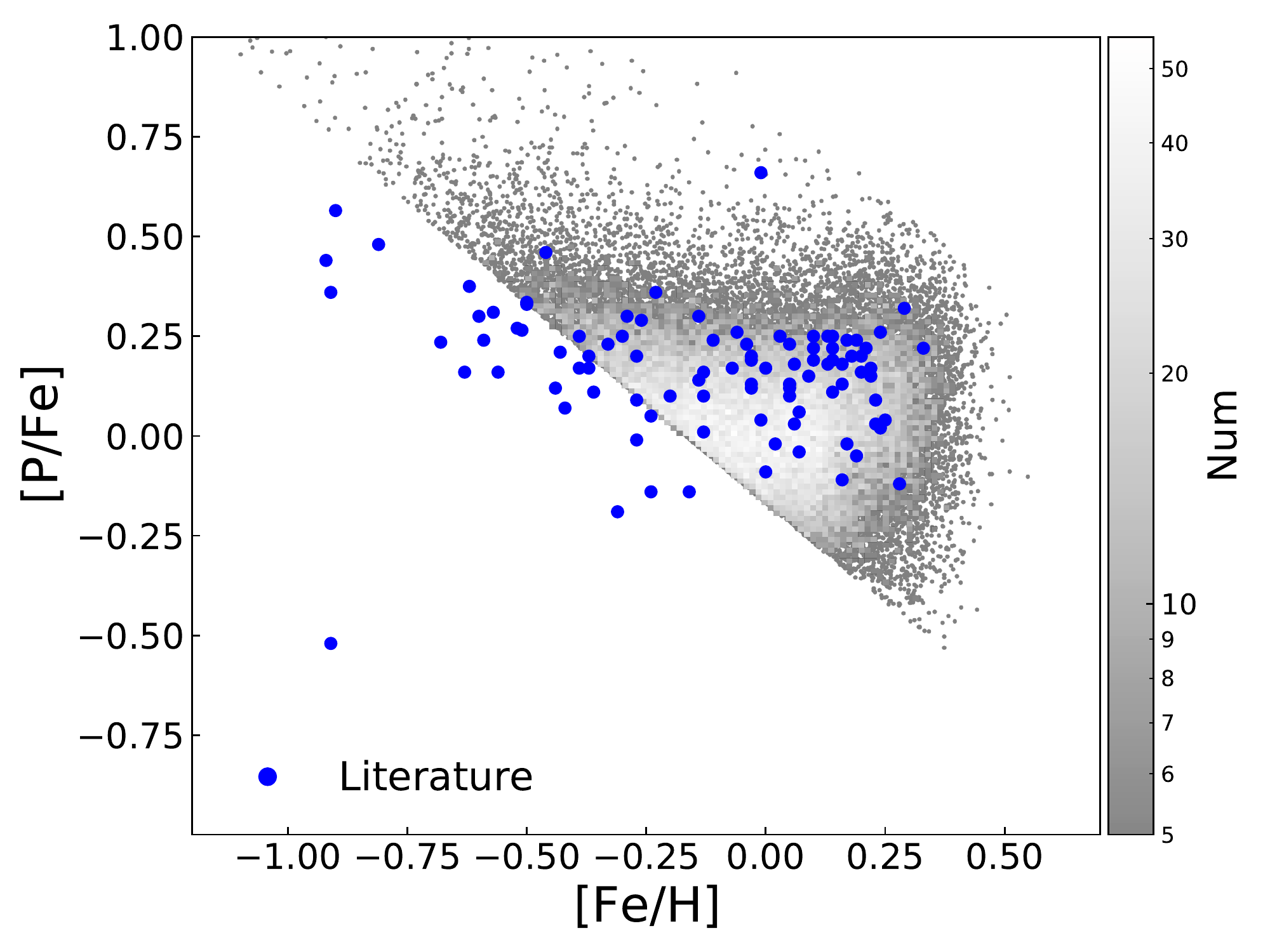}
  \caption{\xfe{P} vs. metallicity, \xh{Fe} with color-coding as in Figure \ref{fig:xfe_litcomp}.}
  \label{fig:pfe_litcomp}
\end{figure}

The chemical abundances from this literature compilation are shown over our derived abundance distributions in Figures \ref{fig:xfe_litcomp} and \ref{fig:pfe_litcomp}.  In general, we see that there is a fairly good agreement between the chemical abundance patterns measured in the literature and by our analysis here.  

{We note that, while we have good agreement with the Na abundances from the literature referenced here, there is some disagreement between our results for Na and those of bulge RGB stars from \citet{Johnson2014}, particularly in the metal-poor regime, where they report sub-solar \xfe{Na}.  However, other optical literature abundances do not show sub-solar Na abundances even when considering bulge stars \citep[e.g.,][]{Duong2019}.  Therefore given the good agreement between BAWLAS and the optical literature here, it suggests that there may be some systematics in the \citet{Johnson2014} Na abundance measurements particularly at low metallicities, as suggested by \citet{Duong2019}.}

For elements where we see stars that have significantly elevated abundances like Ce and Nd, we also see high-resolution literature measurements of similarly enhanced stars (albeit fewer of them).  For P we note that there are very few literature measurements, so it is difficult even with the literature to clearly uncover the underlying chemical abundance pattern.  However the literature distribution does seem to reflect a similar pattern to what we see at the metallicities where we are able to measure P.
   
The one element where we see some deviation from the literature is Cu.  Literature measurements show a decreasing \xfe{Cu} abundance with decreasing metallicity at metallicities, \feh{} $\lesssim -0.8$, which we do not find in our sample.  One possible explanation for these differences is that in the H-band, at APOGEE resolution, we are limited in how metal-poor we can measure Cu because of it's weak, blended lines. These limits begin to affect our measured \xfe{Cu} abundance pattern as metal-rich as \feh{} $\sim -0.7$.  This biases our final Cu sample to stars with relatively higher \xfe{Cu} at low metallicities.  Stars with low Cu, such as those in the decreasing literature trend, would only be measureable as upper limits in our analysis.  This combination of effects might cause some of the differences between the Cu patterns we see in our analysis and those in literature.

Another possible factor is that studies have found that nLTE effects should be more significant in optical Cu measurements at lower metallicities, while they are less significant in the H-band \citep{Yan2015,Andrievsky2018}.  The necessary nLTE corrections would push the optically measured, literature, Cu to higher values, bringing the literature into better agreement with the trend we see with BAWLAS.  However, these nLTE corrections are not quite large enough to completely account for the differences.

{Cu is also one of the elements that we measure using the ``wln'' method in BACCHUS for one of its lines.  This method relies on only a single pixel, and therefore could be more subject to systematics or random uncertainties, and could skew our results away from what is found in the literature.  However, the differences in the Cu trend between BAWLAS and the literature are present in the Cu 16005.5 \AA{} measurements which are calculated with the ``chi2'' method, and persist even if we use the ``chi2'' method for the Cu 16006.0 \AA{} line (though the scatter is larger).  Instead, systematic blending, the influence of upper limit flagging and optical nLTE effects are a more likely the source of the discrepancy between the optical literature Cu measurements and BAWLAS.  P is also heavily reliant on the ``wln'' method, but we remark for completeness, that as with Cu, the abundance patterns of P do not change significantly if we use the ``chi2'' method, which is more robust to individual pixel noise.}

We may also be seeing an affect of upper limits in Nd, insofar as we are limited to upper limits at the low metallicity end of our distribution.  This in turn may make our distribution look relatively higher in Nd than the literature at low metallicities, due to the missing stars with lower Nd.  Otherwise, the Nd abundance pattern we measure appears fairly similar to that seen in the literature.

\section{Conclusion}
\label{sec:conclusion}

In this work we use the BACCHUS code to analyze $\sim$ 126,000 high S/N, $> 150$, APOGEE spectra of giant stars.  Using the APOGEE calibrated stellar parameters of these stars we measure their \cisotope{} ratios and chemical abundances in C, N, O, Na, P, S, V, Cu, Ce, Nd.  We provide both the line-by-line abundances measured by BACCHUS and combinations of these line-by-line measurements, where suspicious measurements have been identified and removed using native BACCHUS line flagging, additional spectral flags, and upper limit relations that we calculate for our elements of interest.

Alongside these combined abundances, we also report upper limit measurements and two measures of the uncertainties on our abundance measurements.  (1) We measure uncertainties from line-by-line scatter of the abundances in a given star.  And (2) we report empirical uncertainties derived from the abundance scatter in stars with repeat observations.
  
Because of the flexibility of our analysis and its careful treatment of spectroscopic details, such as blends and upper limits, we are able to provide our BAWLAS catalog that is complementary to the APOGEE DR17 catalog.  More automated and rigid pipelines such as ASPCAP can have difficulty measuring elements that are weak or heavily blended, whereas our more time-intensive but focused analysis allows us to expand upon what is measurable from APOGEE spectra.  This provides improvements for some of the elements that APOGEE already provides like Na, S and Ce, by cleaning suspect measurements, but also significantly improves the measurement of blended V lines, and can even measure elements that APOGEE is unable to like  P, Cu, Nd, and \cisotope{}, providing the largest samples of P and \cisotope{} measured to date.  

In addition to providing improved abundances for elements that were measured by APOGEE, we can now explore other dimensions of chemical space previously unavailable with APOGEE.  This includes examining r- {\it and} s-process abundances for APOGEE stars, studying dredge-up and stellar evolution along the giant branch with \cisotope{}, and investigating the relatively poorly studied P abundance distribution.  There are, of course multiple uses for this data particularly, because this relatively large sample spans a range of stellar parameters, positions in the galaxy, and even extends to some of the more massive satellites of the Milky Way (such as Sagittarius and the LMC).

Several large spectroscopic surveys are beginning soon or are planned, e.g., Milky Way Mapper (MWM) as part of SDSS-V \citep{sdss5}, the William Herschel Telescope Enhanced Area Velocity Explorer \citep[WEAVE;][]{weave}, the 4-metre Multi-Object Spectroscopic Telescope \citep[4MOST;][]{4most}, and Maunakea Spectroscopic Explorer \citep[MSE;][]{mse}.  These surveys will necessarily require more and more efficient pipelines to handle the volume of data that they will observe.  The current epoch of surveys, like APOGEE, GALAH, LAMOST, etc., have demonstrated that a great wealth of information is available, and can be extracted, and many new, quick methods are being developed to help facilitate this kind of information extraction in the next generation of surveys \citep[e.g., The Cannon, StarNet, astroNN, The Payne;][]{cannon, starnet, astronn, payne}.  

However with the present analysis we show that there is more information that can be extracted from spectroscopic surveys, when analyzed with more careful and meticulous methods.  While methods such as ours here are time consuming and often require high quality data (e.g., high S/N spectra, precise stellar parameters, etc.), future surveys may benefit from having smaller scale, complementary analyses like what we present in this work to capitalize on their highest quality data.

\section*{}
DAGH and TM acknowledge support from the State Research Agency (AEI) of the Spanish Ministry of Science, Innovation and Universities (MCIU), and the European Regional Development Fund (FEDER) under grant AYA2017-88254-P. TM acknowledges financial support from the Spanish Ministry of Science and Innovation (MICINN) through the Spanish State Research
Agency, under the Severo Ochoa Program 2020-2023 (CEX2019-000920-S) as well as support from the ACIISI, Consejería de Economía, Conocimiento y Empleo del Gobierno de Canarias and the European Regional Development Fund (ERDF) under grant with reference PROID2021010128.

This research made use of {\texttt{topcat}} \citep{topcat}, Astropy, a community-developed core Python package for Astronomy \citep{astropy}, NASA's Astrophysics Data System, and the SIMBAD database, operated at CDS, Strasbourg, France.

Funding for the Sloan Digital Sky 
Survey IV has been provided by the 
Alfred P. Sloan Foundation, the U.S. 
Department of Energy Office of 
Science, and the Participating 
Institutions. 

SDSS-IV acknowledges support and 
resources from the Center for High 
Performance Computing  at the 
University of Utah. The SDSS 
website is \url{www.sdss.org}.

SDSS-IV is managed by the 
Astrophysical Research Consortium 
for the Participating Institutions 
of the SDSS Collaboration including 
the Brazilian Participation Group, 
the Carnegie Institution for Science, 
Carnegie Mellon University, Center for 
Astrophysics | Harvard \& 
Smithsonian, the Chilean Participation 
Group, the French Participation Group, 
Instituto de Astrof\'isica de 
Canarias, The Johns Hopkins 
University, Kavli Institute for the 
Physics and Mathematics of the 
Universe (IPMU) / University of 
Tokyo, the Korean Participation Group, 
Lawrence Berkeley National Laboratory, 
Leibniz Institut f\"ur Astrophysik 
Potsdam (AIP),  Max-Planck-Institut 
f\"ur Astronomie (MPIA Heidelberg), 
Max-Planck-Institut f\"ur 
Astrophysik (MPA Garching), 
Max-Planck-Institut f\"ur 
Extraterrestrische Physik (MPE), 
National Astronomical Observatories of 
China, New Mexico State University, 
New York University, University of 
Notre Dame, Observat\'ario 
Nacional / MCTI, The Ohio State 
University, Pennsylvania State 
University, Shanghai 
Astronomical Observatory, United 
Kingdom Participation Group, 
Universidad Nacional Aut\'onoma 
de M\'exico, University of Arizona, 
University of Colorado Boulder, 
University of Oxford, University of 
Portsmouth, University of Utah, 
University of Virginia, University 
of Washington, University of 
Wisconsin, Vanderbilt University, 
and Yale University.

This publication makes use of data products from the Two Micron All Sky Survey, which is a joint project of the University of Massachusetts and the Infrared Processing and Analysis Center/California Institute of Technology, funded by the National Aeronautics and Space Administration and the National Science Foundation.

This work has made use of data from the European Space Agency (ESA) mission
{\it Gaia} (\url{https://www.cosmos.esa.int/gaia}), processed by the {\it Gaia}
Data Processing and Analysis Consortium (DPAC,
\url{https://www.cosmos.esa.int/web/gaia/dpac/consortium}). Funding for the DPAC
has been provided by national institutions, in particular the institutions
participating in the {\it Gaia} Multilateral Agreement.

\facilities{Sloan (APOGEE-N spectrograph), Du Pont (APOGEE-S spectrograph)}

\software{BACCHUS \citep{masseron2016}, TOPCAT \citep{topcat}, Astropy \citep{astropy}, Matplotlib (\url{http://dx.doi.org/10.1109/MCSE.2007.55}), Numpy (\url{http://scitation.aip.org/content/aip/journal/cise/13/2/10.1109/MCSE.2011.37})}

\begin{appendix}

\section{BACCHUS Flag and Combination Settings}
\label{app:comb_settings}
 
As mentioned in Section \ref{sec:line_comb}, here we describe the choices we have made to flag individual line measurements and combine them to derive the final elemental abundances that are presented in this paper.  In many cases we have used a set of default settings that work well for most lines across the elements we measure.  The line-by-line settings that we have to choose from fall into the following categories:  (1) which lines did we use in our combination, (2) which BACCHUS measurement method was used for each line, (3) what BACCHUS method flags were use to identify satisfactory measurements, (4) what spectra flags were used for each line, and finally (5) what upper limit threshold was used for each line.  Our default choices for how to flag and combine lines for each of these categories are:
 \begin{enumerate}
     \item Line selection:  use all input lines of a given element
     \item BACCHUS measurement method:  use the \chitwo{} method
     \item BACCHUS method flags:  require that all reported BACCHUS method flags = 1 for a given measured line
     \item Spectra flags:  require that all spectra flags = 1 (if such a flag is present for the star and line in question).
     \item Upper limit threshold:  use the t = 1\% empirical threshold relation to define upper limits
 \end{enumerate}
However, in some cases we have deviated from these default settings in order to better treat specific lines.  Table \ref{tab:settings} lists all of the flag choices and settings that we have used in our line combination, and below we go through each element and describe the choices that differ from the defaults listed above.

 \startlongtable
\begin{deluxetable*}{l c c c c c c c c c c}
\tabletypesize{\scriptsize}
\tablewidth{0pt}
\tablecolumns{11}
\tablecaption{Line Flag and Combination Settings \label{tab:settings}}
\tablehead{\colhead{Line (\AA)} & \colhead{Use Line?} & \colhead{Measurement Method} & \multicolumn{5}{c}{Method Flags} & \multicolumn{2}{c}{Spectra Flags} & \colhead{Upper Limit Threshold} \\ \cline{4-8} \cline{9-10} \colhead{} & \colhead{} & \colhead{} & \colhead{\syn{}} & \colhead{\eqw{}} & \colhead{\intmethod{}} & \colhead{\chitwo{}} & \colhead{\wln{}} & \colhead{blend} & \colhead{cont} & \colhead{}}
\startdata
\cutinhead{C}
15578.0 & Yes & \chitwo{} & 1 & 1 & 1 & 1 & 1 & - & - & BACCHUS \intmethod{} limit\\
15775.5 & Yes & \chitwo{} & 1 & 1 & 1 & 1 & 1 & - & - & BACCHUS \intmethod{} limit\\
15783.9 & Yes & \chitwo{} & 1 & 1 & 1 & 1 & 1 & - & - & BACCHUS \intmethod{} limit\\
15978.7 & Yes & \chitwo{} & 1 & 1 & 1 & 1 & 1 & - & - & BACCHUS \intmethod{} limit\\
16004.9 & Yes & \chitwo{} & 1 & 1 & 1 & 1 & 1 & - & - & BACCHUS \intmethod{} limit\\
16021.7 & Yes & \chitwo{} & 1 & 1 & 1 & 1 & 1 & - & - & BACCHUS \intmethod{} limit\\
16185.5 & Yes & \chitwo{} & 1 & 1 & 1 & 1 & 1 & - & - & BACCHUS \intmethod{} limit\\
16397.2 & Yes & \chitwo{} & 1 & 1 & 1 & 1 & 1 & - & - & BACCHUS \intmethod{} limit\\
16481.5 & Yes & \chitwo{} & 1 & 1 & 1 & 1 & 1 & - & - & BACCHUS \intmethod{} limit\\
16614.0 & Yes & \chitwo{} & 1 & 1 & 1 & 1 & 1 & - & - & BACCHUS \intmethod{} limit\\
16836.0 & Yes & \chitwo{} & 1 & 1 & 1 & 1 & 1 & - & - & BACCHUS \intmethod{} limit\\
16890.4 & Yes & \chitwo{} & 1 & 1 & 1 & 1 & 1 & - & - & BACCHUS \intmethod{} limit\\
17063.0 & Yes & \chitwo{} & 1 & 1 & 1 & 1 & 1 & - & - & BACCHUS \intmethod{} limit\\
17448.6 & Yes & \chitwo{} & 1 & 1 & 1 & 1 & 1 & - & - & BACCHUS \intmethod{} limit\\
17456.0 & Yes & \chitwo{} & 1 & 1 & 1 & 1 & 1 & - & - & BACCHUS \intmethod{} limit\\
\cutinhead{N}
15119.0 & Yes & \chitwo{} & 1 & 1 & 1 & 1 & 1 & - & - & BACCHUS \intmethod{} limit\\
15210.2 & Yes & \chitwo{} & 1 & 1 & 1 & 1 & 1 & - & - & BACCHUS \intmethod{} limit\\
15222.0 & Yes & \chitwo{} & 1 & 1 & 1 & 1 & 1 & - & - & BACCHUS \intmethod{} limit\\
15228.8 & Yes & \chitwo{} & 1 & 1 & 1 & 1 & 1 & - & - & BACCHUS \intmethod{} limit\\
15242.5 & Yes & \chitwo{} & 1 & 1 & 1 & 1 & 1 & - & - & BACCHUS \intmethod{} limit\\
15251.8 & Yes & \chitwo{} & 1 & 1 & 1 & 1 & 1 & - & - & BACCHUS \intmethod{} limit\\
15309.0 & Yes & \chitwo{} & 1 & 1 & 1 & 1 & 1 & - & - & BACCHUS \intmethod{} limit\\
15317.6 & Yes & \chitwo{} & 1 & 1 & 1 & 1 & 1 & - & - & BACCHUS \intmethod{} limit\\
15363.5 & Yes & \chitwo{} & 1 & 1 & 1 & 1 & 1 & - & - & BACCHUS \intmethod{} limit\\
15410.5 & Yes & \chitwo{} & 1 & 1 & 1 & 1 & 1 & - & - & BACCHUS \intmethod{} limit\\
15447.0 & Yes & \chitwo{} & 1 & 1 & 1 & 1 & 1 & - & - & BACCHUS \intmethod{} limit\\
15462.4 & Yes & \chitwo{} & 1 & 1 & 1 & 1 & 1 & - & - & BACCHUS \intmethod{} limit\\
15466.2 & Yes & \chitwo{} & 1 & 1 & 1 & 1 & 1 & - & - & BACCHUS \intmethod{} limit\\
15495.0 & Yes & \chitwo{} & 1 & 1 & 1 & 1 & 1 & - & - & BACCHUS \intmethod{} limit\\
15514.0 & Yes & \chitwo{} & 1 & 1 & 1 & 1 & 1 & - & - & BACCHUS \intmethod{} limit\\
15581.0 & Yes & \chitwo{} & 1 & 1 & 1 & 1 & 1 & - & - & BACCHUS \intmethod{} limit\\
15636.5 & Yes & \chitwo{} & 1 & 1 & 1 & 1 & 1 & - & - & BACCHUS \intmethod{} limit\\
15659.0 & Yes & \chitwo{} & 1 & 1 & 1 & 1 & 1 & - & - & BACCHUS \intmethod{} limit\\
15706.9 & Yes & \chitwo{} & 1 & 1 & 1 & 1 & 1 & - & - & BACCHUS \intmethod{} limit\\
15708.5 & Yes & \chitwo{} & 1 & 1 & 1 & 1 & 1 & - & - & BACCHUS \intmethod{} limit\\
15825.7 & Yes & \chitwo{} & 1 & 1 & 1 & 1 & 1 & - & - & BACCHUS \intmethod{} limit\\
\cutinhead{O}
15373.5 & Yes & \chitwo{} & 1 & 1 & 1 & 1 & 1 & - & - & BACCHUS \intmethod{} limit\\
15391.0 & Yes & \chitwo{} & 1 & 1 & 1 & 1 & 1 & - & - & BACCHUS \intmethod{} limit\\
15569.0 & Yes & \chitwo{} & 1 & 1 & 1 & 1 & 1 & - & - & BACCHUS \intmethod{} limit\\
15719.7 & Yes & \chitwo{} & 1 & 1 & 1 & 1 & 1 & - & - & BACCHUS \intmethod{} limit\\
15778.5 & Yes & \chitwo{} & 1 & 1 & 1 & 1 & 1 & - & - & BACCHUS \intmethod{} limit\\
16052.9 & Yes & \chitwo{} & 1 & 1 & 1 & 1 & 1 & - & - & BACCHUS \intmethod{} limit\\
16055.5 & Yes & \chitwo{} & 1 & 1 & 1 & 1 & 1 & - & - & BACCHUS \intmethod{} limit\\
16650.0 & Yes & \chitwo{} & 1 & 1 & 1 & 1 & 1 & - & - & BACCHUS \intmethod{} limit\\
16704.8 & Yes & \chitwo{} & 1 & 1 & 1 & 1 & 1 & - & - & BACCHUS \intmethod{} limit\\
16714.5 & Yes & \chitwo{} & 1 & 1 & 1 & 1 & 1 & - & - & BACCHUS \intmethod{} limit\\
16872.0 & Yes & \chitwo{} & 1 & 1 & 1 & 1 & 1 & - & - & BACCHUS \intmethod{} limit\\
16909.4 & Yes & \chitwo{} & 1 & 1 & 1 & 1 & 1 & - & - & BACCHUS \intmethod{} limit\\
\cutinhead{Na}
16373.9 & Yes & \wln{} & 1 & 1 & 1 & 1 & 1 & - & 1 & 1\% \\
16388.8 & Yes & \chitwo{} & 1 & 1 & 1 & 1 & 1 & 1 & 1 & 1\% \\
\cutinhead{P}
15711.6 & Yes & \wln{} & 1 & 1 & 1 & 1 & 1 & 1 & - & 3\% \\
16482.9 & Yes & \wln{} & 1 & 1 & 1 & 1 & 1 & - & 1 & 3\% \\
\cutinhead{S}
15403.5 & Yes & \chitwo{} & 1 & 1 & 1 & 1 & 1 & 1 & - & 1\% \\
15422.3 & No & \chitwo{} & 1 & 1 & 1 & 1 & 1 & 1 & - & 1\% \\
15469.8 & Yes & \chitwo{} & 1 & 1 & 1 & 1 & 1 & 1 & - & 1\% \\
15475.6 & No & \chitwo{} & 1 & 1 & 1 & 1 & 1 & 1 & - & 1\% \\
15478.5 & Yes & \chitwo{} & 1 & 1 & 1 & 1 & 1 & 1 & - & 1\% \\
\cutinhead{V}
15924.8 & Yes & \chitwo{} & 1 & 1 & 1 & 1 & 1 & - & 1 & 3\% \\
16137.3 & No & \chitwo{} & 1 & 1 & 1 & 1 & 1 & - & - & 1\% \\
16200.2 & Yes & \chitwo{} & 1 & 1 & 1 & 1 & 1 & - & - & 1\% \\
\cutinhead{Cu}
16005.5 & Yes & \chitwo{} & 1,3 & all & all & 1 & 1 & 1 & - & 3\% \\
16006.0 & Yes & \wln{} & all & all & all & 1 & 1 & 0,1 & - & 3\% \\
\cutinhead{Ce}
15784.8 & Yes & \chitwo{} & 1 & 1 & 1 & 1 & 1 & 1 & - & 1\% \\
16376.5 & Yes & \chitwo{} & 1 & 1 & 1 & 1 & 1 & - & - & 1\% \\
16595.2 & Yes & \chitwo{} & 1 & 1 & 1 & 1 & 1 & - & - & 1\% \\
16722.5 & Yes & \chitwo{} & 1 & 1 & 1 & 1 & 1 & - & - & 2\% \\
\cutinhead{Nd}
15368.1 & Yes & \chitwo{} & 1 & 1 & 1 & 1 & 1 & 0,1 & 0,1 & 2\% \\
16053.6 & Yes & \wln{} & 1,3 & 1,3 & all & 1 & 1 & - & 1 & 1\% \\
16262.0 & Yes & \chitwo{} & 1 & 1 & 1 & 1 & 1 & - & - & 3\% \\
\cutinhead{\cisotope{}}
15641.7 & Yes & \wln{} & 1,2,3 & 1,2,3 & all & 0,1 & 1,2 & - & 1 & \multirow{8}{2.25cm}{\centering Line-by-line lower limit of \cisotope{} $= 49$, require that line 16530 \AA{} is measured, and final lower limits based on a relation for 16530 \AA{}.} \\
16121.4 & Yes & \chitwo{} & 1,2,3 & 1,2,3 & all & 0,1 & 1,2 & - & 1 &  \\
16323.4 & Yes & \wln{} & 1,2,3 & 1,2,3 & all & 0,1 & 1,2 & - & 1 &  \\
16326.0 & Yes & \wln{} & 1,2,3 & 1,2,3 & all & 0,1 & 1,2 & - & 1 &  \\
16327.3 & Yes & \wln{} & 1,2,3 & 1,2,3 & all & 0,1 & 1,2 & - & 1 &  \\
16530.0 & Yes & \chitwo{} & 1,2,3 & 1,2,3 & all & 0,1 & 1,2 & - & 1 &  \\
16741.2 & Yes & \wln{} & 1,2,3 & 1,2,3 & all & 0,1 & 1,2 & - & 1 &  \\
16744.7 & Yes & \wln{} & 1,2,3 & 1,2,3 & all & 0,1 & 1,2 & - & 1 &  \\
\enddata
\tablecomments{When multiple flags (comma separated) are given for a specific method, any lines flagged with those values were considered in line combination.}
\tablecomments{Lines marked with a ``-'' in their blend or cont spectra flag column indicate that no such flag was defined for that line.}
\end{deluxetable*}

\subsection{Carbon, Nitrogen, and Oxygen (C, N, and O)}
For C, N, and O, since these were used primarily for fitting blends, we have used a fairly simple selection of settings to combine these elements.  For C, N, and O we use our default settings for the line selection (all lines), measurement method (\chitwo{}), and the method flag choices (flag = 1 for all methods).  No spectra flags were defined or used for C, N, and O. 

For the upper limits of C, N, and O, because these are dominantly measured in molecular features, they are therefore sensitive to each other's relative abundance.  For instance typical stars have abundance ratios with C $<$ O, but the molecular features in a star strongly differ when they have abundance ratio C $>$ O.  However, our empirical upper limit relations are derived assuming solar abundances of all elements except the element in question, therefore driving a changing abundance ratio relative to other elements.  Therefore for C, N, and O the empirical relations we derive are not appropriate because they are derived from spectra with strongly changing [C/N], [C/O], etc. and do not reflect the ratios seen in typical stars, leading to unrealistic limits.

Instead we have used the native BACCHUS derived \intmethod{} method upper limit for each line in each star.  While these upper limits can occasionally extend to very high values, but because C, N, and O have quite strong features, it is rare that a line is simultaneously flagged as well measured by BACCHUS and below these upper limits.

\subsection{Sodium (Na)}
\begin{figure}
  \centering
  \includegraphics[angle=-90,scale=0.3,trim = 0.in 0.in 0.in 0.in, clip]{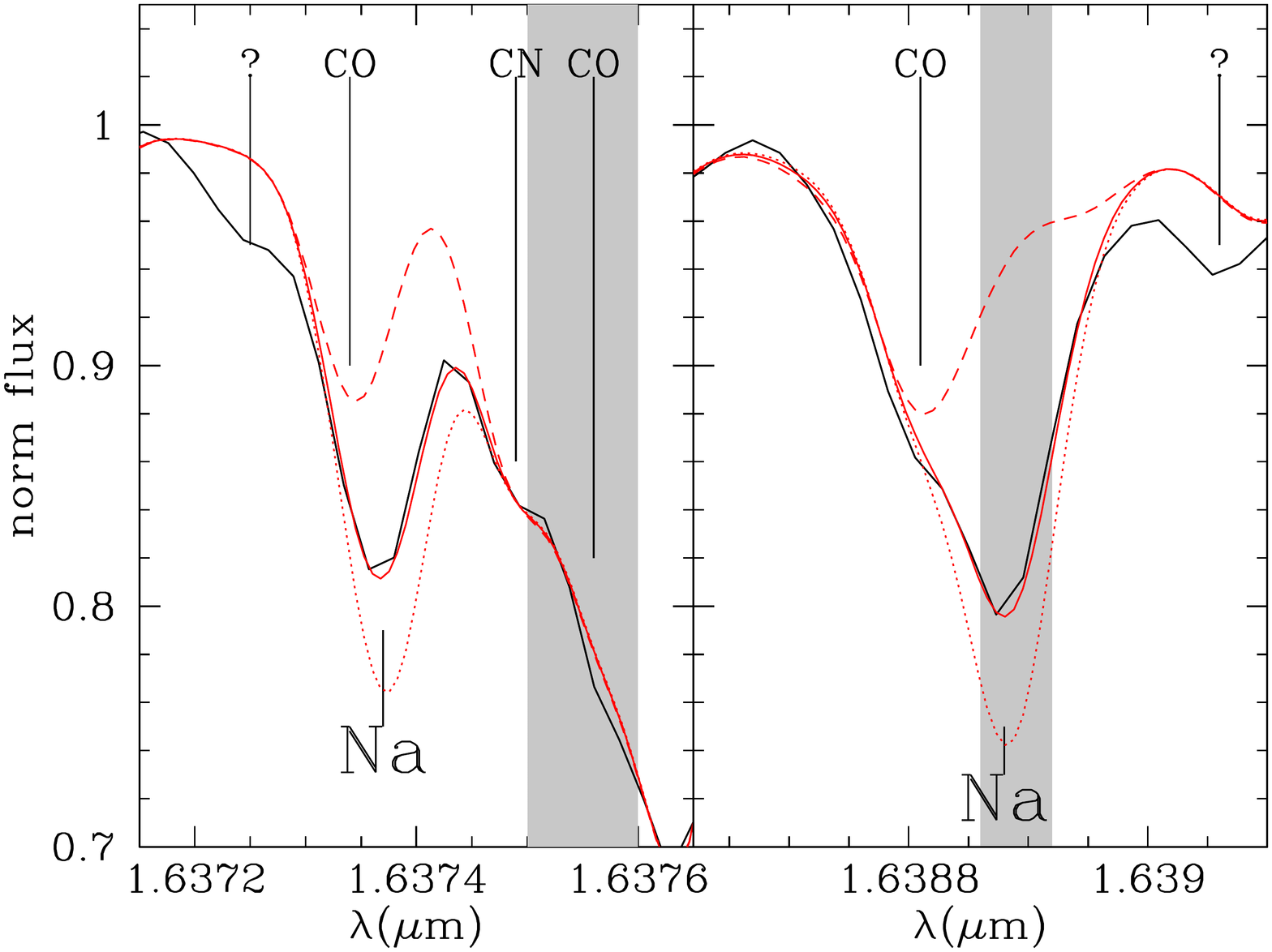}
  \caption{Na lines in the observed spectrum of 2M06370368+0624303 (solid black line) compared to the best fit model (solid red line), the best fit model without the Na lines (red dashed line) and the best fit model $+0.3$ dex in Na abundance (red dotted line). Both Na lines are affected by underlying CO lines, most particularly for cool stars. The shaded areas highlight the regions used to define the Na continuum flags to ensure that the modelling CO lines synthesis reach a satisfying level.}
  \label{fig:Nalines}
\end{figure}

An example of the two Na lines is shown in Figure ~\ref{fig:Nalines}.  For Na we use the \wln{} method instead of \chitwo{} for the Na 16373 \AA{} line because it is a weak line with both blue-ward and red-ward blends (with the red-ward CO blend nearly coincident with the Na 16373 \AA{} line) and the \wln{} method allows us to measure Na from a point where the Na line is maximal while minimizing contamination from any poorly fit blends.

\subsection{Phosphorus (P)}
\begin{figure}
  \centering
  \includegraphics[angle=-90,scale=0.3,trim = 0.in 0.in 0.in 0.in, clip]{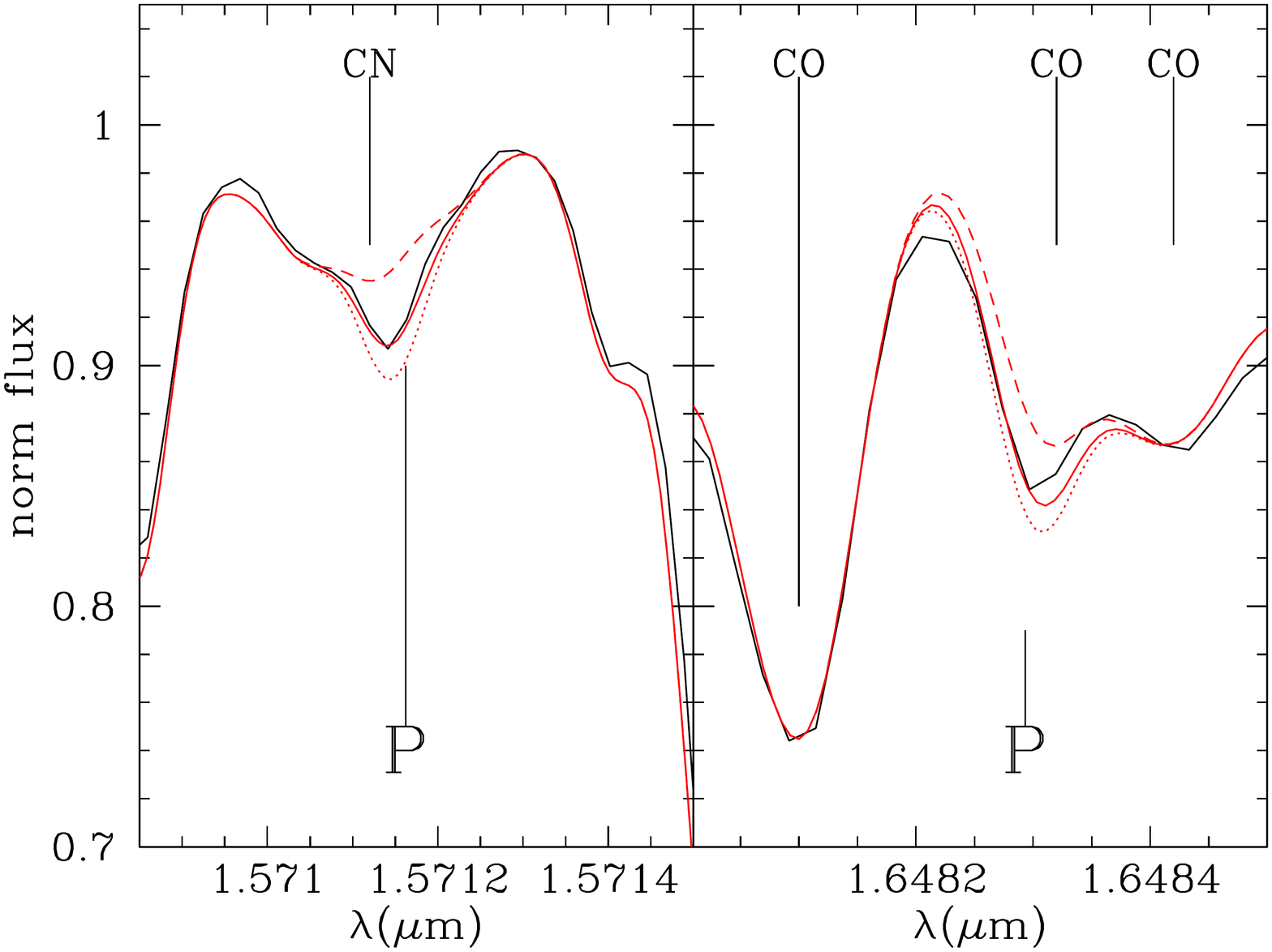}
  \caption{P lines in observed spectrum of 2M23200098+3508030 compared to the best fit model with the best fit P abundance, no P, and the best fit P abundance $+0.3$ dex (as color-coded in Figure \ref{fig:Nalines}). The P abundance measurement is dependent on the quality of the fit of CN and CO blending features, for which we have developed specific blend flags.  }
  \label{fig:Plines}
\end{figure}

An example of the two P lines is shown in Figure ~\ref{fig:Plines}.  Both of the P lines that we measure are weak and blended, so we have chosen to use the \wln{} method to measure them where the lines are strongest and minimize noise and contamination from incompletely fit blends (because even small mismatches in the underlying blend can have a large impact on calculated the abundances from weak P lines).  Similarly because of the strong blends with the weak P lines we have raised the upper limit threshold to the 3\% level which has been verified with by-eye investigation of borderline measurements.

\subsection{Sulfur (S)}
\begin{figure*}
  \centering
  \includegraphics[scale=0.65,trim = 0.in 0.in 0.in 3.2in, clip]{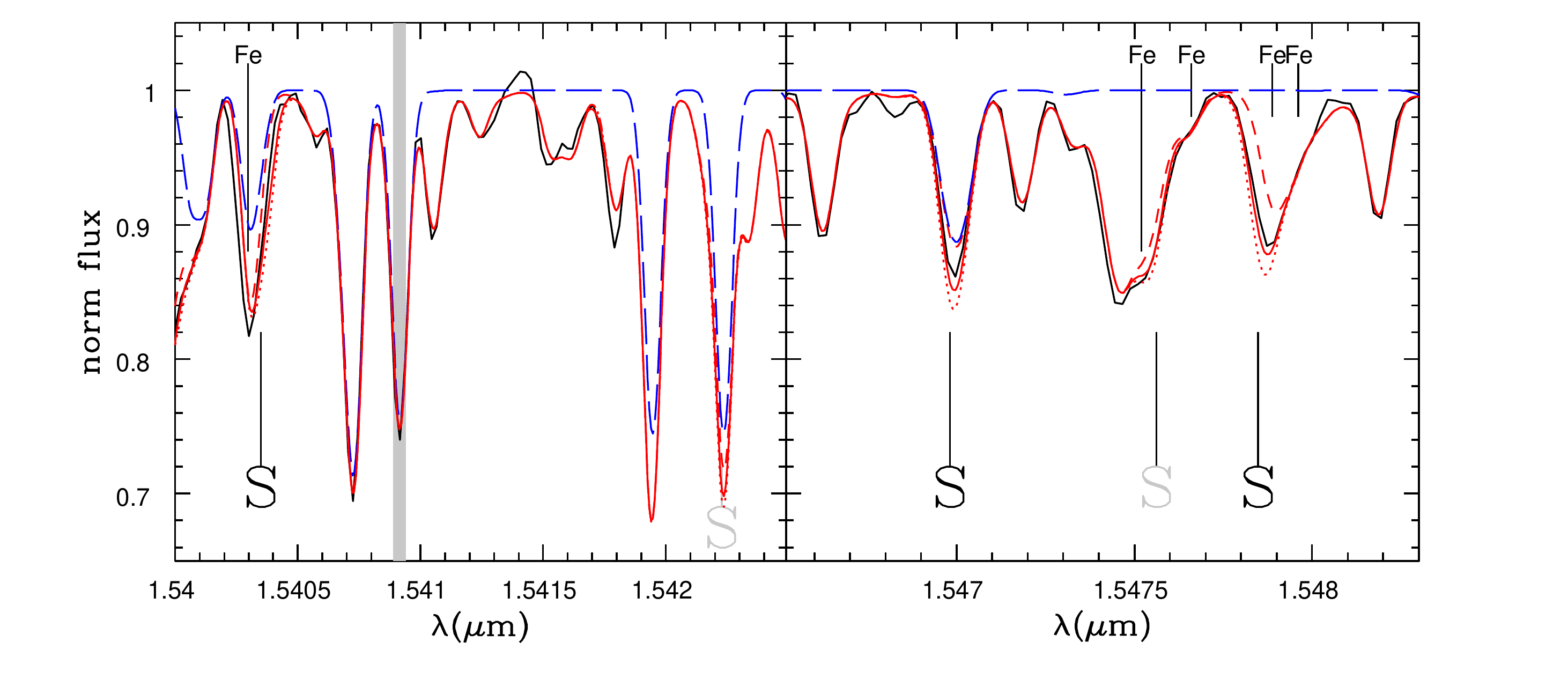}
  \caption{S lines in the observed spectrum of 2M19371509+4007494 compared to the best fit model with the best fit S abundance, no S, and the best fit S abundance $+0.3$ dex (as color-coded in Figure \ref{fig:Nalines}). Over-plotted we also include a synthesis of only the OH lines of the best fit model (dashed blue line).  Two of the S lines are affected by underlying OH lines, and in the case of particularly cool and metal-poor stars, these lines can completely dominate the S line. The shaded areas highlight the region used to define the S blend flags, by requiring that the OH lines should not reach a certain strength. Some of the S lines are also affected by Fe lines. Note that although there are five S detectable lines in the APOGEE spectra that we analyzed, we use only three of them. }
  \label{fig:Slines}
\end{figure*}

An example of the five possible S lines analyzed here is shown in Figure ~\ref{fig:Slines}.  While we use the default settings for our S lines, after inspection of the output line-by-line abundance ratios we have chosen to not include the measurements of lines 15422 \AA{} and 15475 \AA{}.  These lines both show strong temperature trends even once removing stars that have been flagged and have chemical abundance patterns that differ from each other and the other three lines.  

\subsection{Vanadium (V)}
\begin{figure*}
  \centering
  \includegraphics[scale=0.65,trim = 0.in 0.in 0.in 3.2in, clip]{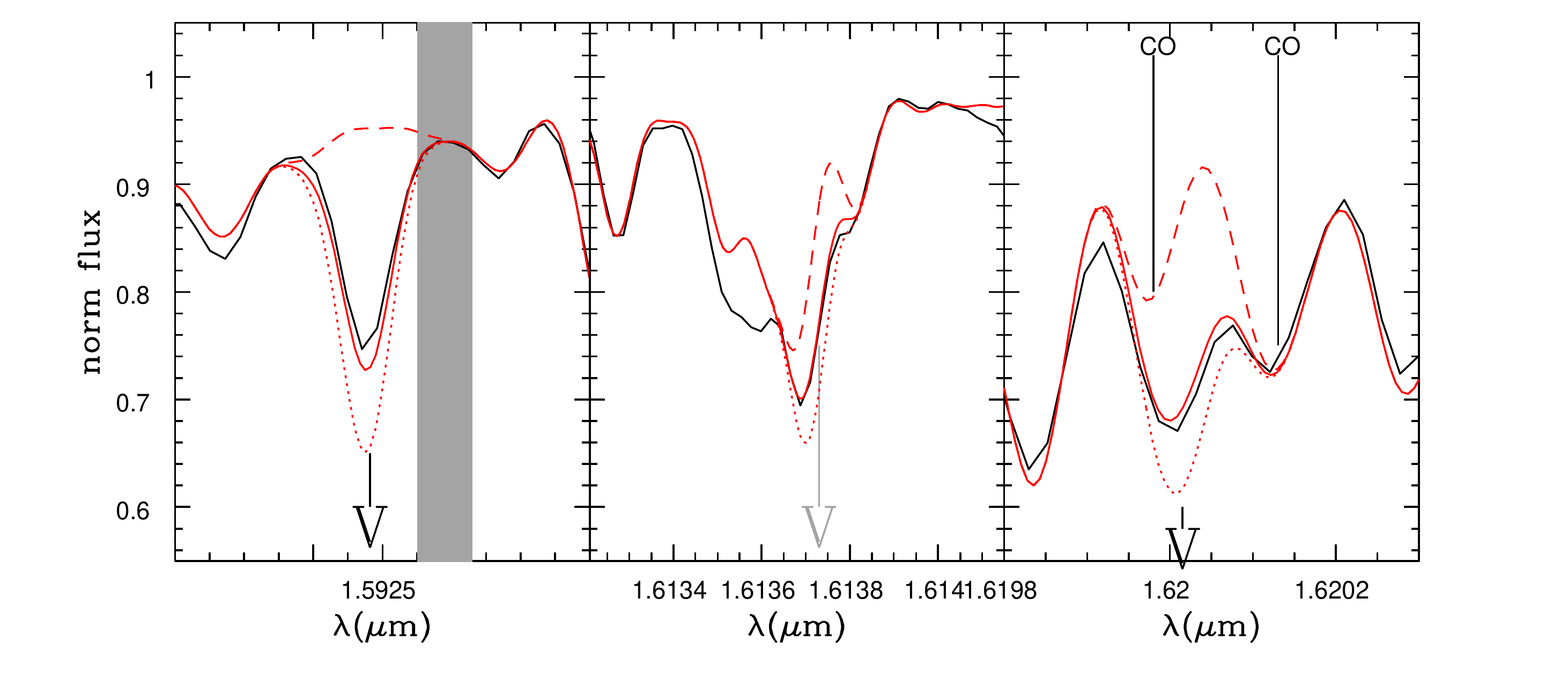}
  \caption{The V lines in the observed spectrum of 2M19054118+3927539 compared to the best fit model with the best fit V abundance, no V, and the best fit V abundance $+0.3$ dex (as color-coded in Figure \ref{fig:Nalines}). The shaded area shows the region used to estimate the quality of the pseudo-continuum placement and set the corresponding continuum flag. Note that we show the 16137$\rm \AA$ line but we do not use it in our line combination due to its heavy multiple blends.}
  \label{fig:Vlines}
\end{figure*}

An example of the three V lines we analyze is shown in Figure ~\ref{fig:Vlines}.  Similar to S, for V we have not included the V 16137 \AA{} line in our combination, because it is a slightly weaker and more blended line, and shows strong temperature trends and abundance patterns that deviate from the other two (likely due to misfit blends).  We also impose an upper limit threshold of t = 3\% to remove stars that end up with anomalously low V abundances that should be below our detection limit.  This was implemented in addition to the continuum flag designed for the V 15924 \AA{} line, to only consider measurements of lines that are sufficiently strong to dominate over errors from the spectral normalization around this line (see Section \ref{sec:v15924contflag} for why these errors may arise).

\subsection{Copper (Cu)}
\begin{figure}
  \centering
  \includegraphics[angle=-90,scale=0.3,trim = 0.in 0.in 0.in 0.in, clip]{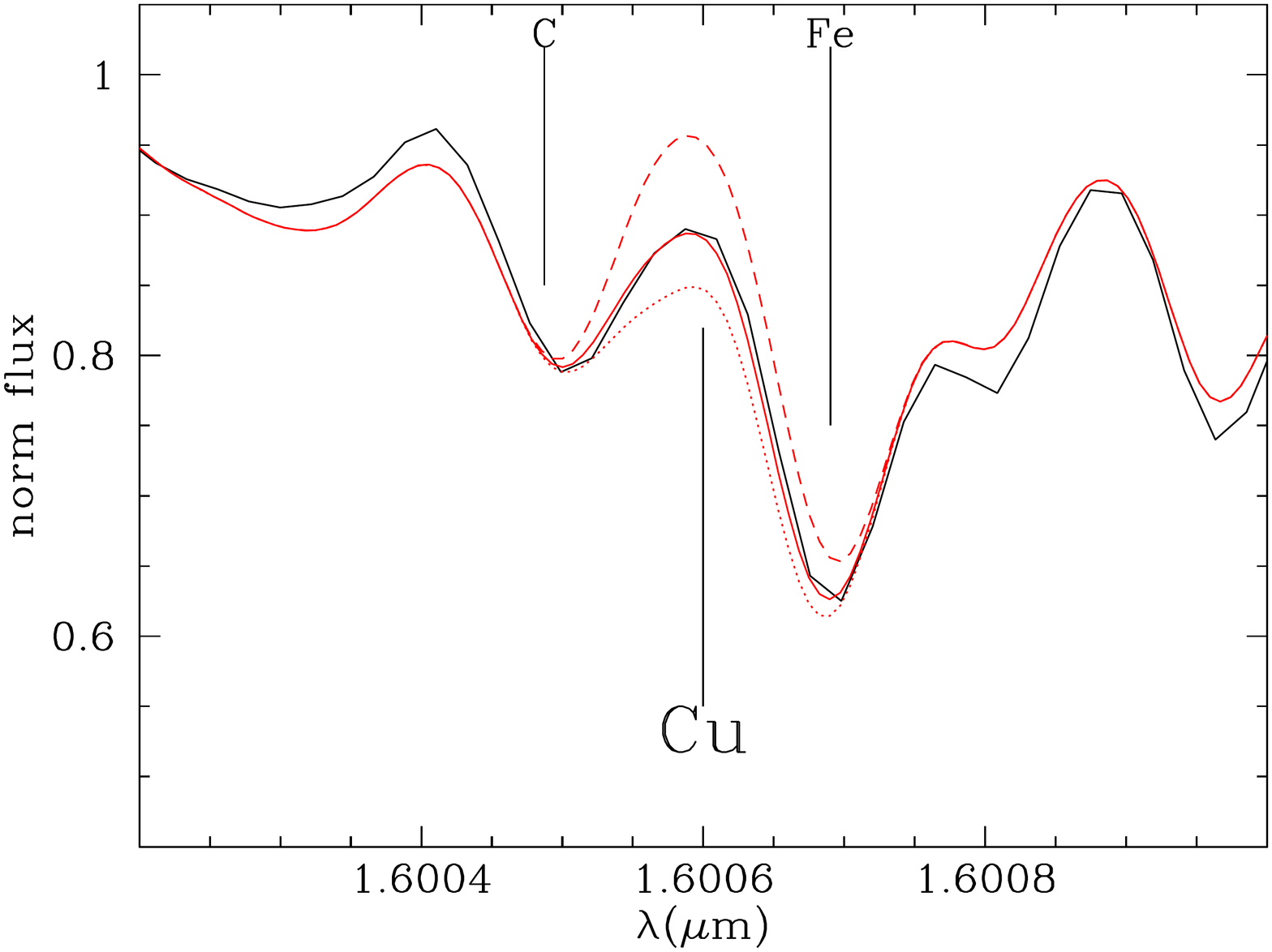}
  \caption{The Cu lines (blended into one feature) in the observed spectrum of 2M05123340-4010546 compared to the best fit model with the best fit Cu abundance, no Cu, and the best fit Cu abundance $+0.3$ dex (as color-coded in Figure \ref{fig:Nalines}). }
  \label{fig:Culine}
\end{figure}

An example of the single Cu feature (a blend of two Cu lines) is shown in Figure ~\ref{fig:Culine}. The Cu 16005 \AA{} and 16006 \AA{} lines both are significantly blended both blue-ward by a combination of \ion{C}{1}, \ion{Ti}{2} and \ion{Fe}{1} lines, and red-ward by \ion{C}{1} and \ion{Fe}{1} lines.  The blue-ward blend is weaker, so for the Cu 16005 \AA{} line, we can still use the \chitwo{} method with minimal contamination from the blue-ward blends, but the red-ward blend is typically stronger and can contaminate wings of the redder Cu 16006 \AA{} line.  So for the Cu 16006 \AA{} line we can minimize this contamination by measuring the the Cu abundance at a point with minimal contribution from blends by using the \wln{} method.

The blends also cause many method flags to be thrown by BACCHUS, so we have modified our flag selection for the Cu lines.  For the \eqw{} and \intmethod{} flags, we accept all flags (i.e., we ignore these flags or do not exclude a measurement for any specific flag from these methods) for both of the Cu lines.  Of the \syn{} flags, we allow any flags for the Cu 16006 \AA{} line, but because the blends are weaker for the Cu 16005 \AA{} line we only use \syn{} flag = 1 or 3 (many stars have \syn{} flag = 3 because the Cu line falls in between two blends and does not produce a local minimum in flux, and are therefore flagged despite decent fits by BACCHUS).

To further reduce the impact of poorly fit blends that can be interpreted by BACCHUS as enhancements in Cu, we have also used a higher upper limit threshold of t = 3\%.

Finally, while we defined a blend flag for the Cu 16006 \AA{} line we did not cut any stars based on this flag since there were no clear offsets in the Cu measurements of these (few) stars from the rest of the population.  However all stars with blend flag = 0 were removed through other quality and flag cuts when calculating combined Cu abundances.

\subsection{Cerium (Ce)}
\begin{figure*}
  \centering
  \includegraphics[scale=0.65,trim = 0.in 0.in 0.in 3.8in, clip]{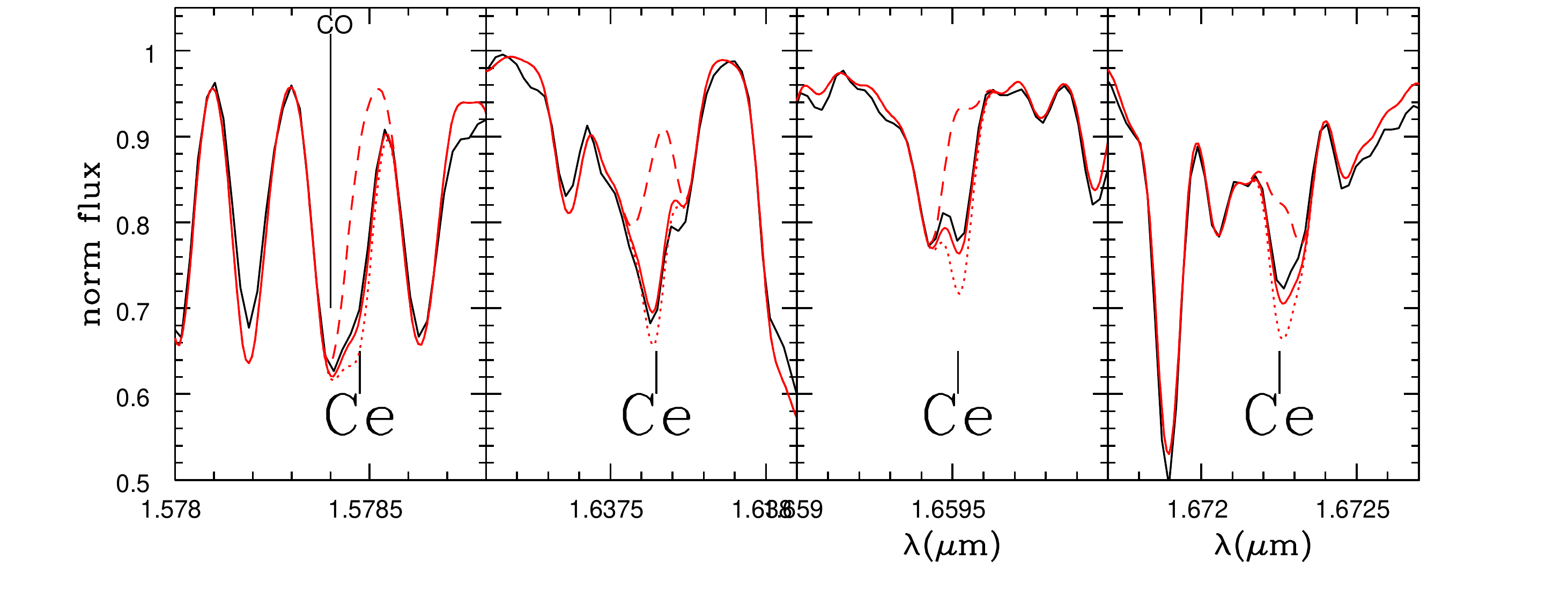}
  \caption{The four Ce lines analyzed in this work in observed spectrum of 2M20583420+6433429 compared to the best fit model with the best fit Ce abundance, no Ce, and the best fit Ce abundance $+0.3$ dex (as color-coded in Figure \ref{fig:Nalines}). }
  \label{fig:Celines}
\end{figure*}

An example of the four Ce lines used in this study is shown in Figure ~\ref{fig:Celines}.  For Ce we use the default settings for line flagging and combination, except for raising the upper limit threshold slightly to t = 2\% for the Ce 16722 \AA{} line.  This helps account for stars that visually appear to be upper limits but are considered detections in warm stars at the t = 1\% threshold due to mismatches in the nearby \ion{Si}{1} blend.  

\subsection{Neodymium (Nd)}
\label{app:Nd}

\begin{figure*}
  \centering
  \includegraphics[scale=0.65,trim = 0.in 0.in 0.in 3.2in, clip]{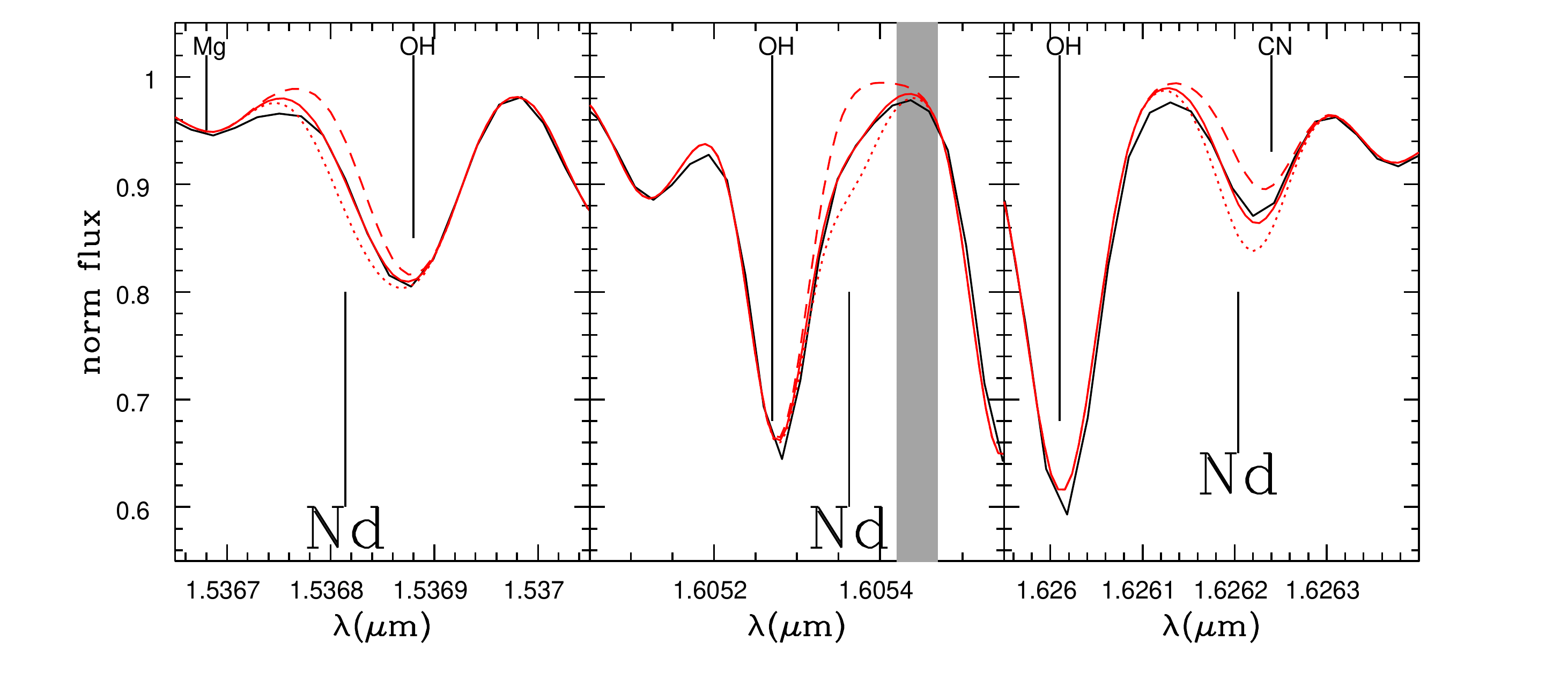}
  \caption{The three Nd lines analyzed in this work as seen in the observed spectrum of 2M18384250-3452180 compared to the best fit model with the best fit Nd abundance, no Nd, and the best fit Nd abundance $+0.3$ dex (as color-coded in Figure \ref{fig:Nalines}). The shaded area shows the region used to estimate the quality of the continuum placement and set the corresponding continuum flag. }
  \label{fig:Ndlines}
\end{figure*}

An example of the three Nd lines used in this study is shown in Figure ~\ref{fig:Ndlines}.  The Nd 16053 \AA{} line lies in the wings of an OH line, which is frequently included in a larger BACCHUS window around the Nd 16053 \AA{}.  To avoid adding more noise to our Nd measurements and avoid contamination in cases of poorer fits to this OH feature, we have chosen to use the \wln{} method for this line, which appears to provide more consistent measurements and is fitting the Nd 16053 \AA{} at a wavelength that is minimally blended with any other lines.

In addition to using the \wln{} method for Nd 16053 \AA{}, BACCHUS tends to focus on the OH line core as the nearest local minimum within it's fitting window.  Therefore it will report \syn{} and \eqw{} flag = 3 for many stars when the fits to the Nd line are otherwise good, so we include \syn{} and \eqw{} flag = 1 or 3 when assessing Nd measurements of this line.  Again, because BACCHUS focuses on the core of the OH blend for its \intmethod{} method, the \intmethod{} flags are inappropriate for the overall quality of the BACCHUS fit to the Nd line, so we allow any \intmethod{} flag for the Nd 16053 \AA{} line.

As with the weak lines of some other elements, we've increased the upper limit threshold for the weaker Nd 15368 \AA{} and 16262 \AA{} lines because of coincident blends and continuum placement that can lead BACCHUS to derive good measurements where visual inspection reveals that the fits are likely upper limits.  This particularly occurs for hotter stars where our linear empirical upper limit threshold relations may not be complex enough to encompass the change in line depths from synthetic spectra as a function of temperature.  Particularly we've raised the thresholds to 2\% and 3\% for Nd 15368 \AA{} and 16262 \AA{} respectively, to flag hotter stars as much as possible (where the upper limits are not quite high enough when inspecting individual stars), while minimizing the number of cooler stars that are flagged with Nd abundances that do appear to be measurable.

Finally, for Nd, while we created a blend and continuum flag for the Nd 15368 \AA{} line, we decided not to use these flags as they primarily flagged stars with parameters where the Nd measurements should be more reliable (the blend flag flags nearly the whole sample and the continuum flag flagged only cool stars leaving things below 4800 K where visual inspection suggests that these measurements are predominantly upper limits).  This does, however, suggest that this line (or its blends) may be more poorly fit on average, possibly the source of the large zero-point offset relative to \citet{Grevesse2007} of 0.372.  However, this line does show similar abundance trends to the other Nd lines so it has been retained.

\subsection{\cisotope{} Isotopic Ratio}

As mentioned in Section \ref{sec:method_flags}, because BACCHUS's flags are tuned typical abundances measurements of individual lines, the flags are not quite appropriate for measuring $^{13}$C features.  Typically the flags are designed for narrow features, and for the broader features of \cisotope{}, especially those with a variety of blends, BACCHUS will often throw \syn{} flag = 3 and \eqw{} = 3. 

These flags are not necessarily a description of the quality of the fit for \cisotope{}, so we also allow these flags when combining \cisotope{} measurements.  Furthermore, since the \intmethod{} method attempts to measure abundances from line core intensities, it is not particularly suited to the, typically broader, $^{13}$C features, so we allow any \intmethod{} flags in our combination.

In general for \cisotope{}, because they have a relatively broad wavelength coverage and are therefore blended with multiple lines, we have chosen to use the \wln{} method for our measurements, measuring the most intense parts of these features.  While the \chitwo{} method could provide more robust measurements by averaging multiple pixels, it is also more susceptible to blends which are more frequent when taking a larger wavelength range.  The only two lines that we use the \chitwo{} method for are the $^{13}$C$^{16}$O 16121 \AA{} and 16530 \AA{} features, which are the strongest and least blended features in our list.

Since $^{13}$C features have increasing depth with decreasing \cisotope{} values, measurement limits are actually lower limits, and our empirical relations are not appropriate for the \cisotope{} values we derive.  Instead because our input range maxes out at \cisotope{} $= 50$ we put a default line-by-line lower limit of 49 (which can cut off stars that sit near the upper synthesis range) on the measurements of all lines. To add another safe guard that helps ensure the \cisotope{} measurements are reasonable, for all of the stars that we report a final, combined \cisotope{} value, we require that they have good measurements from the $^{13}$C$^{16}$O 16530 \AA{} feature which is the strongest, least blended line.  So, if this line has not been measured, we posit that the other lines too shouldn't be measurable.

Finally to provide useful lower limits, we have derived a lower limit relation based on the strongest $^{13}$C feature we analyze, the $^{13}$C$^{16}$O 16530 \AA{} feature.  The form of our \cisotope{} ratio lower limit relation is $\log_{10}($\cisotope$) = A/T_{\rm eff} + {\rm [C/H]} + B$, where \xh{C} is that calculated by BACCHUS. By adding a factor of \xh{C}, this limit essentially is a limit on \xh{$^{13}$C}.  

In order to determine the value of the coefficient $A$, we synthesize the $^{13}$C$^{16}$O 16530 \AA{} region as described in Section \ref{sec:upper_limits_relation} for \cisotope{} ratios of: 1, 3, 5, 7, 10, 15, 20, 30, 50, and 70.  Based on the temperature dependence of these syntheses we derive a value of $A = 37200$ K.  We then set the coefficient $B = -7.0$ from a data driven determination, based on the cont flag for $^{13}$C, which provides a limit threshold at the flux level.

Nominally the cont flag indicates when a $^{13}$C line should provide a measurement (cont flag = 1) or if it can only provide a limit (cont flag = 0).  In theory we would expect that for a given temperature and \xh{C}, at some threshold, stars with higher \cisotope{} should be lower limits, and below that threshold we have measurements.  However we find that there is some overlap, such that BACCHUS measures similar \cisotope{} ratios for stars of the same temperature and \xh{C} but with a mix of cont = 0 and 1 set.

So we have chosen to set the lower limit relationship coefficient $B$ such that there are approximately an equal number of stars with cont = 1 but \cisotope{} ratios that are above the lower limit relation as there are stars with cont = 0 but BACCHUS \cisotope{} ratios for the $^{13}$C$^{16}$O 16530 \AA{} feature that lie below the lower limit relation.  For the cont = 1 stars, these are stars where, based on the line flux, we should be able to measure \cisotope{} ratios, but according to our abundance based relation should be lower limits.  On the other hand, for the cont = 0 stars, these are stars that the flux limit indicates should be lower limits, but BACCHUS's measurements suggest that they would have low enough \cisotope{} ratios that we should be able to measure them according to our lower limit relation.  By setting the coefficient $B$ such that the number of stars in each of these cases is approximately the same, we have an equal number of stars where we believe these lower limits should be over- and underestimated.

We then use this relation to set lower limits for our sample.  Any stars with combined \cisotope{} ratios below this relation are replaced with these limits, and all stars that have cont = 0 for the 16530 \AA{} line, but no other bad method flags, are given lower limits according to this relation.  Finally, the \cisotope{} measurements are not populated for any stars without C measurements, because it is critical to have measured C in order to provide a reliable \cisotope{} ratio. 

\subsection{Zn, Ge, Rb and Yb}
\begin{figure}
  \centering
  \includegraphics[angle=-90,scale=0.3,trim = 0.in 0.in 0.in 0.in, clip]{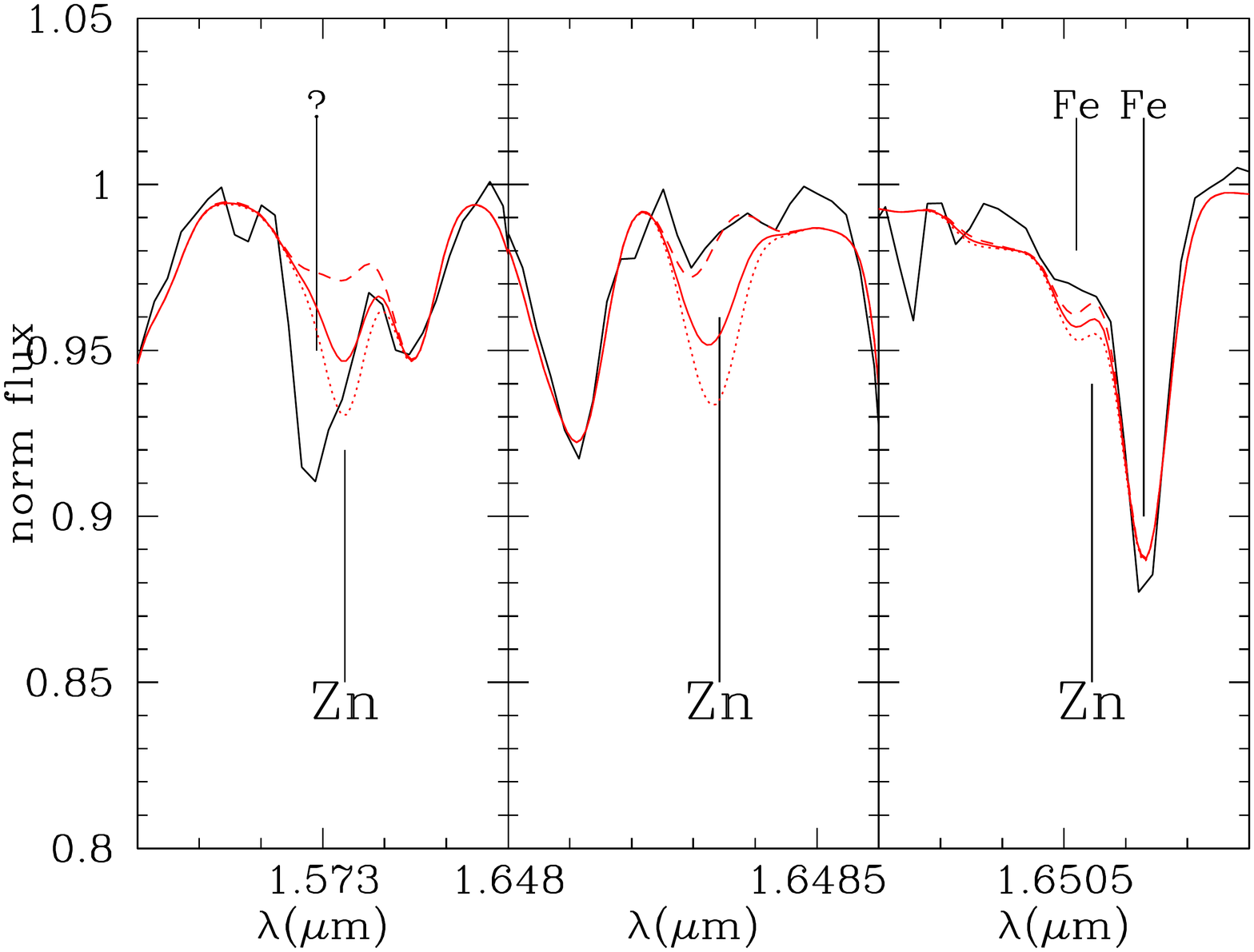}
  \caption{The strongest Zn main lines in the observed spectrum of 2M19220197+6121429 compared to the best fit model with the best fit Zn abundance, no Zn, and the best fit Zn abundance $+0.3$ dex (as color-coded in Figure \ref{fig:Nalines}).  However, the best fit Zn abundance does not seem appropriate for all of the features, and it appears that none of these line can be used to reliably measure Zn abundances in the whole sample. }
  \label{fig:Znlines}
\end{figure}

There are three Zn lines at 15730.4, 16483.4 and 16505.5 $\rm \AA$ in the APOGEE spectra that are theoretically strong enough to allow abundance measurement. However, as shown in Fig.~\ref{fig:Znlines}, all those lines suffer from major caveats which do not allow any measurement. The 15730.4 $\rm \AA$ line is the strongest but is severely blended by an unknown feature, which may have lead to an overestimation of this line's astrophysical $\log{gf}$. We note that this feature appears at all temperature, including for T$\rm _{eff} > 5000$ K which strongly suggests that it is an atomic line. The 16483.4 $\rm \AA$ line is cleaner but appear to have a largely overestimated strength in the models, such that it is not detectable in the spectra. Finally, the 16505.5 $\rm \AA$ line is the weakest and blended by two Fe lines.

Concerning Ge, Rb and Yb, the strongest known lines in APOGEE spectra -- and more generally in the H-band -- are located at respectively 16759.8, 15289.5 and  16498.4 $\rm \AA$. Unfortunately, those lines are very weak and following our upper limits procedure such as described in  Sec.~\ref{sec:upper_limits_relation}, very few stars pass the upper limits flag even at solar metallicity with the lowest intensity threshold. So we do not consider these elements further. We still stress that those elements can become clearly measurable in neutron-capture enhanced stars (particularly Yb), but as they represent a minority of stars in this general purpose catalog, and cannot be calibrated or treated in a similar way to the other elements, we have decided not to provide their abundances.

\section{Spectra Flag Definitions}
\label{app:spectra_flags}

\subsection{Sodium (Na)}
\subsubsection{Na Cont Flag}
\label{sec:na16388contflag}

As illustrated in Fig.~\ref{fig:Nalines}, both Na lines are affected by a CO blend. In order to control whether the modelling of the CO lines is satisfactory, we build an identical flag for both Na lines such that the difference between the modelling and the observation of the nearest CO clean line (at 16375 \AA{}) is less than 20\% of the depth of the 16388 \AA{} Na line.  This flag implies that the weaker the Na lines are, the better the modelling of the CO lines needs to be, for us to measure accurate Na abundances.

\subsubsection{Na 16388  \AA{} Blend Flag}
\label{sec:na16388blendflag}

At certain radial velocities ($\sim$ $-110$ -- $-60$ \kms{}) a relatively strong sky line aligns with the Na 16388 \AA{} line.  In many stars at these velocities this sky line is either subtracted so that the Na 16388 \AA{} line can be recovered or sky line is strong enough that these pixels have been masked in the APOGEE pipeline.  However, in some cases the sky line has been over-subtracted and not masked.  When this occurs, the flux uncertainty in the affected pixels are sometimes increased to reflect the higher uncertainty from the sky line subtraction, but BACCHUS does not account for the flux uncertainty in each pixel, so it will find an enhanced Na abundance from these sky contaminated lines.

To flag stars where this occurs, we have defined a blend flag for this line that attempts to identify cases where the uncertainties at the Na 16388 \AA{} line are larger than typical, or when stars have improperly subtracted sky lines at other pixels.  To flag cases where the flux uncertainties per pixel have been raised, we identify stars whose flux uncertainty is greater than 5\%.  

Because not all affected stars have elevated flux uncertainties, we also use two other points in the spectra that have near continuum flux and are at wavelengths where strong sky sit in stars at the same radial velocities where Na 16388 \AA{} is affected.  At these two wavelengths, 15431 \AA{} and 16128 \AA{}, we compare the observed flux to that of a BACCHUS model.  If the sky line at Na 16388 \AA{} is erroneously subtracted, we may also see significant differences between observations and the model at these wavelengths from other poorly subtracted sky lines.  To evaluate this, we calculate a mean and standard deviation of the differences between model and observed spectra at these wavelengths for stars with velocities between 0$-$100 \kms{} (which won't be affected by sky contamination), and flag all stars (at all radial velocities) whose differences at either of these wavelengths deviate from the calculated mean differences by more than 5$\sigma$.

Combining these two methods we create a final blend flag for Na 16388 \AA{} that records whether the uncertainty at the line center is large or if other spectral pixels that may be affected by sky contamination deviate significantly from their BACCHUS model.  Stars that may be negatively impacted by sky lines in Na 16388 \AA{} have blend = 0 and those that pass the flagging have blend = 1.

\subsection{Phosphorus (P)}
Phosphorus is among the most challenging elements to be measured in the APOGEE spectra because the lines are particularly weak and blended as illustrated in Fig.~\ref{fig:Plines}. 

\subsubsection{P 15711 \AA{}  Blend Flag}
The 15711 \AA{} line is blended by a CN line at 15706.9 \AA{}. To improve modelling of that CN line, we adjust the N abundance to this specific line before letting the code determine the P abundance of that line. The blend flag of the 15711 P line is raised as bad (= 0) only when the difference between the N abundance from the 15706.9 \AA{} CN line and the N abundance from an average of several good N lines is larger than 0.5 dex. In other words, this flag indicates when there has been a local adjustment of the blend.

\subsubsection{P 16482 \AA{}  Blend Flag}
The 16482 \AA{} P line is blended by a CO line. The procedure to control the potential impact of a poorly modelled CO is quite similar to the 15711 \AA{} line.  The carbon abundance is locally adjusted to match the CO feature at 16481 \AA{} before the P abundance is determined. We then create a blend flag for the 16482 \AA{} P line in a similar fashion as for the 15711 \AA{} blend flag, i.e. it is raised as bad (= 0) when the difference between carbon abundance of the 16481 \AA{} CO line and the mean carbon abundance derived from other good lines is larger than 0.2 dex. However, it turned out that this flag has never been raised all over the sample (implying that the 16481 \AA{} CO line is a good representative of the C abundance)  and we did not find useful to provide it.

\subsubsection{P 16482 \AA{}  Cont Flag}
Depending on the radial velocity of the star, the 16482 \AA{} line can near the edge of the CCD, where the quality of the spectra can be poor. The cont flag of this line is set to 0 (i.e., bad) when the line is within 15 \AA{} of the chip edge.

\subsection{Sulfur (S)}
\subsubsection{S Blend Flag}
As illustrated in Fig.~\ref{fig:Slines}, two of three lines used in this study are blended by an underlying OH line. While OH lines are negligible if present at all in the hottest stars ($\approx$ 5000~K), the strength of OH increases rapidly with decreasing temperature such that at 4000~K it dominates the absorption, preventing any S abundance measurement in nearly all such cool stars. In addition, this effect is also increased for low metallicity stars, where the O abundance is relatively higher than at solar metallicities.

In order to measure the S abundance only in cases where the contribution from OH lines is small, we have created a blend flag for the 15403 \AA{} and 15470 \AA{} lines such that the the flag is set as bad (= 0) if the observed flux at 15409 \AA{} (which corresponds to the wavelength of a clean OH line) is more than 5\% of the continuum.

Regarding the 15478 \AA{}, it is blended by Fe lines that we assumed to be always be well modelled. Consequently, we did not implement any special flagging for this line (hence always =1).

\subsection{Vanadium (V)}

\subsubsection{V 15924 \AA{} Cont Flag}
\label{sec:v15924contflag}
As illustrated in Fig.~\ref{fig:Vlines}, the 15924 \AA{} line is the best line (in contrast to the 16137 \AA{}  severely blended and not used in our final combination, and the 16200 \AA{} line quite strong, but blended by CO lines).
Nevertheless, in hotter stars or very metal-poor stars, the V 15924 \AA{} line is relatively weak, so a slight mismatch in the BACCHUS normalization can introduce errors in the continuum placement that lead to erroneous V 15924 \AA{} measurements.  We have therefore defined a continuum flag to identify when the local observed continuum deviates significantly from the BACCHUS model, such that the poor normalization may lead to errors in the derivation of V abundances for this line. This flag is set as bad (= 0) if the average of the difference between the observed and model flux at 15924.9 $\pm$ 0.4 is more than 15\% of the depth of the observed flux in the core the the V 15924 \AA{} line.

\subsubsection{Cu Blend Flag}
Despite the fact that the Cu feature is blended on both its red and blue side (see Fig.~\ref{fig:Culine}) by a C and Fe line respectively, the derivation of Cu abundances is possible in many of the stars as long as the blends are appropriately modeled. To assure this, we have implemented a procedure as follows: the C atomic blend is locally adjusted by setting the carbon abundance to the ``\intmethod{}'' method value of the C 16004.9 \AA{} line. If this local carbon abundance is more than 0.4 dex different from the mean carbon abundance derived from the average of the many lines we use, then we consider this local carbon measurement unreliable.  And therefore we consider the the model fitting for Cu suspect, and set the Cu blend flag to 0.  

\subsection{Cerium (Ce)}
Four Ce lines have been used in this study (Fig.~\ref{fig:Celines}). All of the lines are blended, but from our line-by-line comparison tests, it appears that for three of them the blends are sufficiently weak and/or well-modelled so that the Ce measurements are not severely affected. However, the fourth line, Ce 15784 \AA{}, is affected in some cases, for which we have built a specific procedure and flag.

\subsubsection{Ce 15784 \AA{} Blend Flag}
As shown in the left panel of Fig.~\ref{fig:Celines}, the Ce 15784 \AA{} line is blended on its blue side by a CO line. To minimize the impact of an inaccurate modelling of this line, we adjust the carbon abundance to match the core of the 15783.9  \AA{} CO line with the ``\wln{}'' method. The use of the ``\wln{}'' method is particularly relevant for this CO line as its core is clearly distinct from the Ce line 15784 \AA{} and thus offer the best leverage to accurately constrain the fit of the CO line. Nevertheless, whenever the carbon abundance differs from the mean carbon abundance by more than 0.3 dex, then the Ce 15784 \AA{} blend flag is set to 0.

\subsection{Neodymium (Nd)}
Along with P, Nd is a very challenging element to measure precisely because the lines are weak and blended. Fortunately, there are several lines present in the APOGEE spectra, which has allowed some robustness in the Nd measurement in our sample (Fig.~\ref{fig:Ndlines}). While we developed a similar procedure for locally fitting the blending feature such as for P, Cu or Ce, the blends do not appear to significantly impact the results, implying that the blends were reasonably well modeled, so we have not implemented any blend flags for these Nd lines like these other elements. Nevertheless, we have still developed a procedure to control for the quality of the continuum placement of the 16054 \AA{} line.

\subsubsection{Nd 16053 \AA{} Cont Flag}
Similar to what has been done for the Na and V cont flags: if the difference in the synthesis and the observed spectrum in the pseudo-continuum region 16054.2-16054.7 \AA{} is more than 30\% of the average flux at the wavelength of the 16053 \AA{} Nd line, then the cont flag is set to 0 (i.e., the fit is likely to be poor). 

\subsection{\cisotope{} Isotopic Ratio}
\subsubsection{\cisotope{} Cont Flags}

\begin{deluxetable}{l c c c c}
\tablewidth{0pt}
\tablecolumns{5}
\tablecaption{$\rm ^{13}C$ lines flux threshold} \label{tab:13Clinesthreshold}
\tablehead{\colhead{Wavelength ($\rm \AA$)} & \colhead{Threshold} }
\startdata
15641.7 & 0.85 \\
16121.3 & 0.97 \\
16323.4 & 0.7  \\
16326.0 & 0.98 \\
16327.3 & 0.985 \\
16530.0 & 0.982 \\
16741.2 & 0.97 \\
16744.7 & 0.98 \\
\enddata
\end{deluxetable}

In addition to an upper limit relation, we incorporate a flux limit flagging into the continuum flags for $\rm ^{13}C$.  For each line we use a flux threshold value for each of the $\rm ^{13}C$ features when considering whether to flag each line.  If the observed, normalised flux around the central wavelength (averaged within a window of $\pm$~0.5~$\AA)$) is higher than the threshold, the corresponding $^{13}C$ continuum flag is raised (flag = 0), and if the flux is lower than the threshold, the continuum flag for that feature is set to flag = 1.  As indicated in Appendix \ref{app:comb_settings} we require that cont flag = 1 for all $\rm ^{13}C$, providing limit flagging at the flux level rather than at the abundances.  This, however, means that for stars where all $\rm ^{13}C$ features are flagged as not passing the flux limits, no \cisotope{} limits are provided.

\section{Upper Limit Relations}
\label{app:upper_limits}

\startlongtable
\begin{deluxetable*}{l c c c c c c c c c c c}
\tabletypesize{\scriptsize}
\tablewidth{0pt}
\tablecolumns{12}
\tablecaption{Upper Limit Relation Constants \label{tab:upper_limits}}
\tablehead{\colhead{Element} & \colhead{Line} & \colhead{$A_1$} & \colhead{$B_1$} & \colhead{$A_2$} & \colhead{$B_2$} & \colhead{$A_3$} & \colhead{$B_3$} & \colhead{$A_4$} & \colhead{$B_4$} & \colhead{$A_5$} & \colhead{$B_5$} \\ \colhead{} & \colhead{\AA{}} & \colhead{dex/10$^{4}$ K} & \colhead{} & \colhead{dex/10$^{4}$ K} & \colhead{} & \colhead{dex/10$^{4}$ K} & \colhead{} & \colhead{dex/10$^{4}$ K} & \colhead{} & \colhead{dex/10$^{4}$ K} & \colhead{}}
\startdata
C & 15578.0 & 14.912 & 0.825 & 15.490 & 0.818 & 28.277 & -4.128 & 42.768 & -9.773 & 57.914 & -15.676 \\
C & 15775.5 & 11.826 & 1.902 & 14.417 & 1.055 & 15.009 & 0.947 & 14.754 & 1.153 & 19.638 & -0.712 \\
C & 15783.9 & 5.979 & 4.358 & 6.444 & 4.438 & 7.192 & 4.335 & 7.543 & 4.331 & 7.534 & 4.441 \\
C & 15978.7 & 11.197 & 2.097 & 13.251 & 1.468 & 14.614 & 1.051 & 14.729 & 1.109 & 14.122 & 1.439 \\
C & 16004.9 & 0.523 & 6.252 & 1.028 & 6.381 & 1.232 & 6.535 & 1.146 & 6.738 & 0.976 & 6.943 \\
C & 16021.7 & 1.517 & 6.020 & 2.350 & 6.018 & 2.364 & 6.189 & 2.217 & 6.392 & 2.240 & 6.513 \\
C & 16185.5 & 10.737 & 2.310 & 12.488 & 1.798 & 13.844 & 1.381 & 14.314 & 1.293 & 14.332 & 1.367 \\
C & 16397.2 & 10.992 & 2.212 & 12.616 & 1.773 & 13.793 & 1.450 & 14.065 & 1.457 & 13.657 & 1.717 \\
C & 16481.5 & 14.272 & 1.447 & 24.747 & -2.390 & 38.484 & -7.691 & 58.048 & -15.131 & 80.308 & -23.545 \\
C & 16614.0 & 11.002 & 2.279 & 12.364 & 1.976 & 13.000 & 1.895 & 13.004 & 2.025 & 13.032 & 2.122 \\
C & 16836.0 & 11.091 & 2.350 & 12.615 & 2.017 & 12.755 & 2.147 & 12.611 & 2.343 & 13.145 & 2.245 \\
C & 16890.4 & 0.246 & 6.017 & -0.121 & 6.372 & -0.485 & 6.725 & -0.807 & 7.053 & -1.037 & 7.325 \\
C & 17063.0 & 11.096 & 2.448 & 12.509 & 2.190 & 12.096 & 2.545 & 11.531 & 2.915 & 17.408 & 0.686 \\
C & 17448.6 & 12.033 & 2.718 & 42.485 & -9.100 & 77.479 & -22.768 & 124.107 & -40.439 & 170.862 & -58.125 \\
C & 17456.0 & 41.070 & -8.492 & 123.426 & -39.866 & 213.908 & -72.611 & 304.937 & -104.691 & 394.770 & -136.108 \\
N & 15119.0 & 8.484 & 2.636 & 8.359 & 3.000 & 7.671 & 3.490 & 7.495 & 3.715 & 7.627 & 3.775 \\
N & 15210.2 & 8.863 & 2.565 & 8.143 & 3.181 & 7.678 & 3.582 & 7.772 & 3.689 & 8.070 & 3.675 \\
N & 15222.0 & 8.495 & 2.633 & 8.394 & 2.984 & 7.702 & 3.473 & 7.512 & 3.702 & 7.634 & 3.765 \\
N & 15228.8 & 8.848 & 2.564 & 8.201 & 3.149 & 7.733 & 3.552 & 7.811 & 3.667 & 8.091 & 3.664 \\
N & 15242.5 & 8.576 & 2.648 & 8.153 & 3.142 & 7.551 & 3.598 & 7.512 & 3.764 & 7.734 & 3.785 \\
N & 15251.8 & 8.777 & 2.557 & 8.335 & 3.051 & 7.755 & 3.496 & 7.711 & 3.664 & 7.938 & 3.682 \\
N & 15309.0 & 8.679 & 2.678 & 7.862 & 3.345 & 7.485 & 3.711 & 7.669 & 3.781 & 7.975 & 3.768 \\
N & 15317.6 & 8.774 & 2.516 & 8.618 & 2.879 & 7.968 & 3.346 & 7.795 & 3.568 & 7.927 & 3.626 \\
N & 15363.5 & 8.561 & 2.574 & 8.667 & 2.826 & 7.996 & 3.301 & 7.745 & 3.557 & 7.801 & 3.649 \\
N & 15410.5 & 8.748 & 2.796 & 7.844 & 3.513 & 8.119 & 3.604 & 8.506 & 3.584 & 8.658 & 3.631 \\
N & 15447.0 & 8.676 & 2.742 & 7.683 & 3.487 & 7.517 & 3.762 & 7.835 & 3.774 & 8.141 & 3.758 \\
N & 15462.4 & 8.687 & 2.731 & 7.720 & 3.464 & 7.535 & 3.749 & 7.836 & 3.768 & 8.140 & 3.753 \\
N & 15466.2 & 8.687 & 2.731 & 7.720 & 3.464 & 7.538 & 3.747 & 7.841 & 3.765 & 8.143 & 3.752 \\
N & 15495.0 & 8.654 & 2.746 & 7.685 & 3.479 & 7.476 & 3.771 & 7.768 & 3.791 & 8.092 & 3.766 \\
N & 15514.0 & 8.604 & 2.669 & 8.056 & 3.214 & 7.507 & 3.646 & 7.525 & 3.783 & 7.781 & 3.787 \\
N & 15581.0 & 8.980 & 2.501 & 8.377 & 3.061 & 7.944 & 3.448 & 8.022 & 3.565 & 8.293 & 3.566 \\
N & 15636.5 & 8.690 & 2.787 & 7.655 & 3.548 & 7.668 & 3.744 & 8.063 & 3.719 & 8.365 & 3.698 \\
N & 15659.0 & 8.699 & 2.799 & 7.683 & 3.555 & 7.787 & 3.713 & 8.203 & 3.679 & 8.474 & 3.671 \\
N & 15706.9 & 8.101 & 3.275 & 8.209 & 3.577 & 8.742 & 3.544 & 8.826 & 3.646 & 8.542 & 3.883 \\
N & 15708.5 & 8.101 & 3.275 & 8.209 & 3.577 & 8.742 & 3.544 & 8.826 & 3.646 & 8.541 & 3.884 \\
N & 15825.7 & 8.546 & 2.630 & 8.441 & 2.980 & 7.819 & 3.438 & 7.677 & 3.646 & 7.820 & 3.701 \\
O & 15373.5 & 5.668 & 6.015 & 7.566 & 5.373 & 8.231 & 5.161 & 8.779 & 4.994 & 9.328 & 4.828 \\
O & 15391.0 & 5.416 & 6.079 & 7.112 & 5.486 & 8.125 & 5.152 & 8.665 & 4.974 & 9.205 & 4.795 \\
O & 15569.0 & 5.827 & 5.819 & 6.470 & 5.702 & 7.785 & 5.243 & 8.291 & 5.092 & 8.761 & 4.936 \\
O & 15719.7 & 5.832 & 5.829 & 6.624 & 5.650 & 7.898 & 5.211 & 8.384 & 5.063 & 8.871 & 4.902 \\
O & 15778.5 & 5.665 & 5.927 & 6.763 & 5.610 & 7.952 & 5.210 & 8.454 & 5.050 & 8.956 & 4.887 \\
O & 16052.9 & 5.807 & 5.813 & 6.303 & 5.772 & 7.594 & 5.326 & 8.149 & 5.154 & 8.586 & 5.014 \\
O & 16055.5 & 5.807 & 5.813 & 6.303 & 5.772 & 7.594 & 5.326 & 8.149 & 5.154 & 8.586 & 5.014 \\
O & 16650.0 & 5.781 & 5.918 & 7.393 & 5.392 & 8.259 & 5.112 & 8.845 & 4.921 & 9.430 & 4.731 \\
O & 16704.8 & 5.673 & 5.872 & 6.225 & 5.808 & 7.476 & 5.381 & 8.070 & 5.191 & 8.479 & 5.062 \\
O & 16714.5 & 5.686 & 5.861 & 6.210 & 5.812 & 7.456 & 5.387 & 8.066 & 5.192 & 8.487 & 5.060 \\
O & 16872.0 & 5.710 & 5.820 & 6.056 & 5.859 & 7.218 & 5.459 & 7.978 & 5.208 & 8.349 & 5.089 \\
O & 16909.4 & 5.666 & 5.887 & 6.302 & 5.784 & 7.594 & 5.345 & 8.117 & 5.180 & 8.547 & 5.046 \\
Na & 16373.9 & 5.295 & 3.106 & 5.243 & 3.418 & 4.766 & 3.806 & 4.694 & 3.985 & 4.844 & 4.046 \\
Na & 16388.8 & 4.650 & 3.060 & 5.310 & 3.112 & 5.397 & 3.252 & 5.185 & 3.468 & 4.962 & 3.672 \\
P & 15711.6 & 1.227 & 4.558 & 0.254 & 5.250 & 0.023 & 5.522 & -0.116 & 5.796 & -0.215 & 6.036 \\
P & 16482.9 & 2.198 & 3.866 & 1.280 & 4.673 & 0.651 & 5.088 & 0.648 & 5.229 & 0.664 & 5.384 \\
S & 15403.5 & -0.946 & 6.361 & -1.481 & 6.912 & -2.438 & 7.583 & -2.881 & 7.998 & -3.139 & 8.299 \\
S & 15422.3 & -1.311 & 6.127 & -1.426 & 6.604 & -1.915 & 7.054 & -2.736 & 7.606 & -3.344 & 8.052 \\
S & 15469.8 & -0.158 & 6.154 & -1.072 & 6.845 & -1.574 & 7.350 & -1.721 & 7.634 & -1.975 & 7.914 \\
S & 15475.6 & -0.374 & 5.970 & -0.626 & 6.423 & -1.479 & 7.012 & -2.053 & 7.472 & -2.360 & 7.792 \\
S & 15478.5 & -0.374 & 5.970 & -0.626 & 6.423 & -1.479 & 7.012 & -2.053 & 7.472 & -2.360 & 7.792 \\
V & 15924.8 & 11.592 & -2.154 & 12.350 & -2.180 & 12.732 & -2.148 & 12.675 & -1.988 & 12.375 & -1.754 \\
V & 16137.3 & 11.045 & -1.471 & 11.321 & -1.255 & 10.790 & -0.859 & 10.661 & -0.686 & 10.733 & -0.615 \\
V & 16200.1 & 11.768 & -2.062 & 12.548 & -2.073 & 12.489 & -1.859 & 12.040 & -1.537 & 11.768 & -1.320 \\
Cu & 16005.5 & 3.512 & 1.393 & 2.944 & 1.950 & 3.655 & 1.819 & 4.139 & 1.750 & 4.144 & 1.870 \\
Cu & 16006.0 & 3.512 & 1.393 & 2.944 & 1.950 & 3.655 & 1.819 & 4.139 & 1.750 & 4.144 & 1.870 \\
Ce & 15784.8 & 9.574 & -3.934 & 10.135 & -3.901 & 9.918 & -3.624 & 9.893 & -3.477 & 10.021 & -3.422 \\
Ce & 16376.5 & 10.590 & -4.227 & 10.393 & -3.855 & 10.529 & -3.730 & 10.880 & -3.742 & 11.066 & -3.707 \\
Ce & 16595.2 & 10.405 & -3.947 & 10.526 & -3.695 & 11.003 & -3.706 & 11.116 & -3.611 & 11.005 & -3.450 \\
Ce & 16722.5 & 9.727 & -3.563 & 9.638 & -3.219 & 9.990 & -3.169 & 10.155 & -3.096 & 10.165 & -2.988 \\
Nd & 15368.1 & 8.264 & -2.352 & 8.376 & -2.055 & 7.537 & -1.499 & 7.176 & -1.213 & 7.313 & -1.166 \\
Nd & 16053.6 & 9.867 & -3.079 & 9.474 & -2.616 & 8.941 & -2.203 & 8.918 & -2.056 & 9.238 & -2.072 \\
Nd & 16262.0 & 9.355 & -2.801 & 8.989 & -2.331 & 8.424 & -1.899 & 8.373 & -1.736 & 8.691 & -1.754 \\
\enddata
\end{deluxetable*}

In Table \ref{tab:upper_limits} we present the constants used in our upper limit relations of the form \logeps{X}$_{\rm lim} = A_{\rm t} \cdot T_{\rm eff} + B_{\rm t}$ where t is the chosen \% threshold of the continuum.  We have calculated these relations for continuum threshold levels of 1, 2, 3, 4, and 5\%, and have reported the $A_{\rm t}$ in dex per 10$^4$ K in Table \ref{tab:upper_limits}.

\end{appendix}
\end{document}